\documentclass[fleqn,usenatbib,useAMS]{mnras}

\usepackage{graphicx}	% Including figure files
\usepackage{amsmath}	% Advanced maths commands
\usepackage{amssymb}	% Extra maths symbols
\usepackage{multicol}        % Multi-column entries in tables
\usepackage{bm}		% Bold maths symbols, including upright Greek
\usepackage{pdflscape}	% Landscape pages
\usepackage{xcolor,ulem}
\usepackage{adjustbox}
\newcommand{\be}{\begin{equation}}
\newcommand{\ee}{\end{equation}}
\newcommand{\ba}{\begin{eqnarray}}
\newcommand{\ea}{\end{eqnarray}}

\newcommand\restr[2]{{% we make the whole thing an ordinary symbol
  \left.\kern-\nulldelimiterspace % automatically resize the bar with \right
  #1 % the function
  \vphantom{\big|} % pretend it's a little taller at normal size
  \right|_{#2} % this is the delimiter
  }}
  
\definecolor{blazeorange}{rgb}{1.0, 0.4, 0.0}
\definecolor{seagreen}{rgb}{0.18, 0.55, 0.34}
\definecolor{rufous}{rgb}{0.66, 0.11, 0.03}
\definecolor{royalfuchsia}{rgb}{0.79, 0.17, 0.57}
\definecolor{scarlet}{rgb}{1.0, 0.13, 0.0}
\definecolor{royalpurple}{rgb}{0.47, 0.32, 0.66}
\definecolor{darkblue}{rgb}{0, 0, 0.66}

\usepackage[T1]{fontenc}
\usepackage{ae,aecompl}
\usepackage{newtxtext,newtxmath}
\usepackage{ulem}

\title[Off-axis sGRBs]{X-ray Afterglow limits on the viewing angles of short gamma-ray bursts}

\author[B. O'Connor, P. Beniamini and R. Gill]{
Brendan O'Connor,$^{1}$\thanks{E-mail: boconno2@andrew.cmu.edu}\thanks{McWilliams Fellow}
Paz Beniamini,$^{2,3,4}$ 
\&
Ramandeep Gill$^{5,3}$
\\
\\
$^{1}$McWilliams Center for Cosmology and Astrophysics, Department of Physics, Carnegie Mellon University, Pittsburgh, PA 15213, USA \\ 
$^2$Department of Natural Sciences, The Open University of Israel, P.O Box 808, Ra'anana 43537, Israel\\
$^3$Astrophysics Research Center of the Open university (ARCO), The Open University of Israel, P.O Box 808, Ra'anana 43537, Israel  \\
$^4$Department of Physics, The George Washington University, Washington, DC 20052, USA \\
$^5$ Instituto de Radioastronom\'ia y Astrof\'isica, Universidad Nacional Aut\'onoma de M\'exico, Antigua Carretera a P\'atzcuaro $\#$ 8701,  Ex-Hda. San Jos\'e de la \\ Huerta, Morelia, Michoac\'an, C.P. 58089, M\'exico 
}

\date{Accepted XXX. Received YYY; in original form ZZZ}

\pubyear{2024}

\begin{document}
\label{firstpage}
\pagerange{\pageref{firstpage}--\pageref{lastpage}}
\maketitle

\begin{abstract}
The behavior of a short gamma-ray burst (sGRB) afterglow lightcurve can reveal the angular structure of the relativistic jet and constrain the observer's viewing angle $\theta_\textrm{obs}$. The observed deceleration time of the jet, and, therefore, the time of the afterglow peak, depends on the observer's viewing angle. A larger viewing angle leads to a later peak of the afterglow and a lower flux at peak. We utilize the earliest afterglow detections of 58 sGRBs detected with the \textit{Neil Gehrels Swift Observatory} X-ray Telescope to constrain the ratio of the viewing angle $\theta_\textrm{obs}$ to the jet's core $\theta_\textrm{c}$. We adopt a power-law angular jet structure in both energy $E(\theta)\propto\theta^{-a}$ and Lorentz factor $\Gamma(\theta)\propto\theta^{-b}$ beyond the core. We find that either sGRBs are viewed within $\theta_\textrm{obs}/\theta_\textrm{c}<1$ or the initial Lorentz factor of material in their jet's core is extremely high ($\Gamma_0>500$). If we consider tophat jets, we constrain 90\% of our sample to be viewed within $\theta_\textrm{obs}/\theta_\textrm{c}<1.06$ and 1.15 for our canonical and conservative afterglow scenarios. For a subset of events with measurements of the jet break, we can constrain $\Gamma_0\theta_\textrm{c}\gtrsim 30$. This confirmation that cosmological sGRBs are viewed either on-axis or very close to their jet's core has significant implications for the nature of the prompt gamma-ray production mechanism and for the rate of future sGRB detections coincident with gravitational waves (GWs), implying that they are extremely rare.
\end{abstract}

%%%%%%%%%%%%%%%%%%%%%%%%%%%%%%%%%%%%%%%%%%%%%%%%%%

%%%%%%%%%%%%%%%%% BODY OF PAPER %%%%%%%%%%%%%%%%%%
%%%%%%%%%%%%%%%%%%%%%%%%%%
\section{Introduction}
\label{sec:Intro}

It is now well established that the merger of two neutron stars is capable of launching a relativistic jet \citep{Abbott+17-GW170817A-MMO,Savchenko2017,Goldstein2017}. Internal processes (such as shocks or magnetic reconnection; see, e.g., \citealt{Piran2004,KumarZhang2015,Pe'er2015} for reviews) within the jet produce a gamma-ray burst (GRB). Short duration gamma-ray bursts (sGRBs) are brief signals of high energy radiation canonically lasting for $<$\,$2$ s \citep{Kouveliotou1993}, though recent studies have highlighted longer duration events (main emission phases of $\sim$\,$10$\,$-$\,$40$ s) 
likely arising from compact object mergers \citep[e.g., GRBs 060614, 211211A and 230307A;][]{DellaValle2006,Galyam2006,Yang2015,Rastinejad2022,Troja2022,Yang2022kn211211A,Gompertz2023,Levan2023,Yang2023,Gillanders2023,Dichiara2023}. These longer duration events show some similarities to the sample of extended emission sGRBs ($\lesssim$\,$100$ s) known for almost two decades \citep{Norris2006,Gehrels2006,Norris2010,Bostanci2013,Kaneko2015}, but do not display the same initial short ($<$\,$2$ s) pulse and instead have longer lasting main emission phases.  

The interaction between the relativistic jet and its surrounding environment produces a broadband (radio to very high energy gamma-rays) counterpart known as the afterglow. Early theoretical modeling of GRB afterglows found that they were consistent with arising from highly collimated ``tophat''-like jets \citep{Rhoads1999,Frail2001,Panaitescu2002,Yost2003,Bloom2003}. Despite this, many early studies expanded the GRB afterglow theory to include jets viewed outside of their core (i.e., off-axis), and introduced an angular structure to the energy and Lorentz factor profiles that modified the overall lightcurve behavior \citep{Meszaros1998,Rossi02,Granot2002jet,Panaitescu2003,Kumar2003}. The ground-breaking discovery of GW170817 \citep{Abbott+17-GW170817A-MMO,Savchenko2017,Goldstein2017}  
conclusively demonstrated a complex angular jet structure that is not adequately modeled by the tophat assumption, and was well matched by theoretical predictions for an off-axis structured jet \citep{Troja2017,Lamb2017jet,Lazzati2018,Resmi2018,Mooley2018,D'Avanzo2018,Alexander2018,Xie2018,Margutti2018,GG2018,Ghirlanda2019,Troja2020}. This discovery opened an entirely new avenue to studying the GRB phenomenon.

While GW170817 was a revolutionary breakthrough for the field of GRBs and multi-messenger astronomy, there still remain open questions. One of the most puzzling aspects of the discovery was the detection of a short-lived prompt gamma-ray signal (GRB 170817A; \citealt{Savchenko2017,Goldstein2017}) despite the far off-axis observer viewing angle ($15$\,$-$\,$30$ deg; \citealt{Mooley2018,Mooley2022,Ghirlanda2022,Hotokezaka+19,Fernandez2022,Makhathini2021ApJ,Hajela2022,Balasubramanian2022,Govreen-Segal2023,Ryan2023}). While it has been argued that this signal was produced by less energetic (line-of-sight) material due to the angular structure (the wings) of the GRB jet  \citep{Ioka2018,Kathirgamaraju2018,Lamb2017jet,Granot2017}, such an interpretation would potentially require that a significant fraction of the observed sGRB population are viewed from angles far outside of their cores. However, this leads to inconsistencies with several key observed properties of those GRBs \citep{BeniaminiNakar2019,Beniamini2019structuredjet}. Therefore, it is plausible that GRB 170817A was produced by a different mechanism than in a typical sGRB, e.g., cocoon shock breakout \citep{Kasliwal2017,Lazzati2017,Salafia2018,Bromberg2018,Gottlieb2018,Nakar2018,Matsumoto2019a}. This has significant implications for the rate of joint detections of GRBs and gravitational wave (GW) events \citep{Beniamini2019structuredjet,Howell2019,Mogushi2019,Saleem2020,Yu2021,Colombo2022,Ronchini2022,Patricelli2022,Salafia2023,Bhattacharjee2024}.

In the small angle limit, the key afterglow observables depend only on the ratio of the viewing angle to the jet's core half-opening angle, $\theta_\textrm{obs}/\theta_\textrm{c}$ \citep{BGG2020,Nakar-Piran-21}. 
In this work, we utilize a sample of sGRB X-ray lightcurves from the \textit{Neil Gehrels Swift Observatory} (hereafter \textit{Swift}; \citealt{Gehrels2004}) in order to investigate this ratio. This method allows us to measure the maximum off-axis observer angle ($\theta_\textrm{obs}/\theta_\textrm{c}$) for a sample of cosmological sGRBs. We consider both tophat and structured jets. 
We then perform X-ray and gamma-ray detectability simulations of a population of sGRBs for comparison to the off-axis detection of GRB 170817A. 

The paper is organized as follows. In \S \ref{sec: obs}, we present our sample of sGRBs and their selection criteria.
In \S \ref{sec: methods}, we introduce the theoretical constraints to the viewing angle that can be derived based on the observations. The results for our sample are then discussed (\S \ref{sec: results}), and their implications for the rate of sGRBs and the production of gamma-rays from GRB 170817A are introduced (\S \ref{sec: discussion}). We then state our conclusions regarding the typical viewing angle for cosmological sGRBs in \S \ref{sec: conclusions}. 
Throughout the work we adopt flat cosmology with parameters $H_0=67.4$ and $\Omega_\Lambda=0.685$ \citep{Planck2018}. 

\section{Observations: Our Sample}
\label{sec: obs}

%\subsection{Sample} %% RG: DO WE NEED TO HAVE A LONE SUB-SECTION HERE?

%\subsubsection{Short GRBs}

\begin{figure}
    \centering
\includegraphics[width=\columnwidth]{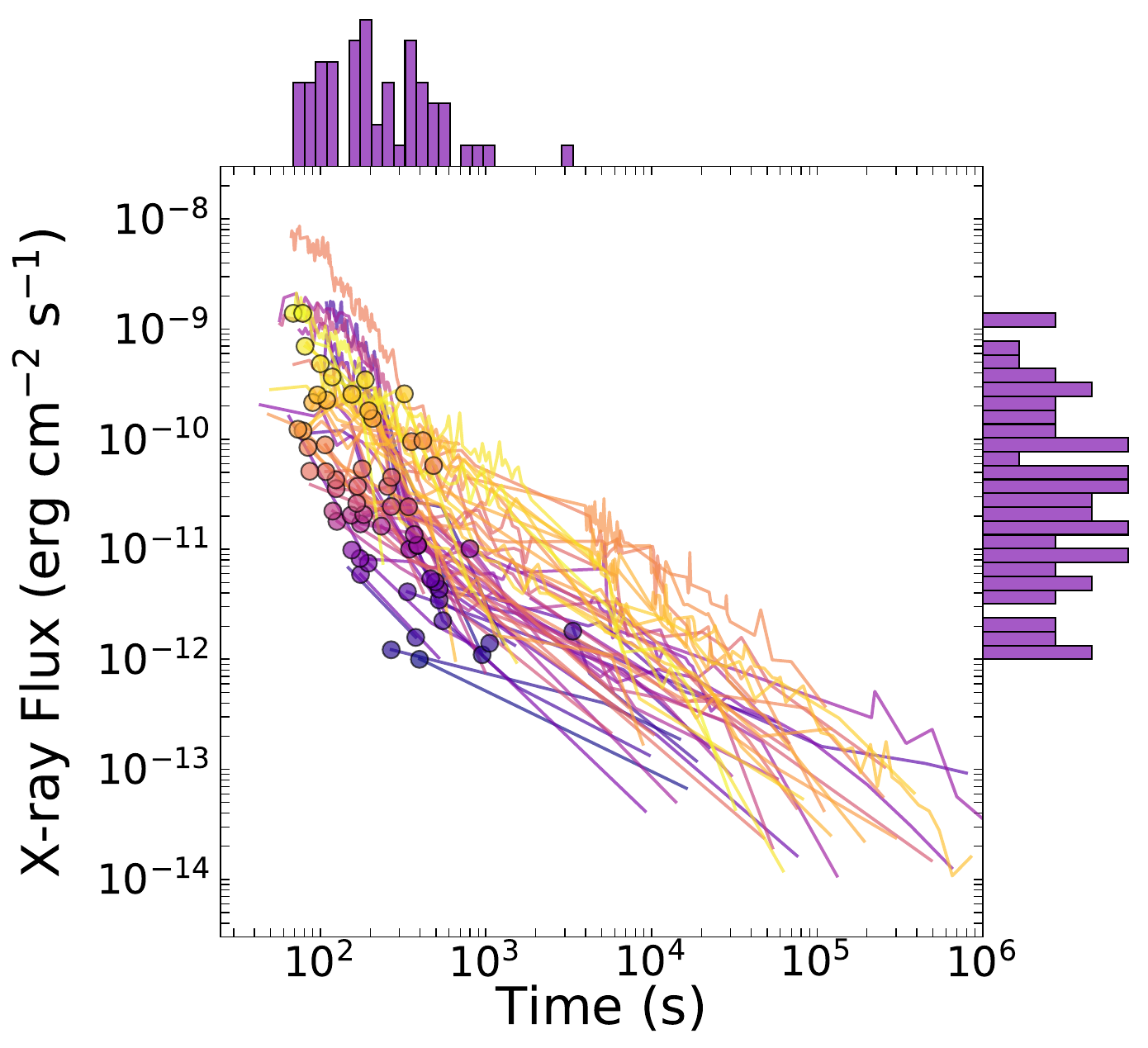}
\vspace{-0.5cm}
    \caption{Observer-frame X-ray lightcurves ($0.3$\,$-$\,$10$ keV) for \textit{Swift} sGRBs in our sample. The filled circles represent our constraints on the time $t_\textrm{o}$ (upper limit) and flux $F_\textrm{o}$ (lower limit), and the histograms on both axes track the distribution of these values for our sample. In general our sample has $t_\textrm{o}$\,$\lesssim$\,$3600$ s and $F_\textrm{o}$\,$\gtrsim$\,$10^{-12}$ erg cm$^{-2}$ s$^{-1}$.
    }
    \label{fig:lightcurves}
\end{figure}

Following \citet{OConnor2020}, we select our sample of sGRBs from those detected by the \textit{Swift} Burst Alert Telescope \citep[BAT;][]{Barthelmy2005} with $T_{90}<2$ s as of January 1, 2024. 
We require that each event is also detected by the \textit{Swift} X-ray Telescope \citep[XRT;][]{Burrows2005}. 
We utilized the \textit{Swift}-XRT GRB Lightcurve Repository\footnote{\url{https://www.swift.ac.uk/xrt_curves/}} to retrieve their X-ray lightcurves, and have selected against events contaminated by early steep decline or internal plateaus as such phenomena likely require emission radii that are significantly smaller than the external shock radius, from which the bulk of the afterglow is produced (see also \citealt{OConnor2020} for selection criteria). Each of these events is seen to decay following the initial detection by  \textit{Swift}/XRT. 
We select the first X-ray detection that can be robustly classified as forward shock emission \citep{Meszaros1997,Sari1998,Wijers1999}, and conservatively adopt the time $t_\textrm{o}$ and flux $F_\textrm{o}$ as our upper limit on the peak time of the afterglow $t_\textrm{p}$ and lower limit to the peak flux $F_\textrm{p}$ (Table \ref{Table_SGRB_DATA}). We have corrected the observed flux for absorption by using the energy conversion factor derived by the automatic \textit{Swift}/XRT spectral fitting in the \textit{Swift}-XRT GRB Spectrum Repository\footnote{\url{https://www.swift.ac.uk/xrt_spectra/}}. The distribution of $t_\textrm{o}$ and $F_\textrm{o}$ for our sample of events is shown in Figure \ref{fig:lightcurves}. We further retrieved the gamma-ray fluence $\phi_\gamma$ for each event from the \textit{Swift}/BAT GRB Catalog \citep{Lien2016}, see Table \ref{Table_SGRB_DATA}. 

%%%%%%%%%%%%%%%%%%%%%%%%%%%%%%%%
%%%%%%%%%%%%%%%%%%%%%%%%%%%%%%%%

In total we identify 58 events matching our selection criteria (see Table \ref{Table_SGRB_DATA}). 
In our sample, 30 events ($52\%$) have a measured host galaxy spectroscopic redshift (Table \ref{Table_SGRB_DATA}) with a median value of $\langle z\rangle$\,$\approx$\,$0.55$.
This is consistent with the median of $\sim0.6$ derived from recent sGRB population studies  \citep{OConnor2022,Nugent2022,Fong2022}. An additional 5 events have robust photometric redshifts from their host galaxies \citep{OConnor2022,Nugent2022,Fong2022}. For the 23 events in our sample without a measured redshift we adopt values in the range of $z$\,$=$\,$0.1$ to $0.5$. This choice is discussed further in \S \ref{sec: redshiftassumption}.

\section{Methods}
\label{sec: methods}

In the following section we discuss how early afterglow observations of sGRBs can be used to constrain the viewing angle. For clarity, we begin the discussion with the idealized case of tophat jets (\S \ref{methods: tophat jets}) before moving to the more realistic case of structured jets (\S \ref{methods: jet structure}).

\subsection{Off-axis tophat jets}
\label{methods: tophat jets}

In \citet{OConnor2020}, we assumed that all sGRBs in our sample were viewed within the core of their jets (i.e., on-axis), where the observer's viewing angle $\theta_\textrm{obs}$ is less than the half-opening angle of the jet's core $\theta_\textrm{c}$. In this work, we extend our methodology to consider off-axis tophat jets \citep{Rossi02,Dalal2002,Granot2002jet,EichlerGranot2006,vanEerten2011,Granot2018,Gill2019}.
 
Observing a tophat jet outside of its core ($\theta_{\rm obs}>\theta_c$) impacts 
the emission in a few different ways. The prompt gamma-ray fluence of the GRB is strongly Doppler 
suppressed as $\theta_{\rm obs}$ becomes increasingly larger than $\theta_c$ \citep[e.g.,][]{Kasliwal2017,Ioka2018}. In addition, since 
the emission is now Doppler de-boosted in the observer frame, the peak of the 
spectrum shifts to lower energies as compared to the on-axis one. During the afterglow 
phase, the main impact is to push the time of the lightcurve peak later, and to decrease 
the flux at the peak. For a far off-axis observer (with $\Delta\theta$\,$\gg$\,$\theta_\textrm{c}$, 
where $\Delta\theta = \theta_{\rm obs}-\theta_\textrm{c}$), the time of the peak 
further depends on whether lateral spreading of the jet is included \citep[e.g.,][]{Nakar2002}, 
whereas the peak flux is independent of the prescription for lateral spreading 
\citep[see, e.g.,][and \S \ref{sec:spreading}]{Lamb2021rev}. 
% These values likewise 
% depend strongly on the assumed angular structure of the jet. In \S \ref{sec:peaktimemethods} 
% and \S \ref{sec:peakflux} we consider tophat jets and in \S \ref{methods: jet structure} 
% we apply our methods to (power-law) structured jets. 
The following sections include  further details of these effects, where we present useful formulae that we employ in our analysis.

\subsubsection{Kinetic energy of the jet's core} 
\label{sec:energyconv}

Tophat jets are defined to have a constant energy per solid angle in a core of half-opening angle $\theta_\textrm{c}$ with no energetic material outside of this region: 
\begin{eqnarray}
\label{eqn:tophat}
E(\theta) = E_{0} \left\{ \begin{array}{ll} 1 & \theta \leq \theta_\textrm{c}, \\
0 &  \theta > \theta_\textrm{c}. 
\end{array} \right. 
\end{eqnarray}
Here $E_0$ represents the (isotropic equivalent) core energy and can refer to either kinetic or gamma-ray energy. 

For some assumed (or measured) redshift, the isotropic equivalent gamma-ray energy is derived using the observed fluence $\phi_\gamma$  ($15$\,$-$\,$150$ keV) from \textit{Swift}/BAT \citep{Lien2016} and converted to the $1$\,$-$\,$10,000$ keV energy range using a bolometric correction $k_\textrm{bol}$, as in \citet{OConnor2020}, and adopting the average values for the Band function \citep{Band1993} spectral fit obtained by \citet{Nava2011} for short GRBs.

When using the gamma-ray fluence to infer the jet energy, 
it is typically assumed that the observer is on-axis ($\theta_{\rm obs}\leq\theta_c$), 
in which case the emission is dominated by material along the line of sight. 
This assumption breaks down for off-axis observers, and we must consider the Doppler suppression of the line-of-sight gamma-rays relative to the core in order to truly test the viability of an off-axis interpretation. 
A similar Doppler de-boosting occurs for the spectral peak energy $E_p$. In order to be fully consistent with the off-axis assumption we utilize the observed peak energies for sGRBs as the observed (de-boosted) values \citep{Nava2011}. 

The Doppler suppression of the gamma-ray fluence becomes very large once $\Delta \theta$\,$\gg$\,$\Gamma_0^{-1}$ and can be approximated as \citep[e.g.,][]{Kasliwal2017,Granot2017,Ioka2018,Kathirgamaraju2018} 
\begin{eqnarray}
\label{eqn:supressed}
E_{\gamma,0}\approx E_{\gamma,\textrm{obs}} \left\{ \begin{array}{ll} 1 & \theta_\textrm{obs} \leq \theta_\textrm{c}, \\
q^{4} &  \theta_\textrm{c} < \theta_\textrm{obs} \leq 2\theta_\textrm{c}, \\
q^{6} (\theta_\textrm{c}\Gamma_0)^{-2} &  2\theta_\textrm{c} < \theta_\textrm{obs}, 
\end{array} \right. 
\end{eqnarray}
where $E_{\gamma,0}$ and $E_{\gamma,\textrm{obs}}$\,$=$\,$E_\gamma(\theta_\textrm{obs})$ are the core gamma-ray energy and observed gamma-ray energy, respectively. We have defined  $q$\,$=$\,$\Delta \theta\, \Gamma_0$, where $\Gamma_0$ is the initial bulk Lorentz factor of material at the jet's core.  
Thus for even modest values of $\Delta \theta$ the inferred core energy increases substantially. For angles $\theta_\textrm{obs}$\,$\approx$\,$\theta_\textrm{c}$ the small angle approximation is no longer valid and we utilize the full expression for the Doppler suppression $(1+q^2)^2$\,$\approx$\,$q^4$. 

\begin{figure}
    \centering
\includegraphics[width=\columnwidth]{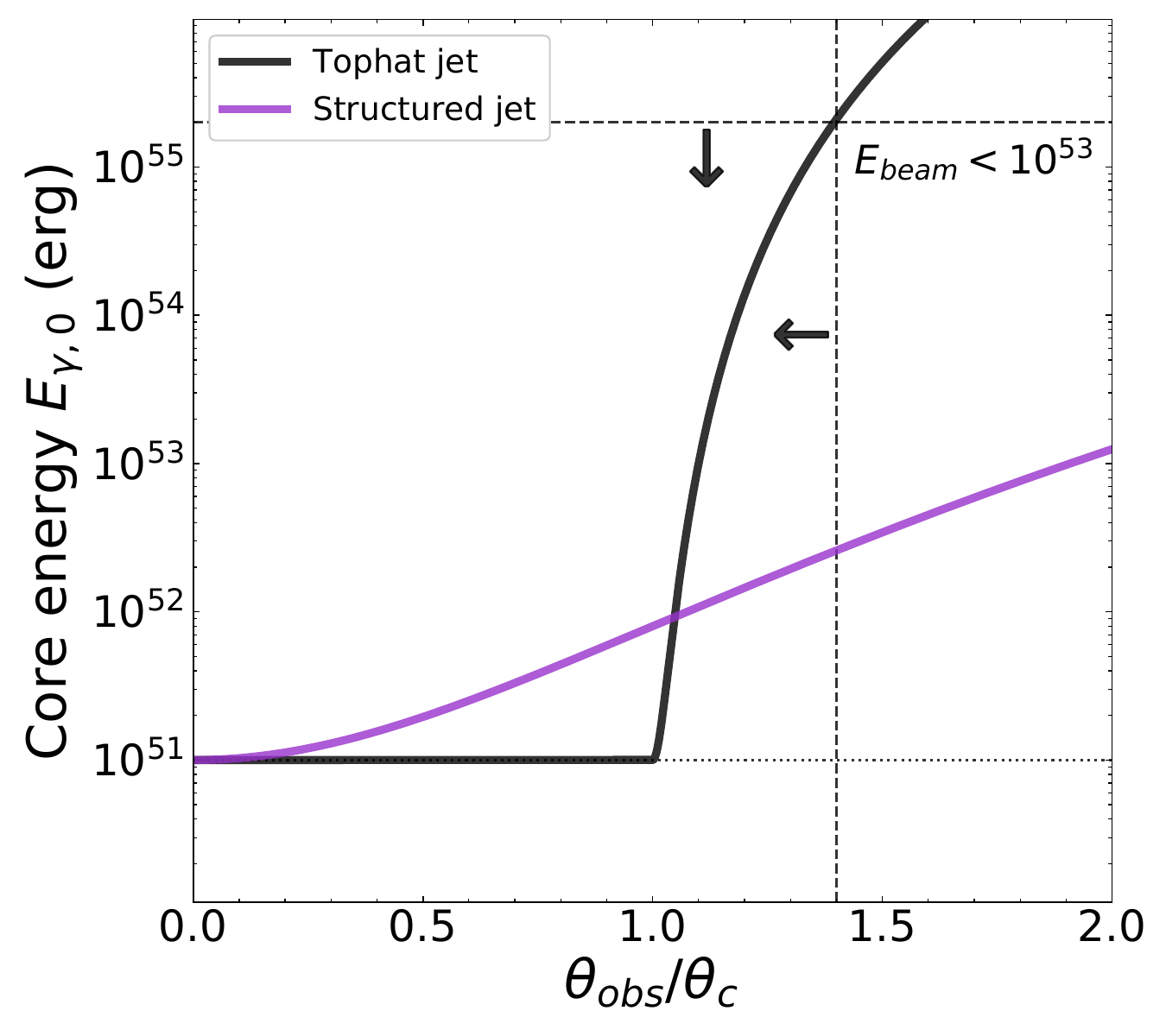}
\vspace{-0.5cm}
    \caption{The inferred isotropic equivalent gamma-ray energy at the jet's core for an observed (line-of-sight) energy of $E_\gamma(\theta_\textrm{obs})$\,$=$\,$10^{51}$ erg. We have adopted $\Gamma_0$\,$=$\,$300$ and $\theta_\textrm{c}$\,$=$\,$0.1$ rad. The assumed structured jet model \citep{GG2018,Gill2019} is a power-law that asymptotically approaches $\theta^{-a_\gamma}$, where $a_\gamma$\,$=$\,$6$ is assumed (for further details see \S \ref{methods: jet structure}). The horizontal dashed line marks a beaming corrected energy of $10^{53}$ erg, which we consider as an upper bound. In the case of this assumed line-of-sight energy, it leads to the requirement that $\theta_\textrm{obs}/\theta_\textrm{c}$\,$\lesssim$\,$1.4$ for the tophat jet model.  
    } 
    \label{fig:energyconv}
\end{figure}

Once we have converted the line-of-sight (observed) value to the core, we can compute the (on-axis) isotropic equivalent kinetic energy of the blastwave $E_\textrm{kin,0}$, which is related to the (isotropic equivalent) gamma-ray energy released during the prompt phase, $E_{\gamma,0}$, via
\begin{align}
E_\textrm{kin,0}=E_{\gamma,0} \, \Big(\frac{1-\varepsilon_{\gamma,0}}{\varepsilon_{\gamma,0}}\Big)=E_{\gamma,0}\,\xi_\gamma,
\end{align}
where $\varepsilon_{\gamma,0}$ is the fraction of the total jet energy released in gamma-rays. We apply a gamma-ray efficiency of $\varepsilon_{\gamma,0}$\,$=$\,$0.15$ at the core \citep{Nava2014,Beniamini2015,Beniamini2016corr}. 
In the following calculations, for every value of $\theta_\textrm{obs}/\theta_\textrm{c}$ we convert the line-of-sight energy to the value at $\theta_\textrm{obs}$\,$=$\,$0$. %, and apply a 15\% gamma-ray efficiency to compute $E_\textrm{kin,0}$. 

At large angles, Equation \ref{eqn:supressed} would dictate an extreme core energy (Figure \ref{fig:energyconv}). In order to avoid completely unreasonable energies, we require that the beaming corrected kinetic energy is $<$\,$10^{53}$ erg. Such an energy is at the extreme end of the GRB phenomena, and 
%For example, GRB 221009A, the brightest and most energetic GRB of all time, had an isotropic equivalent gamma-ray energy of $10^{55}$ erg \citep{Lesage2023,Fredericks2023,Burns2023,An2023,Ripa2023} and an inferred beaming corrected kinetic energy as high as $10^{53}$ erg \citep{OConnor2023}, see also \citep{GG2023,Laskar2023,Williams2023,Kann2023,Sato2023,Ren2023}. \pb{I think it might be better to remove the discussion about the BOAT, since this is a long GRB. We could mention the most extreme energetic constraints relevant to short GRBs (maybe the Ricci paper can be used here). These would naturally be quite a bit lower (thus also demonstrating why this choice is conservative).}
is beyond what can be reasonably produced by, for example, a magnetar central engine \citep{Usov1992,Thompson04,Metzger2011,Metzger2018}, accounting for the fact that only ejecta released with a large enough energy per baryon can be used to power a highly relativistic GRB \citep{BGM2017}. %In defining this total energy budget we are agnostic to the progenitor system. 
While sGRBs likely have significantly smaller (beaming corrected) energy budgets ($10^{49-51}$ erg), for the purposes of this constraint on the allowed angle, setting a larger value for the total energy is more conservative (i.e., less a-priori restrictive of the allowed angles). Similar constraints to the viewing angle could be obtained by setting an upper bound to the peak energy $E_{p,c}$ at the jet's core \citep{Kasliwal2017,Fan2023}, which also strongly depends on Doppler beaming.  
%$E_{\rm pk}(1+z)=\delta_DE_{\rm pk}'\approx 2\Gamma(\theta)E_{\rm pk}'/[1+(\Gamma(\theta)\tilde\theta)^2]$

\begin{table}
\centering
\caption{Useful notations used throughout this work. 
}
\label{tab:notations}
\begin{tabular}{|c|c|c|}
\hline
\hline
Notation & Definition & Equation   \\
\hline
$E_\textrm{kin,0}$ & Core isotropic equivalent kinetic energy & -- \\
$E_{\gamma,0}$ & Core isotropic equivalent gamma-ray energy  & -- \\
$\Gamma_0$ & Initial bulk Lorentz factor of jet's core & -- \\
$\theta_\textrm{c}$ & Half-opening angle of jet's core & --\\
$\theta_\textrm{obs}$ & Observer's viewing angle & -- \\
$\Delta\theta$ & $\theta_\textrm{obs}-\theta_\textrm{c}$ & -- \\
$q$ & $\Delta\theta\Gamma_0$ & -- \\
$\theta_\textrm{obs}/\theta_\textrm{c}$ & $1$\,$+$\,$\Delta\theta/\theta_\textrm{c}$ & -- \\
\hline
$t_\textrm{dec,0}$ & On-axis deceleration time & \ref{eqn:tdec} \\
$t_\textrm{j}$ & On-axis jet break time & \ref{eqn:tjeteqn} \\
$t_\textrm{p}$ & Time of the lightcurve peak  & \ref{eqn:peaktimeeqn} \\
$t_\textrm{o}$ & Observed upper limit to the peak time  & -- \\
%$\Delta\theta_{t_\textrm{o}}$ & Upper limit to the viewing angle based on $t_\textrm{o}$ & \ref{eqn:deltathetatime} \\
\hline
$L_\textrm{dec,0}$ & Luminosity at (on-axis) deceleration & \ref{eqn:Ldec} \\
%$L_\textrm{j}$ & Luminosity at the (on-axis) jet-break & \ref{eqn:Ljetbreak} \\
$L_\textrm{p}$ & Peak luminosity of the afterglow  & -- \\ %\ref{eqn:peaklum} 
$L_\textrm{o}$ & Observed lower limit to the peak luminosity  & -- \\
$F_\textrm{o}$ & Observed lower limit to the peak flux  & -- \\
%$\Delta\theta_{F_\textrm{o}}$ & Upper limit to the viewing angle based on $F_\textrm{o}$ & \ref{eqn:deltathetaflux} \\
\hline
 & Structured jets & \\
\hline
$\Theta$ & $\sqrt{1+\Big(\frac{\theta}{\theta_\textrm{c}}\Big)^2}$ & \ref{eqn:appjetstuc} \\
$E_\textrm{kin}(\theta)$& Angular profile of kinetic energy: $\Theta^{-a_\textrm{kin}}$ & \ref{eqn:appjetstuc} \\
$E_\gamma(\theta)$& Angular profile of gamma-ray energy: $\Theta^{-a_\gamma}$ & \ref{eqn:appjetstuc} \\
$\Gamma(\theta)$ & Angular profile of Lorentz energy: $\Theta^{-b}$ & \ref{eqn:appjetstuc} \\
%$\varepsilon_\gamma(\theta)$& & \\
$a_\textrm{kin}$& Power-law slope of kinetic energy profile & --\\
$a_\gamma$& Power-law slope of gamma-ray energy profile & --\\
$b$& Power-law slope of Lorentz factor profile & -- \\
%$c$ & & \\
$t_\textrm{dec}(\theta)$ & Deceleration time for material at $\theta$ & \ref{eqn:structdec}\\
$L_\textrm{dec}(\theta)$ & Luminosity at  $t_\textrm{dec}(\theta)$ & \ref{eqn:structdec2}\\
%& & \\
%& & \\
\hline
\end{tabular}
\label{tab: Table_Equation_Summary}
\end{table}

\subsubsection{The time of the afterglow peak}
\label{sec:peaktimemethods}

From rapid X-ray observations of sGRBs we have robust upper limits on the time of the afterglow lightcurve peak $t_\textrm{p}$, which we refer to as $t_\textrm{o}$ (see Figure \ref{fig:lightcurves}). As the peak time is dependent on  viewing angle, we can use these upper limits to constrain the viewing angle for a population of short GRBs. We define the viewing angle as measured from the near side of the jet's core $\Delta \theta$\,$=$\,$\theta_{\rm obs}$\,$-$\,$\theta_\textrm{c}$ (see also Table \ref{tab:notations} for a list of useful notations). 
%For very small angular offsets, $\Delta \theta$\,$<$\,$\Gamma_0^{-1}$ (where $\Gamma_0$ is the initial bulk Lorentz factor of material at the jet's core), the observer is within the beaming cone of material in the jet's core and therefore the peak time and flux of the afterglow light-curve will be approximately the same as for an on-axis GRB (i.e. a deceleration peak), see, e.g., \citet{Sari1999,Molinari2007,Ghisellini2010,Ghirlanda2012,Nava2013,Nappo2014,Ghirlanda2018}. 

Short GRBs are generally found far from the center of their host galaxies \citep[e.g.,][]{Fong2013,Fong2022,OConnor2022} in low density interstellar medium (ISM) environments \citep[e.g.,][]{Sari1999,Granot2002,Soderberg2006,OConnor2020}. Furthermore, their compact binary progenitors do not produce strong stellar winds, as thought to surround many long GRBs \citep[e.g.,][]{ChevalierLi2000,Panaitescu2002,Granot2002,Schulze2011,Gompertz2018grb,Srinivasaragavan2020,Chrimes2022}. 
We consider only the case of a blastwave expanding into a uniform external environment $\rho$\,$=$\,$A r^{-k}$ with $k$\,$=$\,$0$. 
The observed (on-axis) jet deceleration time in a uniform density medium ($k$\,$=$\,$0$) is \citep{Sari1999,Molinari2007,Ghisellini2010,Ghirlanda2012,Nava2013,Nappo2014,Ghirlanda2018} 
\begin{align}
t_{\rm dec,0}=(1+z)\Big(\frac{17}{64\pi}\frac{E_\textrm{kin,0}}{c^5\, \Gamma_0^8\, m_\textrm{p}\, n}\Big)^{1/3},
\label{eqn:tdec}
\end{align}
where $c$ is the speed of light, $m_\textrm{p}$ is the mass of the proton, and $n$ is the density of the surrounding environment. 

For very small angular offsets, $\Delta \theta$\,$\ll$\,$\Gamma_0^{-1}$, the observer is within the beaming cone of material in the jet's core and therefore the peak time and flux of the afterglow light-curve will be approximately the same as for an on-axis GRB (i.e. a deceleration peak; Equation \ref{eqn:tdec}). %, see, e.g., \citet{Sari1999,Molinari2007,Ghisellini2010,Ghirlanda2012,Nava2013,Nappo2014,Ghirlanda2018}. 
For larger values of $\Delta \theta$\,$\gg$\,$\Gamma_0^{-1}$, and a uniform external medium ($k$\,$=$\,$0$), the lightcurve will peak when $\Gamma$\,$=$\,$\Delta \theta^{-1}$ and since $\Gamma$\,$\propto$\,$t^{-3/8}$ \citep{Blandford1976}, the peak time $t_\textrm{p}$ is \citep{Nakar2002,Granot2002,Matsumoto2019a,Duque2020,BGG2020}: 
\begin{eqnarray}
\label{eqn:peaktimeeqn}
 t_{\rm p} = \left\{ \begin{array}{ll} t_\textrm{dec,0} & \Delta \theta < j\Gamma_0^{-1}, \\
3.6\, t_{\rm j} \left( \frac{\Delta\theta}{\theta_\textrm{c}} \right)^{8/3} & j\Gamma_0^{-1} < \Delta \theta \ll \theta_\textrm{c}, %\\
%2.7\, t_{\rm j} \left( \frac{\Delta\theta}{\theta_\textrm{c}} \right)^{2} &  \Delta \theta \gg \theta_\textrm{c}.
\end{array} \right.
\end{eqnarray}
where $t_\textrm{j}$ is the on-axis time of the jet break. %Throughout the manuscript both $t_\textrm{j}$ and $t_\textrm{p}$ are in the observer frame. 
The factor of 3.6 out front is calibrated based on numerical simulations \citep{Granot2018}. 
We further require that $t_\textrm{p}$ is a continuous function such that $t_\textrm{p}$\,$=$\,$t_\textrm{dec,0}$ at $\Delta\theta/\theta_\textrm{c}$\,$=$\,$j/(\Gamma_0\theta_\textrm{c})$. The numerical factor $j$\,$\approx$\,$0.75$ is an order unity correction (Chand et al. in prep.). %Thus for each choice of $\Gamma_0$ and $\theta_\textrm{c}$, the minimum constraint that can be placed is $\Delta\theta/\theta_\textrm{c}$\,$<$\,$j/(\Gamma_0\theta_\textrm{c})$. 

The (on-axis) time of the jet break $t_\textrm{j}$ is related to the jet's opening angle $\theta_\textrm{c}$ as\footnote{We have adopted the definition reported by \citet{Nakar2002,Lamb2021rev}, which utilizes $t$\,$\sim$\,$R/(2\Gamma^2c)$. This leads to a factor of $\sim$2 larger jet break time compared to other definitions in the literature \citep[e.g.,][]{SariPiranHalpern1999,Zhang2009,Wang2018,RoucoEscorial2022}. This has no significant impact on our results for tophat jets, decreasing  $\Delta\theta$ by a factor of $\sim$1.3 (order unity), which when converted to $\theta_\textrm{obs}/\theta_\textrm{c}$ modifies the viewing angle limits by $\lesssim$\,$2$\,$-$\,$5\%$ on average. Furthermore, it does not modify our constraints on $\Gamma_0\theta_\textrm{c}$ in \S \ref{sec: compactness}. 
} \citep{SariPiranHalpern1999,Rhoads1999,Frail2001,Nakar2002,Lamb2021rev} 
\begin{equation}
 t_\textrm{j} = 2.3\,(1+z)\, \theta_{\textrm{c},-1}^{8/3}\,n_{-2}^{-1/3}\,E_{\textrm{kin,0},52}^{1/3} \; \rm{day},
 %\theta_\textrm{c} = 0.055\, \Big(\frac{1}{1+z}\frac{t_{\rm j}}{\textrm{day}}\Big)^{3/8} \Big(\frac{n}{10^{-3}\, \textrm{cm}^{-3}}\Big)^{1/8}\Big(\frac{E_\textrm{kin}}{10^{52}\,\textrm{erg}}\Big)^{-1/8}  \; \textrm{rad}.
    \label{eqn:tjeteqn}
\end{equation}
where we have adopted the convention $Q_\textrm{x}$\,$=$\,$Q/(10^\textrm{x}\,\textrm{cgs})$ to denote the units of each parameter such that $\theta_{\textrm{c},-1}=\theta_\textrm{c}/(0.1$ rad$)$. 

By inverting Equation \ref{eqn:peaktimeeqn} and inserting our upper limit to the peak time ($t_\textrm{p}$\,$<$\,$t_\textrm{o}$), we find that the observer's viewing angle is determined by \citep{Rossi02,Panaitescu2003,vanEerten2010,BGG2020} 
\begin{equation}
\label{eqn:deltathetatime}
    \Delta\theta < 0.004\, (1+z)^{-3/8}\,t_{\rm o, 200 s}^{3/8} \,n_{-2}^{1/8}\,E_{\textrm{kin,0},52}^{-1/8} \; \textrm{rad},
\end{equation}
%Therefore, $\Delta\theta$ is independent of the jet's half-opening angle  $\theta_\textrm{c}$. 
where $t_{\rm o, 200 s}$\,$=$\,$t_{\rm o}/(200\,\textrm{s})$ is a typical observed upper limit on the peak time for bursts in our sample (Figure \ref{fig:lightcurves}). 
The most direct constraint that can be derived is on the angular offset relative to the core's opening angle, $\Delta \theta/\theta_\textrm{c}$\,$\lesssim$\,$(t_\textrm{o}/t_\textrm{j})^{3/8}$, leading to
\begin{equation}
\label{eqn:deltathetatimediv}
    \Delta\theta/\theta_\textrm{c} < 0.04\, (1+z)^{-3/8}\,t_{\rm o, 200 s}^{3/8} \,n_{-2}^{1/8}\,E_{\textrm{kin,0},52}^{-1/8} \,\theta_{\textrm{c},-1}^{-1}.
\end{equation}

Thus far we have yet to consider the impact of Doppler beaming on the inferred core energy (\S \ref{sec:energyconv}). Inserting $E_\textrm{kin,0}$\,$=$\,$\xi_\gamma E_{\gamma,\textrm{obs}} (1+q^2)^2$ into Equation \ref{eqn:deltathetatime} we find
\begin{equation}
    \Delta\theta[1+(\Delta\theta\Gamma_0)^2]^{2/8} < 0.004 \Big(\frac{t_{\rm o, 200 s}}{1+z}\Big)^{3/8} \Big(\frac{n_{-2}}{\xi_\gamma E_{\gamma,\textrm{obs},52}}\Big)^{1/8},
\end{equation}
where we inserted $q$\,$=$\,$\Delta\theta\Gamma_0$. Asymptotically this becomes
\begin{eqnarray}
 \Delta\theta < \left\{ \begin{array}{ll} 0.004 \Big(\frac{t_{\rm o, 200 s}}{1+z}\Big)^{3/8} \Big(\frac{n_{-2}}{\xi_\gamma E_{\gamma,\textrm{obs},52}}\Big)^{1/8} \; \textrm{rad} & \! q \ll 1,\\
0.0055 \Big(\frac{t_{\rm o, 200 s}}{1+z}\Big)^{1/4} \Big(\frac{n_{-2}}{\xi_\gamma E_{\gamma,\textrm{obs},52}}\Big)^{1/12} \Gamma_{0,2}^{-1/3} \; \textrm{rad} & \!  q \gg 1.  
\end{array} \right.
\end{eqnarray}
In other words, this method provides extremely strong limits (which are also very weakly dependent on model parameters) on the possibility that short GRBs are produced by tophat jets observed off-axis.

For an assumed core half-opening angle, we can further derive the ratio $\theta_\textrm{obs}/\theta_\textrm{c}$\,$=$\,$1$\,$+$\,$\Delta\theta/\theta_\textrm{c}$, 
which for the parameters in Equation \ref{eqn:deltathetatime} and \ref{eqn:deltathetatimediv} leads to an upper limit on the observer's viewing angle with respect to the jet's core of $\theta_\textrm{obs}/\theta_\textrm{c}$\,$<$\,$1.03$. This limit can be calculated for each set of values \{$z$, $t_\textrm{o}$, $\Gamma_0$, $\theta_\textrm{c}$, $n$, $E_\textrm{kin,0}$\}. 
We adopt $\Gamma_0$\,$\approx$\,$300$ \citep[e.g.,][]{Ghirlanda2018} and $\theta_\textrm{c}$\,$\approx$\,$0.1$ rad (6 deg; \citealt{RoucoEscorial2022}) as our canonical values. However, we consider a range of possible values between $\Gamma_0$\,$=$\,$\{50,1000\}$ and $\theta_\textrm{c}$\,$=$\,$\{0.05,0.3\}$ rad. We likewise utilize a range of values for the density of the surrounding environment $n$\,$=$\,$\{10^{-4},10^{-1}\}$ cm$^{-3}$ \citep{Fong2015,OConnor2020}. As a higher density leads to less restrictive limits on the viewing angle, we conservatively adopt values for the density that are likely larger than the majority of the short GRB population \citep{Fong2015,OConnor2020}. 
The two main cases presented throughout this work, deemed our ``canonical'' and ``conservative'' cases are listed in Table \ref{tab: Results_Summary}.

\subsubsection{The peak afterglow flux }
\label{sec:peakflux}

The standard afterglow model 
\citep{Meszaros1997,Sari1998,Wijers1999} is described by a set of five parameters $\{E_\textrm{kin,0}, n, \varepsilon_e, \varepsilon_B, p\}$, where $\varepsilon_e$ and $\varepsilon_B$ are the fractions of total internal energy of the shocked material in electrons and magnetic fields, 
%,  is the fraction of energy in magnetic fields, 
and $p$ is the slope of the electron's power-law energy distribution. 
These five parameters dictate a few characteristic frequencies that define the emission regimes \citep{Granot2002}. 
The synchrotron frequencies relevant to this work are the electron injection frequency $\nu_\textrm{m}$ and the synchrotron cooling frequency $\nu_\textrm{c}$. 
At X-ray wavelengths, short GRBs are found to be in either the regime $\nu_\textrm{m}$\,$<$\,$\nu_\textrm{X}$\,$<\nu_\textrm{c}$ (PLS G) or $\nu_\textrm{c}$\,$<\nu_\textrm{X}$ (PLS H).  Here, and elsewhere, we refer to the different synchrotron power-law segments (PLS) by their definitions in \citet{Granot2002}. We adopt a uniform density environment ($k$\,$=$\,$0$) throughout this work. 

Based on early XRT observations we have identified lower limits to the peak X-ray luminosity $L_\textrm{o}=4\pi d_\textrm{L}(z)^2 F_\textrm{o}$ of a sample of sGRBs (Table \ref{tab:notations}), where $d_\textrm{L}(z)$ is the luminosity distance at redshift $z$ and $F_\textrm{o}$ is the lower limit to the $0.3$\,$-$\,$10$ keV X-ray flux at $t_\textrm{o}$. Therefore, we require that the peak X-ray luminosity is higher than this limit ($L_{\rm p}$\,$\geq$\,$L_{\rm o}$). 

For PLS G and $j\Gamma_0^{-1}$\,$<$\,$\Delta \theta$\,$\ll$\,$\theta_\textrm{c}$, the peak X-ray luminosity of the afterglow is $L_\textrm{p}$\,$\propto$\,$\Delta\theta^{2(1-p)}E_\textrm{kin,0}$, where $E_\textrm{kin,0}$\,$\propto$\,$\Delta\theta^4$. For a typical value of $p$\,$=$\,$2.2$ this becomes $L_\textrm{p}$\,$\propto$\,$\Delta\theta^{1.6}$ such that the value of $L_\textrm{p}$ increases with $\Delta\theta$ for a fixed observed X-ray flux under the assumption of a tophat jet. Therefore, for all viewing angles a tophat jet automatically satisfies the constraint $L_{\rm p}$\,$\geq$\,$L_{\rm o}$. This means that a lower limit on $L_\textrm{p}$ provides no constraint on $\Delta\theta$ in this scenario. We therefore exclude this as a relevant limit for the rest of our analysis of tophat jets, but discuss how these constraint on the peak X-ray flux can still be used in the structured jet case below (\S \ref{methods: jet structure}).

%For each set of parameters $\{z, F_\textrm{o}, \theta_\textrm{c}, \Gamma_0, n, E_\textrm{kin,0}, \varepsilon_e, \varepsilon_B, p\}$, we self-consistently compute the synchrotron regime (see Appendix \ref{appendix:PLS}), including IC effects \citep[e.g.,][]{Beniamini2015,Jacovich2020,McCarthy2024}, and apply the proper flux expressions from \citet{Granot2002}. Throughout the manuscript we adopt a canonical value of $p$\,$=$\,$2.2$ and $\varepsilon_{e}$\,$=$\,$10^{-1}$ \citep{BeniaminiVanDerHorst,Duncan2023}. We consider a range of values for the fraction of energy in magnetic fields $\varepsilon_B$\,$=$\,$\{10^{-5},10^{-1}\}$ \citep{Barniol2014,Santana2014,Zhang2015}, as well as $\Gamma_0$ and $n$ as outlined in \S \ref{sec:peaktimemethods}. 

\begin{table}
\centering
\caption{The canonical and conservative afterglow parameter cases assumed throughout this work.  
}
\begin{tabular}{cccccccc}
\hline
\hline
 Label & $\theta_\textrm{c}$ & $\Gamma_0$ & p & $n$  & $\varepsilon_e $& $\varepsilon_B$  & $\varepsilon_\gamma$ \\
 & (rad) & & & (cm$^{-3}$) & & & \\
\hline
 Canonical & 0.1 & 300 & 2.2 & $10^{-2}$ & $10^{-1}$ & $10^{-3}$ & 0.15 \\
 Conservative&0.05 & 100 & 2.2 & $10^{-1}$ & $10^{-1}$ & $10^{-1}$ & 0.15\\
\hline
\end{tabular}
\label{tab: Results_Summary}
\end{table}

\subsubsection{Impact of assumed redshift}
\label{sec: redshiftassumption}

For the 23 bursts without a measured redshift (Table \ref{Table_SGRB_DATA}) we adopt either $z$\,$=$\,$0.1$ or $0.5$. These calculations (Equations \ref{eqn:deltathetatime}) are maximized for $z$\,$\rightarrow$\,$0$ such that a lower redshift provides a more conservative (less restrictive) constraint on the viewing angle. However, the dependence on redshift is quite weak (Equation \ref{eqn:deltathetatime}): 
%\begin{align}
%    & %\Delta\theta_{t_\textrm{o}}/\theta_\textrm{c}\,\propto\,k_\textrm{bol}^{-1/8}d_L^{-1/4}(1+z)^{-3/8}. 
    %& \Delta\theta_{F_\textrm{o}}/\theta_\textrm{c}\,\propto\,k_\textrm{bol}^{5/12}(1+z)^{-1/4}.
%\end{align}
\begin{eqnarray}
 \frac{\Delta\theta}{\theta_\textrm{c}} \propto \left\{ \begin{array}{ll} k_\textrm{bol}^{-1/8}d_\textrm{L}^{-1/4}(1+z)^{-3/8} & q \ll 1, \\ \\[-3mm] 
k_\textrm{bol}^{-1/12}d_\textrm{L}^{-1/6}(1+z)^{-1/4} & q \gg 1.  
\end{array} \right.
\end{eqnarray}
Varying the redshift changes our constraint on $\Delta\theta/\theta_\textrm{c}$ by $\sim$\,$66\%$ between $z$\,$=$\,$0.1$ and $0.5$ for $q$\,$\ll$\,$1$ (and even less for $q$\,$\gg$\,$1$). 
As we report our constraints in terms of $\theta_\textrm{obs}/\theta_\textrm{c}$\,$=$\,$1+\Delta\theta/\theta_\textrm{c}$ the impact is significantly reduced. As an example (for $q$\,$\ll$\,$1$), for our canonical parameters (Table \ref{tab: Results_Summary}) and a typical limit of $t_\textrm{o}$\,$=$\,$200$ s for a burst with fluence $\phi_\gamma$\,$=$\,$10^{-7}$ erg cm$^{-2}$, the result is only modified by $\sim$\,$3\%$ between using $z$\,$=$\,$0.1$ or $0.5$.

Furthermore, we note that only a single short GRB ($T_{90}$\,$<$\,$2$ s) is measured to have $z$\,$<$\,$0.1$ (i.e., GRB 170817A; \citealt{Abbott2017,Goldstein2017,Savchenko2017}), whereas including likely compact mergers with longer duration signals (i.e., GRBs 211211A and 230307A; \citealt{Rastinejad2022,Troja2022,Levan2023,Yang2023,Gillanders2023}) increases this to three events. Therefore, it is very unlikely that there is a significant number (if any) events in this sample at such low redshifts, and the majority are more likely towards (or past) the median of the observed population ($z$\,$\approx$\,$0.5$; \citealt{OConnor2022,Nugent2022,Fong2022}).

The significant majority of short GRBs lacking either a spectroscopic or photometric redshift are associated to faint host galaxies, which have been suggested to lie above $z$\,$>$\,$1$ \citep{OConnor2022}. These galaxies with dim magnitudes are difficult to measure spectroscopically with ground-based instruments, especially for those lying in the so-called ``redshift desert'' between $1$\,$\lesssim$\,$z$\,$\lesssim$\,$2$ where most optical emission lines are shifted out of the typically observed bands. 
%An alternative possibility for those without redshift is that they are local, off-axis short GRBs \citep{Dichiara2020}. However, in this scenario there would be an obvious, bright, low redshift galaxy and the afterglow behavior would be significantly different. 
As all but a small handful (e.g., GRBs 061201, 091109B, 110112A, 160927A, 201006A) of short GRBs in our sample have identified host galaxies \citep{OConnor2022,Fong2022} % and detected afterglows this scenario is very unlikely. 
it is likely that the events without either photometric or spectroscopic redshifts are drawn from the observed redshift distribution \citep{OConnor2022,Nugent2022,Fong2022}, or potentially are biased towards slightly higher redshifts \citep{OConnor2020,OConnor2022}.  In any case, the choice of redshift (for those without a redshift from their host galaxy; Table \ref{Table_SGRB_DATA}) has only a marginal impact on our results. We again note that a larger redshift would lead to a more restrictive limit, such that our results are consistent with any higher redshift in the range $z$\,$\gtrsim$\,$0.1$\,$-$\,$0.5$.

%\textbf{\citet{Dichiara2020} previously explored the potential for \textit{Swift}/BAT detected short GRBs (with no afterglow detection) to be located within 200 Mpc and observed at potentially off-axis angles. Their study did not produce any conclusive examples due to the lack of afterglow detection for these GRBs, but set an upper limit to the all-sky rate of $\lesssim$\,$2$ yr$^{-1}$, although recent GW observations imply an even lower rate.}

\subsubsection{Impact of jet spreading}
\label{sec:spreading}

For larger viewing angles ($\Delta\theta$\,$\gg$\,$\theta_\textrm{c}$), leading to $t_\textrm{p}$\,$\gg$\,$t_\textrm{j}$, the inclusion of lateral jet spreading \citep{Granot2001,vanEerten2011,Granot2012} modifies the dependence of the peak time on viewing angle \citep[see, e.g., Section 3 of][]{Beniamini2023TDE}. This is because the time of the afterglow peak depends on the (on-axis) time of the jet break, occurring when $\Gamma$\,$=$\,$\theta_\textrm{c}^{-1}$. After the jet break the Lorentz factor of the outflow decreases roughly as $\Gamma$\,$\propto$\,$t^{-1/2}$ \citep{Rhoads1999,SariPiranHalpern1999,Granot2002jet}, whereas prior to the jet break $\Gamma$\,$\propto$\,$t^{-3/8}$ \citep{Blandford1976}.  This leads to $t_{\rm p}/t_\textrm{j}\propto \left( \Delta\theta/\theta_\textrm{c} \right)^{2}$ for far off-axis observers ($\Delta\theta$\,$\gg$\,$\theta_\textrm{c}$). 

This also leads to a modification of the dependence of the peak luminosity such that $L_{\rm p}/L_\textrm{j}\propto \left( \Delta\theta/\theta_\textrm{c} \right)^{-2p}$ regardless of the choice of jet spreading \citep{Nakar2002,BGG2020,Lamb2021rev}. % for both PLS G and H. 
This is due to the change in post-jet-break temporal slope with and without lateral spreading, such that $t^{-3p/4}$ without spreading and $t^{-p}$ for maximal spreading \citep[see, e.g.,][]{Rhoads1999,SariPiranHalpern1999}. 

However, for the timescales considered within this work (e.g., the first X-ray afterglow detection by \textit{Swift}/XRT; ranging from 60 s to 1 hr) the choice of lateral spreading does not impact our results. Simply put, for no reasonable choice of parameters (for a short GRB\footnote{We note the extreme GRB 221009A \citep{Williams2023,Laskar2023,OConnor2023,GG2023} may have had a sufficiently narrow core to allow for a jet-break on extremely early timescales \citep{LHAASO221009A}.}) does $t_\textrm{j}$ occur prior to (or even close to) the first X-ray detection. Therefore, we have presented only the case of observer viewing angles near to the jet's edge  ($\Delta\theta$\,$\ll$\,$\theta_\textrm{c}$), but within our calculations we include the impacts of lateral spreading (smoothing between regimes as calibrated using the numerical simulations of \citealt{Granot2018}; Chand et al. in prep.) in case the viewing angle becomes sufficiently large.

\subsection{Off-axis structured jets}
\label{methods: jet structure}

A tophat jet structure represents an idealized scenario, whereas the true angular structure of the jet is 
likely more complex, having a narrow energetic core surrounded by low-energy wings. This type of angular 
structure is also expected due to the interaction of the incipient jet with the confining 
medium (e.g., merger ejecta and disk winds), as seen in numerical simulations 
\citep[e.g.,][]{Lazzati2017,Gottlieb2021,Nativi2021} and inferred for GW170817 
\citep[e.g.,][]{Lamb2017jet,Lazzati2018,GG2018,Ghirlanda2019,Troja2020}. 
The exact angular profile is sensitive also to the jet magnetization \citep[e.g.,][]{Gottlieb2021magneto,Gottlieb2022}, and at present remains a topic of active study. 
In the structured jet case %\sout{, as there is always emitting material along the line-of-sight, the polar angle that dominates the prompt GRB fluence is determined by two effects, namely Doppler boosting, which is set by the angular profile of the (initial) bulk Lorentz factor, and the angular structure of the gamma-ray energy. }
there is a competition between two effects. On the one hand the radiated gamma-ray energy increases towards the jet's core (and away from the line of sight), while on the other hand the flux that reaches the observer is progressively deboosted as the viewing angle relative to the emitting material increases. Since the first effect typically falls off as a power-law in $(\theta_{\rm obs}/\theta_{\rm c})$ while the second falls off as a power-law in $(\Gamma(\theta)\Delta \theta)$ (see Equation \ref{eqn:supressed}) and since $\Gamma_0^{-1}$\,$\ll$\,$\theta_{\rm c}$, it is almost always the case that for structured jets, material close to the line of sight tends to dominate the observed gamma-rays. 
%\sout{For example, if the two angular profiles, $\Gamma(\theta)$ and $E_\gamma(\theta)$, are described by declining power-laws outside of the core, then it is possible that more energetic material at $\theta$\,$<$\,$\theta_{\rm obs}$, which is not strongly beamed away from the line-of-sight, may dominate the gamma-ray fluence. Such power-law models are generally described in terms of $\theta_\textrm{obs}/\theta_\textrm{c}$, and as the Doppler beaming of emission from the jet's core falls off as $\Gamma_0^{-1}$ (Equation \ref{eqn:supressed}), which is much smaller than $\theta_\textrm{c}^{-1}$, the power-law structure in energy (at angles $\theta_\textrm{obs}$\,$>$\,$\theta_\textrm{c}$) will always dominate over the Doppler suppressed emission from the core.} 
Similarly to the previous sections, we convert the line-of-sight fluence to the gamma-ray energy at the jet's core considering these effects. 

The same considerations imply that during the afterglow phase,   
for an event viewed far outside of the jet's core ($\theta_\textrm{obs}$\,$\gg$\,$\theta_\textrm{c}$) 
the angle that dominates the observed flux evolves over time due to deceleration of 
material at different angles 
(with different energies and initial Lorentz factors) at different times, leading to a modified behavior of the afterglow lightcurve \citep[e.g.,][]{Rossi02,Zhang2002,Granot2002jet,Panaitescu2003,Kumar2003,Rossi2004,Granot05,Lamb2017jet,Lamb2018jet,GG2018,Ryan2020,BGG2020,BGG2022,Takahashi2021,Govreen-Segal2023}. 
%In general, the material dominating the emission at early times is at larger polar angles (i.e., further from the jet's core) with the dominant angle moving towards the core with time \citep[see, e.g., Figure 5 of][]{Ryan2020}. 
As time goes by material moving along any angle is being decelerated. This decreases the amount of Doppler beaming, and can cause either: \textit{i}) emission that is initially dominated by the less energetic line-of-sight material, to gradually become dominated by faster and more energetic material that resides closer to the jet's core; or \textit{ii}) even if the emission is dominated by angles significantly below the line-of-sight (i.e., closer to the jet's core) from the beginning  (valid for steep jets; $a_\textrm{kin}$\,$>$\,$2$, see the definitions in Equation \ref{eqn:appjetstuc}), then over time, as the Doppler factor decreases, the debeaming effect is suppressed and that competes with the overall reduction of luminosity in time that would be seen by an on-axis observer after the deceleration time, and leads to an overall shallower lightcurve evolution. These two scenarios refer to cases 1A and 1B as described in \citet{BGG2020}.

%For example, even for a marginally off-axis event ($\theta_\textrm{obs}/\theta_\textrm{c}$\,$\approx$\,$1.1$), we would observe an initial shallow, plateau-like emission \citep[e.g.,][]{BGG2020}, as opposed to the typical rapid pre-deceleration increase for on-axis events \citep{Sari1999}.

As described above, in the case of structured jets, there exists a particular parameter space of jet structure and viewing angle \cite[see Fig. 8 of][]{BGG2022} in which material along a line-of-sight outside the jet's core decelerates before the emission from the jet's core comes into view of the observer. 
In this case, utilizing the methods of \S \ref{sec:peaktimemethods} could lead us to potentially mistake the peak as emission from the jet's core, as opposed to the deceleration of material along the line-of-sight at $t_\textrm{dec}(\theta_\textrm{obs})$. In such a situation, our constraints on the viewing angle would be smaller than the true viewing angle, though it could not change by much or the lightcurve shape (exhibiting a dip; see Fig. 6 and 11 of \citealt{BGG2020}) becomes completely inconsistent with observations.

\subsubsection{Numerical calculations of off-axis structured jets}

We have explored the impact of jet structure on our constraints on the viewing angle. 
We utilize the numerical calculations of \citet{GG2018} (see also \citealt{Gill2019,GG2023}) in order to perform the full integration over the jet's surface at all times and frequencies. In these studies, the jet structure is assumed to be a power-law form in energy per solid angle d$E/$d$\Omega$ and the initial specific kinetic energy (dictated by the Lorentz factor) such that \citep{GG2018,Gill2019,BGG2020,BGG2022,GG2023} 
\begin{align}
\label{eqn:appjetstuc}
   \;\;\;\;\;\;\;\;\; \;\;\;\;\; &\frac{E_\textrm{kin}(\theta)}{E_\textrm{kin,0}}=\Theta^{-a_\textrm{kin}}, \;\;\;\;\;\;\;\;\; 
    \frac{E_\gamma(\theta)}{E_{\gamma,0}}=\Theta^{-a_\gamma}, \nonumber \\
    &\frac{\Gamma(\theta)-1}{\Gamma_0-1}=\Theta^{-b}, \;\;\;\;\;\;\;\;\;
    \Theta \equiv \sqrt{1+\Big(\frac{\theta}{\theta_\textrm{c}}\Big)^2},
\end{align}
where $E_\textrm{kin,0}$ is the isotropic equivalent kinetic energy at the jet's core, $E_{\gamma,0}$ is the isotropic equivalent gamma-ray energy at the jet's core, and $\Gamma_0$ is the initial Lorentz factor at the jet's core. 
We note that this is a slightly different functional form (Equation \ref{eqn:appjetstuc}) than utilized by other authors for power-law jets, but that \textit{i}) it results in a smoothed transition between the core and the wings, which is particularly important for numerical applications and that \textit{ii}) \citet{BGG2022} fit this form to simulations of sGRB jets \citep[e.g.,][]{Lazzati2017,Gottlieb2021,Nativi2021} finding that using this form $3$\,$\lesssim$\,$a_\textrm{kin}$\,$\lesssim$\,$4$ and $2$\,$\lesssim$\,$b$\,$\lesssim$\,$6$ (see their Table 3). 
We consider this range of power-law slopes here, and for simplicity further assume that the half-opening angle of the Lorentz factor and energy profiles is the same \citep{BGG2020}. %We must also define the quantities $\xi_\textrm{c}$\,$\equiv$\,$(\Gamma_0\theta_\textrm{c})^2$ and $\theta_*$\,$\approx$\,$\theta_\textrm{c}\xi_\textrm{c}^{1/[2(b-1)]}$ as they allow for a separation between different regimes of afterglow lightcurve shapes. The angle $\theta_*$ is the lowest latitude of the jet where material at that latitude is initially beamed towards the observer ($\theta_*\Gamma(\theta_*)$\,$=$\,$1$), and $\xi_\textrm{c}$ represents the degree of compactness of the core (generally on the order of a few tens to hundreds). 

%The major factor dictating whether an observer views a single or double peaked lightcurve is the value of $\theta_*$ (see discussions in \citealt{BGG2020,BGG2022}). If $\theta_\textrm{obs}$\,$>$\,$\theta_*$ then the the observer is continuously receiving emission from decreasing angles until finally observing the core, producing a single peaked lightcurve. The alternative case, where $\theta_\textrm{obs}$\,$<$\,$\theta_*$ produces a double peaked lightcurve as initially the observer is viewing material far away from the core that will decelerate before the observer is able to recieve emission from the core (due to beaming), producing a first peak at approximately $t_\textrm{dec}(\theta_\textrm{obs})$. The secondary peak (occurring at $t_p$ as defined in \S \ref{sec:peakflux}) occurs once the core is de-beamed to the observer, and is the more standard peak for off-axis GRBs in the literature. 

We consider two jet structures: a canonical structure $\{a_\textrm{kin},b\}$\,$=$\,$\{3,4\}$ and a conservative structure $\{a_\textrm{kin},b\}$\,$=$\,$\{4,2\}$, see also Table \ref{tab: structure-summary}. 
These are two situations obtained in numerical simulations of short GRB jets. 
The canonical structure is most similar to the S1 and S2 models from \citet{Gottlieb2021}, and the conservative structure corresponds to the th50 model from \citet{Nativi2021,Lamb2022}. 
A lateral structure with a shallower decline in Lorentz factor with angle is more conservative as it leads material along any given line of sight to decelerate faster. %\sout{, and therefore become de-beamed to the observer while the material is more luminous than at later times. This means that shallower structures lead to similar peak times and fluxes at larger viewing angles compared to a steeper jet (i.e., allowing for a larger viewing angle to match the same observed constraint on $t_\textrm{o}$ and $F_\textrm{o}$), which for our purposes is conservative.}
Therefore, the peak due to the deceleration of the line-of-sight material happens earlier in this case - which means that for a fixed $t_\textrm{o}$ the inferred maximum viewing angle becomes larger. 
Alternatively, a steeper energy profile leads to a smaller kinetic energy along a given line of sight, and therefore, for a given viewing angle, to an earlier deceleration time, also leading to a larger allowed viewing angle. However, the dependence of the deceleration time on the energy is significantly weaker than the Lorentz factor $t_\textrm{dec}$\,$\propto$\,$\Gamma^{-8/3}E_\textrm{kin}^{1/3}$, and therefore the assumption regarding the energy profile only marginally impacts our results.
%\sout{As it is the latter peak that is estimated by Equation \ref{eqn:peaktimeeqn}, this results in a smaller inferred viewing angle relative to the true angle by mis-associating the latter peak.}
%Therefore, the condition $L_{\rm p}$\,$\geq$\,$L_{\rm o}$ leads to a weaker constraint (i.e, a larger upper limit) on $\theta_{\rm obs}$.

For each jet structure we compute the full numerical lightcurve \citep{GG2018} %For each structure we compute the full numerical integration \citep{GG2018} for a range of initial Lorentz factors $\Gamma_0$\,$=$\,$\{100,200,300,500,1000\}$. 
%The lightcurves are computed 
over a range of 64 logarithmically spaced viewing angles $\theta_\textrm{obs}/\theta_\textrm{c}$. The exact range of viewing angles is modified based on the initial Lorentz factor, considered in the range $\Gamma_0$\,$=$\,$\{100,1000\}$, to ensure proper sampling of the peak time periods of interest ($\sim$\,$80$\,$-$\,$3600$ s; Table \ref{Table_SGRB_DATA}). A higher initial Lorentz factor requires computing lightcurves further off-axis, as the on-axis deceleration time is significantly earlier due to $t_\textrm{dec,0}$\,$\propto$\,$\Gamma_0^{-8/3}$ (Equation \ref{eqn:tdec}). 
For each lightcurve we identify the time and flux at the peak, as well as the synchrotron segment for X-rays (by tracking $\nu_\textrm{m}$ and $\nu_\textrm{c}$). For the general times of interest the emission is from the regime $\nu_\textrm{m}$\,$<$\,$\nu_\textrm{X}$\,$<$\,$\nu_\textrm{c}$ (PLS G). 

For simplicity the lightcurves are computed at an observer frequency of $2$ keV for a source at $z$\,$=$\,$0.5$. The canonical afterglow parameters are assumed to be $p$\,$=$\,$2.2$, $\theta_\textrm{c}$\,$=$\,$0.1$ rad, $E_\textrm{kin,0}$\,$=$\,$10^{52}$ erg, $n$\,$=$\,$10^{-2}$ cm$^{-3}$, $\varepsilon_e$\,$=$\,$0.1$, $\varepsilon_B$\,$=$\,$10^{-3}$ (Table \ref{tab: Results_Summary}). We integrate over the synchrotron spectrum to compute the observer frame $0.3$\,$-$\,$10$ keV X-ray flux for comparison to observations (Table \ref{Table_SGRB_DATA}). For each sGRB, the (observer frame) time and flux of the peak are shifted to the corresponding redshift, where $z$\,$=$\,$0.5$ is assumed for bursts without a measured redshift. 

As each sGRB in our sample has a distinct fluence (Table \ref{Table_SGRB_DATA}), the inferred energy differs from burst to burst, even for the same redshift. As $t_\textrm{dec,0}$\,$\propto$\,$(E_\textrm{kin,0}/n)^{1/3}$ and $L_\textrm{dec,0}$\,$\propto$\,$E_\textrm{kin,0} n^{4/5} \varepsilon_B^{4/5}$, we can shift the time and flux of the numerical lightcurve peaks for varying values of the afterglow parameters.  
The values of $\nu_\textrm{m}$ and $\nu_\textrm{c}$ are also shifted in a similar manner to track the synchrotron segment (see Appendix \ref{appendix:PLS}). 
%However, the fluence for a given GRB allows for the calculation of the line-of-sight isotropic equivalent energy, which is not necessarily the core value unless $\theta_\textrm{obs}$\,$=$\,$0$ (or the jet has no angular structure and is adequately described by a tophat). Therefore, in order to shift the lightcurves for different kinetic energies, we first (for every value of $\theta_\textrm{obs}/\theta_\textrm{c}$) convert the line-of-sight gamma-ray energy to the value at $\theta_\textrm{obs}$\,$=$\,$0$, and apply a 15\% gamma-ray efficiency to compute $E_\textrm{kin,0}$. 
Similar to \S \ref{methods: tophat jets}, in order to shift the lightcurves for different kinetic energies, we first (for every value of $\theta_\textrm{obs}/\theta_\textrm{c}$) convert the line-of-sight gamma-ray energy to the value at $\theta_\textrm{obs}$\,$=$\,$0$, and apply a 15\% gamma-ray efficiency at the core to compute $E_\textrm{kin,0}$ (see Figure \ref{fig:energyconv}). For both structures we adopt either a constant gamma-ray efficiency with angle ($a_\textrm{kin}$\,$=$\,$a_\gamma$) or a value of $a_\gamma$\,$=$\,$6>a_{\rm kin}$ (i.e., gamma-ray efficiency decreasing with polar angle). We discuss the impact of this assumption in \S \ref{sec: structjetresults}.  
These methods are also applied to the analytic approximations discussed below. %(Equations \ref{eqn:structdec} and \ref{eqn:freqstruct}) for each burst. %The results of our analysis for structured jets are presented in \S \ref{sec: structjetresults}. 

\begin{table}
\centering
\caption{The canonical and conservative (power-law) structured jet cases assumed throughout this work, see Equation \ref{eqn:appjetstuc}. The kinetic energy profile is dictated by the power-law slope  $a_\textrm{kin}$ and the Lorentz factor by $b$.
}
\begin{tabular}{cccc}
\hline
\hline
 Label &  $a_\textrm{kin}$ & $b$ & Reference \\ 
\hline
 Canonical & 3 &  4 & \citet{Gottlieb2021} \\
 Conservative & 4 & 2 & \citet{Nativi2021} \\
\hline
\end{tabular}
\label{tab: structure-summary}
\end{table}

\subsubsection{Analytic calculations of the (line-of-sight) deceleration peak}

Based on the numerical calculations, we find for these jet structures (Table \ref{tab: structure-summary}) that the first afterglow peak is always due to the deceleration of line-of-sight material over the relevant timescales for our sample (Figure \ref{fig:lightcurves}). 
The time and flux (in PLS G; $\nu_\textrm{m}$\,$<$\,$\nu_\textrm{X}$\,$<\nu_\textrm{c}$) 
due to the deceleration of line-of-sight material can be approximated as \citep{Beniamini2019plateau,BGG2020}
\begin{align}
\label{eqn:structdec}
   &t_\textrm{dec}(\theta_\textrm{obs}) = t_\textrm{dec,0} \Big[1 + \Big(\frac{\theta_\textrm{obs}}{\theta_\textrm{c}}\Big)^2 \Big]^\frac{8b-a_\textrm{kin}}{6} \\
\label{eqn:structdec2}
   &L_\textrm{dec}(\theta_\textrm{obs}) = L_\textrm{dec,0}\Big[1 + \Big(\frac{\theta_\textrm{obs}}{\theta_\textrm{c}}\Big)^2 \Big]^\frac{2b-2bp-a_\textrm{kin}}{2}  %PLS G
   %&L_\textrm{dec,H}(\theta_\textrm{obs}) = L_\textrm{dec,H,0}\Big[1 + \Big(\frac{\theta_\textrm{obs}}{\theta_\textrm{c}}\Big)^2 \Big]^\frac{2b-3bp-a}{3} %PLS H excluding 1+Y corrections...
\end{align}
where $t_\textrm{dec,0}$ and $L_\textrm{dec,0}$ are the on-axis values for  $\theta_\textrm{obs}$\,$=$\,$0$. 

We can also derive the evolution of the characteristic synchrotron frequencies:
\begin{align}
\label{eqn:freqstruct}
   &\nu_\textrm{c,dec}(\theta_\textrm{obs}) = \nu_\textrm{c,dec,0} \Big[1 + \Big(\frac{\theta_\textrm{obs}}{\theta_\textrm{c}}\Big)^2 \Big]^\frac{a_\textrm{kin}-2b}{3} \\
\label{eqn:freqstruct2}
   &\nu_\textrm{m,dec}(\theta_\textrm{obs}) = \nu_\textrm{m,dec,0}\Big[1 + \Big(\frac{\theta_\textrm{obs}}{\theta_\textrm{c}}\Big)^2 \Big]^{-2b}
\end{align}
where $\nu_\textrm{c,dec,0}$ and $\nu_\textrm{m,dec,0}$ are the on-axis values of the characteristic synchrotron frequencies (see Appendix \ref{appendix:PLS} for details).

Using these equations, we compute the emission at the line-of-sight deceleration time in the proper synchrotron segment (Appendix \ref{appendix:PLS}). 
Each of these scalings were verified using numerical simulations \citep{GG2018}, and only break down far off-axis angles ($\Delta\theta$\,$\gg$\,$\theta_\textrm{c}$) when emission from the core of the jet begins to dominate (which depends on the jet structure and initial Lorentz factor; \citealt{BGG2020}). 

We utilize our limits to the peak time ($t_\textrm{o}$) and peak luminosity ($L_\textrm{o}=4\pi d_\textrm{L}(z)^2 F_\textrm{o}$) to constrain the viewing angle for our sample of sGRBs in this structured jet scenario. We require that $t_{\rm dec}(\theta_\textrm{obs})$\,$\leq$\,$t_{\rm o}$ and $L_{\rm dec}(\theta_\textrm{obs})$\,$\geq$\,$L_{\rm o}$. Inserting $t_{\rm dec}(\theta_\textrm{obs})$\,$=$\,$t_{\rm o}$ in Equation \ref{eqn:structdec} gives  
\begin{align}
    \frac{\theta_\textrm{obs}}{\theta_\textrm{c}} < \Big[\Big(\frac{t_\textrm{o}}{t_\textrm{dec,0}} \Big)^{\frac{6}{8b-a_\textrm{kin}}}-1\Big]^{1/2}.
\end{align}
However, we must account for the line-of-sight to core energy conversion, which is not yet included in Equations \ref{eqn:structdec} and \ref{eqn:structdec2}. To do this we apply $E_\textrm{kin}(\theta_\textrm{obs})$\,$=$\,$\xi_\gamma E_{\gamma,\textrm{obs}}\Theta^{a_\gamma-a_\textrm{kin}}$ such that 
\begin{align}
\label{eqn:boomboom}
    \frac{\theta_\textrm{obs}}{\theta_\textrm{c}} < \sqrt{\Bigg(\frac{t_\textrm{o}}{(1+z)} \Big(\frac{64\pi }{17}\frac{ c^5 \Gamma_0^{8} m_p n}{\xi_\gamma E_{\gamma,\textrm{obs}}}\Big)^{1/3}\Bigg)^{\frac{6}{8b+a_\gamma-a_\textrm{kin}}}-1}.
\end{align}
For a constant gamma-ray efficiency with polar angle ($\varepsilon_\gamma(\theta)$\,$=$\,constant), defined by  $a_\gamma$\,$=$\,$a_\textrm{kin}$,  this equation depends only on $\Gamma(\theta)$. As $a_\gamma$\,$\geq$\,$a_\textrm{kin}$ by definition the assumption of a constant gamma-ray efficiency is the most conservative case we can apply (i.e., leads to larger values for the allowed viewing angles). However, more realistic values of $a_\gamma$\,$>$\,$a_\textrm{kin}$ do not modify the limits by much due to the extremely weak dependence on energy. We demonstrate these points further in \S \ref{sec: structjetresults}. 
Adopting our canonical parameters for both jet structure (Table \ref{tab: structure-summary}) and the afterglow (Table \ref{tab: Results_Summary}) 
leads to $\theta_\textrm{obs}/\theta_\textrm{c}$\,$\lesssim$\,$0.4$ for $z$\,$=$\,$0.5$, $t_\textrm{o}$\,$=$\,$200$ s, and $E_{\gamma,\textrm{obs}}$\,$=$\,$10^{52}$ erg.  

%This equation can be approximated asymptotically for $\theta_\textrm{obs}/\theta_\textrm{c}$\,$\gg$\,$1$ which occurs when $t_\textrm{o}$\,$\gg$\,$t_\textrm{dec,0}$
As this equation cannot be easily approximated unless the viewing angle is very large ($\theta_\textrm{obs}/\theta_\textrm{c}$\,$\gg$\,$1$), we instead consider the required value of $t_\textrm{o}$ in order for a limit larger than $\theta_\textrm{obs}/\theta_\textrm{c}$\,$\approx$\,$2$ for both our canonical and conservative jet structures:
\begin{align}
\label{eqnboom}
\frac{t_\textrm{o}}{1+z} \gtrsim \left\{ \begin{array}{ll} 6.8\times10^{5} \, \Bigg(\frac{ \xi_\gamma E_{\gamma,\textrm{obs},52}}{ \Gamma_{0,300}^{8}  n_{-2}}\Bigg)^{1/3} \: \textrm{s},  & \{a_\gamma\!=\!6,\,a_\textrm{kin}\!=\!3,\,b\!=\!4\}, \\
7125 \, \Bigg(\frac{ \xi_\gamma E_{\gamma,\textrm{obs},52}}{ \Gamma_{0,300}^{8}  n_{-2}}\Bigg)^{1/3} \: \textrm{s},  & \, \{a_\gamma\!=\!6,\,a_\textrm{kin}\!=\!4,\,b\!=\!2\}, %\\
%2.7\, t_{\rm j} \left( \frac{\Delta\theta}{\theta_\textrm{c}} \right)^{2} &  \Delta \theta \gg \theta_\textrm{c}.
\end{array} \right.
%57 \, \Bigg(\frac{ \xi_\gamma E_{\gamma,\textrm{obs},52}}{ \Gamma_{0,300}^{8}  n_{-2}}\Bigg)^{1/3} \: \textrm{s}.
\end{align}
As a typical limit for our sample is  $t_\textrm{o}$\,$\approx$\,$200$ s, we are almost never in a regime where $\theta_\textrm{obs}/\theta_\textrm{c}$\,$\gg$\,$1$ (at least for our canonical parameters; see below). As such, the limit based on the peak time generally leads to $\theta_\textrm{obs}/\theta_\textrm{c}$\,$\lesssim$\,$2$. That being said, we can immediately see that for large enough values of $\Gamma_0$ this requirement in Equation \ref{eqnboom} can be satisfied. We discuss this more in \S \ref{sec: structjetresults}.

%\begin{align}
%\frac{\theta_\textrm{obs}}{\theta_\textrm{c}} < \Bigg(\frac{t_\textrm{o,200s}}{(1+z)57} \Big(\frac{ \Gamma_{0,300}^{8}  n_{-2}}{\xi_\gamma E_{\gamma,\textrm{obs},52}}\Big)^{1/3}\Bigg)^{\frac{3}{8b+a_\gamma-a_\textrm{kin}}}.
%\end{align}
%where $\Gamma_{0,300}$\,$=\Gamma/300$. 
%For $a_\gamma$\,$=$\,$6$, $a_\textrm{kin}$\,$=$\,$3$, and $b$\,$=$\,$4$ this becomes 
%\begin{align}
%\frac{\theta_\textrm{obs}}{\theta_\textrm{c}} < 1.1\Big(\frac{t_\textrm{o,200s}}{(1+z)57}\Big)^{3/35}\Big(\frac{ \Gamma_{0,300}^{8}  n_{-2}}{\xi_\gamma E_{\gamma,\textrm{obs},52}}\Big)^{1/35}.
%\end{align}
%}
%\textcolor{red}{ignoring the -1 makes this factor out front *very* wrong...}

Similarly inserting $L_\textrm{o}$ in Equation  \ref{eqn:structdec2} yields (see also Appendix \ref{appendix:PLS})
\begin{align}
\label{eqn:hateit}
    \frac{\theta_\textrm{obs}}{\theta_\textrm{c}} \!<\! \sqrt{\Bigg(\frac{L_\textrm{o}}{\varepsilon_{e}^{6/5}\varepsilon_{B}^{4/5}\xi_\gamma E_{\gamma,\textrm{obs}}n^{4/5}\Gamma_0^{12/5}(1+z)^{2/5}}\Bigg)^{\frac{2}{a_\gamma+2b-2bp-a_\textrm{kin}}}\!-\!1}.\!
\end{align}
%For our canonical scenario this leads to  $\theta_\textrm{obs}/\theta_\textrm{c}$\,$\lesssim$\,$1.9$ for $z$\,$=$\,$0.5$,  $F_\textrm{o}$\,$=$\,$10^{-12}$ erg cm$^{-2}$ s$^{-1}$, and $E_{\gamma,\textrm{obs}}$\,$=$\,$10^{52}$ erg. 

Equation \ref{eqn:hateit} leads to larger viewing angles than Equation \ref{eqn:boomboom}, and therefore can be approximated further. Asymptotically for $\theta_\textrm{obs}/\theta_\textrm{c}$\,$\gg$\,$1$ (and $a_\gamma$\,$=$\,$6$, $a_\textrm{kin}$\,$=$\,$3$, $b$\,$=$\,$4$, and $p$\,$=$\,$2.2$) Equation \ref{eqn:hateit} simplifies to 
\begin{align}
    \frac{\theta_\textrm{obs}}{\theta_\textrm{c}} \!<\! 2.1  \frac{\varepsilon_{e,-1}^{6/33}\varepsilon_{B,-3}^{4/33}(\xi_\gamma E_{\gamma,\textrm{obs},52})^{5/33}n_{-2}^{4/33}\Gamma_{0,300}^{12/33}(1+z)^{2/33}}{L_\textrm{o,45}^{5/33}}.\!
\end{align}
As the flux limit is significantly less restrictive on $\theta_\textrm{obs}/\theta_\textrm{c}$ than the time limit (Equation \ref{eqn:boomboom}), this limit (Equation  \ref{eqn:hateit}) is not further constraining on the allowed parameter space.  
Therefore, for each sGRB we adopt the most constraining limit from each of these cases (which for the case presented here, and in general, is the time limit - which is also less sensitive to uncertain afterglow parameters). We emphasize that for either limit our constraints are independent of an assumed core half-opening angle. The results for our sample are presented in \S \ref{sec: structjetresults}.

\subsubsection{Impact of off-axis angles on the observed prompt emission}

Structured jets also have implications for the observed prompt gamma-ray emission. While it is clear from the decreasing energy profile that the line-of-sight material will be less energetic, leading to less energetic prompt gamma-rays, the decreasing Lorentz factor with increasing angle from the jet's core also conflicts with the gamma-ray production mechanism. There are a few ways in which a low Lorentz factor can decrease the likelihood of efficient non-thermal gamma-ray emission. For instance, this leads to an increase in optical depth at larger viewing angles as the Thomson optical depth goes as $\tau_\textrm{T}(\theta)$\,$\propto$\,$\Gamma(\theta)^{-5}$ \citep[e.g.,][]{Gill2020}. A related issue, is that a lower Lorentz factor results in a decreasing dissipation radius  (approximately $R_\textrm{dis}(\theta)$\,$\propto$\,$\Gamma^2(\theta)$, depending on the details of the dissipation mechanism) such that at large enough angles ($\theta_\textrm{obs}/\theta_\textrm{c}$\,$>$\,$2$) the photospheric radius $R_\textrm{ph}(\theta)$\,$\propto$\,$\Gamma(\theta)^{-3}$ is always larger $R_\textrm{ph}(\theta)$\,$\gg$\,$R_\textrm{dis}(\theta) $\citep[e.g.,][]{LambKobayashi2016,BBPG2020}, meaning that under such conditions the radiation is indeed produced in a region with high optical depth. Furthermore, the gamma-ray producing material along the observer's line-of-sight must also have decelerated before the X-ray afterglow starts declining, which can set a robust and model independent constraint of $\Gamma(\theta_\textrm{obs})$\,$>$\,$50$ \citep{BeniaminiNakar2019}. This result suggests that the Lorentz factor distribution is likely shallower than the energy profile. Each of these independent arguments seems to imply that the efficiency of gamma-ray production decreases sharply outside of $\theta_\textrm{obs}/\theta_\textrm{c}$\,$>$\,$2$ \citep{BeniaminiNakar2019,BBPG2020,Gill2020}.

Therefore, there is likely an additional bias against observing off-axis events (especially in the cosmological sample, where other mechanisms such as shock-breakout are not detectable), in addition to the obvious Malmquist bias against faint gamma-rays, due to the likely lack of gamma-ray production at large angles. However, the results derived in this work are independent of prompt emission models and agree with the overall picture that GRBs cannot be seen far off-axis.

%%%%%%%%%%%%%%%%%%%%%%%%%%%%%%%%%%%%%%%%%%%%%%%%%%

\section{Results}
\label{sec: results}

\subsection{Off-axis Tophat jets}
\label{sec: tophatresults}

\subsubsection{Constraints on viewing angle 
}
\label{sec: viewing_results}

We present the constraints derived from our sample of 58 short GRBs with early limits on the peak time of their X-ray afterglows. For each event we compute the constraint (upper limit) on the viewing angle $\theta_\textrm{obs}/\theta_\textrm{c}$ based on the peak time and based on our adopted constraint on the total (beaming corrected) core energy (\S \ref{sec:energyconv} and Figure \ref{fig:energyconv}). %the peak flux $\Delta\theta_{F_\textrm{o}}$. 
We note that for each burst only the most stringent constraint (either from the time of the first XRT detection or based on energetics) should be adopted, which in general always comes from the time of the first XRT detection (Equation \ref{eqn:deltathetatime}). 
%Therefore, we adopt the most constraining upper limit such that the maximum allowed value for each burst is  $\Delta\theta_\textrm{max}=\min(\Delta\theta_{t_\textrm{o}},\Delta\theta_{F_\textrm{o}})$.  

The upper bound on viewing angle based on burst energetics (beaming corrected energy $<$\,$10^{53}$ erg; Figure \ref{fig:energyconv}) for a tophat jet can be derived using the observed (line-of-sight) gamma-ray energy for every burst in our sample. 
Assuming $\theta_\textrm{c}$\,$=$\,$0.1$ rad leads to the requirement that $\theta_\textrm{obs}/\theta_\textrm{c}$\,$\lesssim$\,$1.7$ for 90\% of bursts in our sample. 
%A more stringent restriction of the allowed energy (e.g., a beaming corrected energy $<$\,$10^{51}$ erg) to agree with more realistic values produced by neutron star mergers, and inferred from the population of sGRBs \citep{Fong2015,RoucoEscorial2022} \pb{Using this last inference is circular - we want to critically examine here the possibility that these jets are viewed off-axis while those studies applied on-axis modeling}, would lead to significantly smaller values for the allowed viewing angles ($\theta_\textrm{obs}/\theta_\textrm{c}$\,$\lesssim$\,$1.2$ or smaller). \textcolor{purple}{BO: agreed. makes sense. }
These allowed angles are significantly larger than those derived based on the time of the first XRT detection which we show below.

We perform this calculation for both a canonical case ($\theta_\textrm{c}$\,$=$\,$0.1$ rad, $n$\,$=$\,$10^{-2}$ cm$^{-3}$,  and $\Gamma_0$\,$=$\,$300$) and a more conservative scenario ($\theta_\textrm{c}$\,$=$\,$0.05$ rad, $n$\,$=$\,$0.1$ cm$^{-3}$, and $\Gamma_0$\,$=$\,$100$). The choice of conservative parameters is driven by the dependence of the viewing angle limits on the unknown afterglow parameters with higher values of $n$ leading to larger allowed angles. In fact, our choice of $n$\,$=$\,$10^{-2}$ cm$^{-3}$ in the canonical case is likewise conservative, as a significant fraction of sGRBs can be found in more rarefied environments \citep{OConnor2020}. The core half-opening angle similarly has a large impact with smaller values of $\theta_\textrm{c}$ leading to larger permitted viewing angles, $\theta_\textrm{obs}/\theta_\textrm{c}$. 

\begin{figure}
    \centering
    \includegraphics[width=\columnwidth]{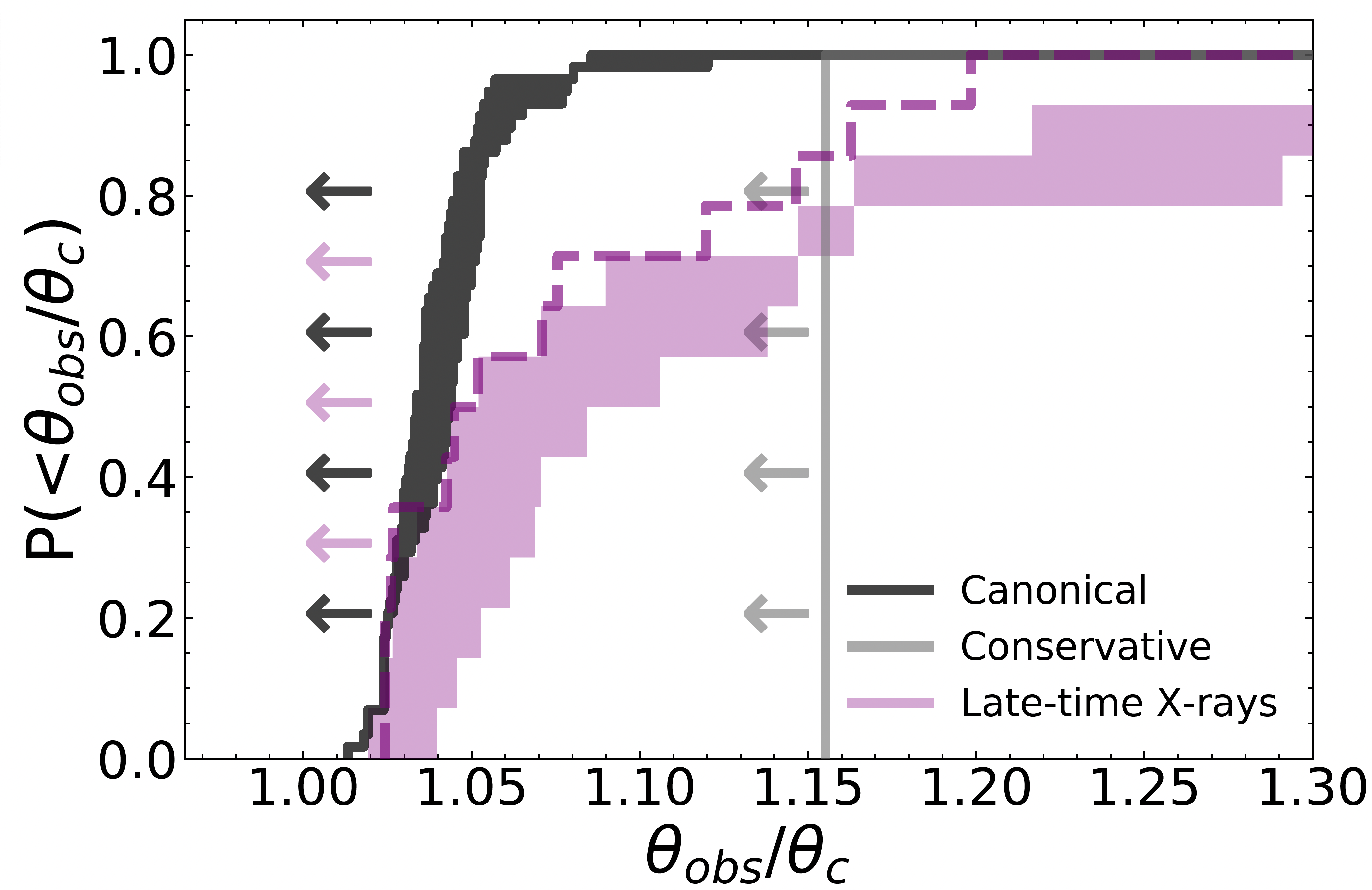}
    \vspace{-0.6cm}
    \caption{
    Cumulative distribution of upper limits on viewing angle, assuming a tophat jet structure, for our sample of 58 short GRBs (black and gray) compared to upper limits derived from a sample with late-time X-ray observations (purple; \citealt{RoucoEscorial2022}). We have adopted the canonical (black) set of parameters $\theta_\textrm{c}$\,$=$\,$0.1$ rad, $n$\,$=$\,$10^{-2}$ cm$^{-3}$, and $\Gamma_0$\,$=$\,$300$, and a conservative case (gray) with $\theta_\textrm{c}$\,$=$\,$0.05$ rad, $n$\,$=$\,$0.1$ cm$^{-3}$, and $\Gamma_0$\,$=$\,$100$. The black shaded region corresponds to varying the redshift between $z$\,$=$\,$\{0.1,0.5\}$ for the 23 events without a measured redshift. The shaded region for the sample with late-time X-ray observations (purple) is due to varying $\Gamma_0$\,$=$\,$\{100,300\}$. The dashed purple line shows how the constraints for the late-time X-ray sample change if we adopt $\theta_\textrm{c}$\,$=$\,$0.1$ rad versus the results for $\theta_\textrm{c}$ from afterglow modeling. 
    }
    \label{fig:bestlim}
\end{figure}

Our results are shown in Figure \ref{fig:bestlim}. 
For the 23 bursts without a known redshift we vary redshift between $z$\,$=$\,$0.1$ and $z$\,$=$\,$0.5$ to compute the shaded region for each scenario. This demonstrates that the assumed redshift has a minimal impact on our calculations (\S \ref{sec: redshiftassumption}), and instead the choice of initial bulk Lorentz factor and core half-opening angle (dictating the minimum value of $\theta_\textrm{obs}/\theta_\textrm{c}$ that can be restricted) has a larger impact. 

The allowed values of $\theta_\textrm{obs}/\theta_\textrm{c}$ are to the left of the cumulative distribution functions (CDFs) in Figure \ref{fig:bestlim}.
The 90th percentile of our sample constrains the viewing angle to $\theta_\textrm{obs}/\theta_\textrm{c}$\,$<$\,$1.06$ and $1.15$ rad for the canonical and conservative parameters, respectively. For reference, this corresponds to $\Delta\theta$\,$<$\,$0.006$ rad and $\Delta\theta$\,$<$\,$0.015$ rad, which is independent of the choice of $\theta_\textrm{c}$. 
Regardless of the adopted parameters, our results demonstrate that if the underlying jets of sGRBs have a tophat structure then the events in our sample must be either on-axis $\theta_\textrm{obs}/\theta_\textrm{c}$\,$<$\,$1$\,$+$\,$j/(\Gamma_0\theta_\textrm{c})$ or (at most) only very slightly off-axis $\theta_\textrm{obs}/\theta_\textrm{c}$\,$<$\,$1.2$.

%\begin{figure}
%    \centering
%    \includegraphics[width=\columnwidth]{figs/MC_error_CDF_all-2.pdf}
%    \vspace{-0.6cm}
%    \caption{
%    }
%    \label{fig:MonteCarloLim}
%\end{figure}

\subsubsection{Events with constraints on the jet-break}

We further consider a sample of events with constraints on the time of their jet break. These constraints are generally derived from late-time ($>$\,$2$-$5$ d) X-ray observations with either \textit{Chandra} or \textit{XMM-Newton} to search for a change in the temporal slope of the X-ray lightcurves \citep[e.g.,][]{Burrows2006,Soderberg2006,Fong2012,Troja2016jetbreak,Lamb2019grb160821B,Troja2019b,OConnor2021kn,Laskar2022,RoucoEscorial2022}, though in some cases they can also be driven by \textit{Hubble Space Telescope} observations \citep[e.g.,][]{Jin2018,Fong2021kn}. The time of the jet break allows for a determination of the jet's half-opening angle $\theta_\textrm{c}$ (Equation \ref{eqn:tjeteqn}). 

Out of 32 sGRBs with sensitive late-time X-ray observations, \citet{RoucoEscorial2022} found that the jet break could only be constrained for 10 events, with the remaining events leading to lower limits on $t_\textrm{j}$. 
Based on this analysis, our sample contains 7 events (out of 10 total)  with a measured jet break, and an additional 7 events with lower limits to the time of their jet break (and therefore to their half-opening angles).  Provided that   $\Delta\theta$/$\theta_\textrm{c}$\,$\lesssim$\,$(t_\textrm{o}/t_\textrm{j})^{3/8}$, we can also use events with lower limits to $t_\textrm{j}$ to compute a maximum allowed value of $\Delta\theta$/$\theta_\textrm{c}$ (which is then converted to $\theta_\textrm{obs}/\theta_\textrm{c}$ independently of any assumption for $\theta_\textrm{c}$). 

The main advantage of this method is that it removes uncertainty regarding our assumptions on afterglow modeling or the afterglow parameters $\{\theta_\textrm{c}, n, E_\textrm{kin,0}, \varepsilon_\gamma$, etc.\}, providing a useful sanity check to the rest of our calculations. 
For each of these 14 events we adopt the jet break time (or the lower limit) from \citet{RoucoEscorial2022}. 
We use these values in conjunction with the first X-ray detection $t_\textrm{o}$ to compute upper limits to their viewing angles (Equation \ref{eqn:deltathetatimediv}) for comparison to our larger sample.  
The results are shown in Figure \ref{fig:bestlim} (purple shaded region). The 90th percentile limit for these events is $\theta_\textrm{obs}/\theta_\textrm{c}$\,$<$\,$1.24$ and $1.35$ for an assumed Lorentz factor of $\Gamma_0$\,$=$\,$300$ and $100$, respectively.

The constraints on the viewing angle above implicitly assume that $t_\textrm{o}$\,$>$\,$t_\textrm{dec,0}$. When this is not the case, the viewing angle constraint becomes $\theta_\textrm{obs}/\theta_\textrm{c}$\,$\lesssim$\,$1+j/(\Gamma_0\theta_\textrm{c})$, see Equation \ref{eqn:peaktimeeqn}. 
To self-consistently compute $t_\textrm{dec,0}$ for these events we utilize $\{\theta_\textrm{c}, E_\textrm{kin,0}, n\}$ from \citet{RoucoEscorial2022}. 
In order to show the impact of our assumption on $\Gamma_0$ and $\theta_\textrm{c}$ (which comes from afterglow modeling), we plot as a dashed purple line in Figure \ref{fig:bestlim} the limit for these events if we use  $\Gamma_0$\,$=$\,$300$ and $\theta_\textrm{c}$\,$=$\,$0.1$ rad, barely changing the limits. Thus, for this sample, the afterglow modeling and derivation of core opening angle has little impact on our population level constraints.  %We note that for a subset of events with lower limits to the jet break the inferred values of $\theta_\textrm{c}$ \citep{RoucoEscorial2022} are biased towards smaller values (e.g., GRB 101219A) due to very early time constraints to the jet break from a break observed by \textit{Swift}/XRT as opposed to late-time detections from \textit{Chandra}. 
We have shown using a broad range of assumed model parameters (Figure \ref{fig:bestlim}), as well as using model independent constraints for a sub-set of 14 sGRBs, that our constraints to $\theta_\textrm{obs}/\theta_\textrm{c}$ are robust. 

\subsubsection{Compactness constraints}
\label{sec: compactness}

\begin{figure}
    \centering
    \includegraphics[width=\columnwidth]{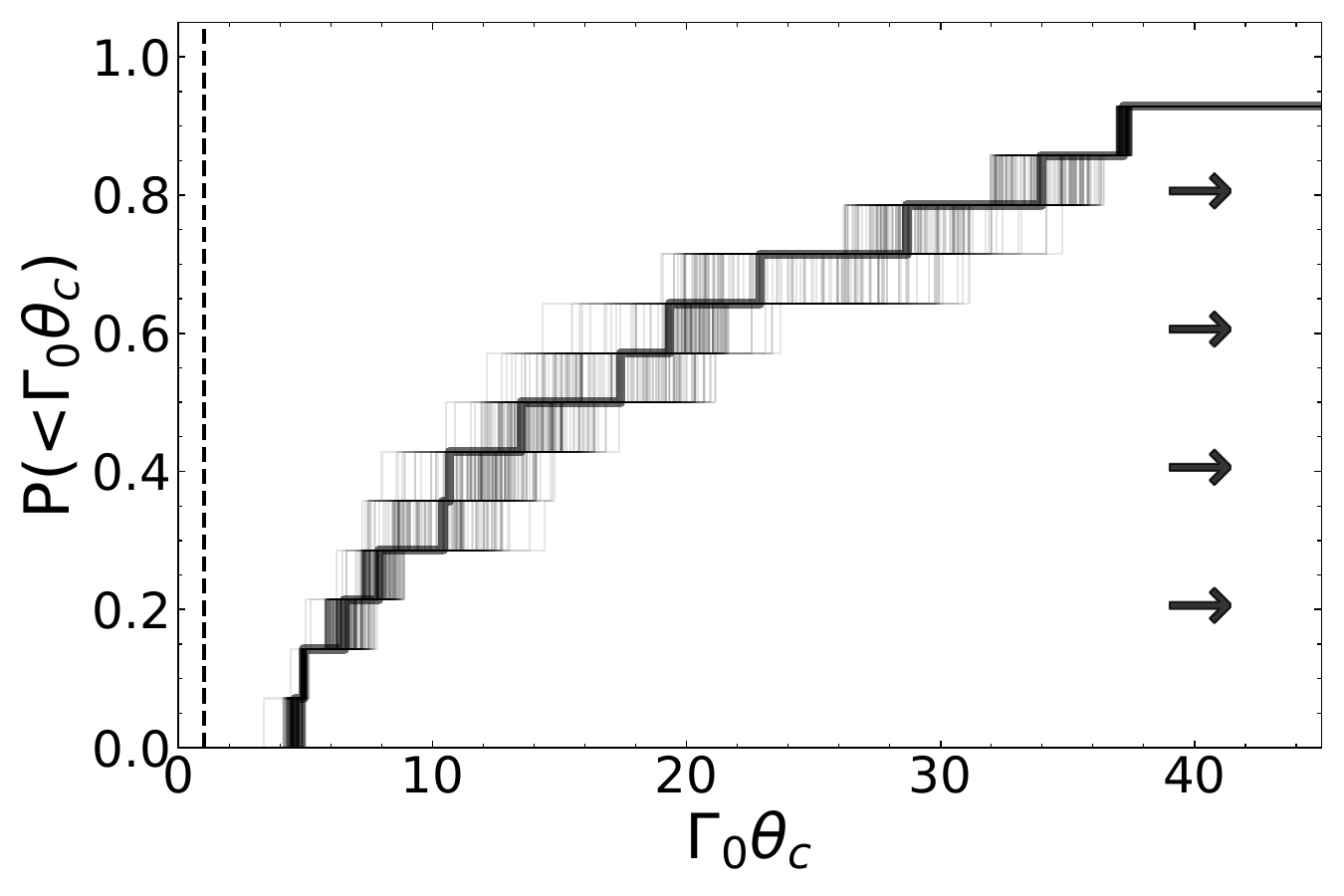}
    \vspace{-0.6cm}
    \caption{Cumulative distribution of lower limits on $\Gamma_0\theta_\textrm{c}$ for the sample of 14 events with late-time X-ray observations constraining their jet break. A dashed line marks $\Gamma_0\theta_\textrm{c}$\,$=$\,$1$. We show 150 realizations of the CDF by sampling from the observed errors on $\{t_\textrm{o},t_\textrm{j}\}$. 
    }
    \label{fig:gammalim}
\end{figure}

The requirement that this sub-sample of sGRBs with jet breaks \citep{RoucoEscorial2022} must be approximately viewed on-axis leads to lower limits on their initial bulk Lorentz factors $\Gamma_0$ and core compactness $\Gamma_0\theta_\textrm{c}$. The quantity that can be most robustly constrained by joint limits on both the  deceleration and jet-break is $\Gamma_0\theta_\textrm{c}$ \citep[e.g.,][]{BGG2020}. To show this we consider that (in a uniform density medium\footnote{We note that the dependence of these parameters on each other depends on the surrounding environment and is modified for a wind medium.}) $t_\textrm{dec,0}$\,$\propto$\,$\Gamma_0^{-8/3}$ (Equation \ref{eqn:tdec}) and $t_\textrm{j}$\,$\propto$\,$\theta_\textrm{c}^{8/3}$ (Equation \ref{eqn:tjeteqn}) such that $t_\textrm{j}/t_\textrm{dec,0}$\,$\propto$\,$(\Gamma_0\theta_\textrm{c})^{8/3}$. As we have direct knowledge of both $t_\textrm{j}$ (either a measurement or a lower limit) and of $t_\textrm{dec,0}$\,$\leq$\,$t_\textrm{o}$, the ratio $\Gamma_0\theta_\textrm{c}$\,$\propto$\,$(t_\textrm{j}/t_\textrm{o})^{3/8}$ is a very well constrained lower limit for each of these 14 events (assuming they are on-axis). In fact, we can write this more clearly as 
\begin{equation}
\label{eqn:gamthet}
    \Gamma_0\theta_\textrm{c} > 14 \, \Big(\frac{t_\textrm{j,day}}{t_\textrm{o,200s}}\Big)^{3/8}
\end{equation}
which is independent of assumptions on $\{\theta_\textrm{c}, E_\textrm{kin,0}, n\}$ from afterglow modeling. 
%Therefore, we can easily derive lower limits to both $\Gamma_0$ and $\Gamma_0\theta_\textrm{c}$, where the limit on $\Gamma_0$ depends weakly on $\{E_\textrm{kin}, n\}$ which we adopt based on the uniform afterglow modeling from \citet{RoucoEscorial2022}. 
%While $\{E_\textrm{kin}, n\}$ are correlated in afterglow modeling, we assume independent errors for the purpose of this investigation, which likely overpredicts the true error on $\Gamma_0$. In order to demonstrate further the effectiveness of utilizing the observed times for $\{t_\textrm{o},t_\textrm{j}\}$ (and their errors), we show also the result when using $\{E_\textrm{kin}, n,\theta_\textrm{c}\}$ (and their errors) from afterglow modeling \citep{RoucoEscorial2022}. 
The results are shown in Figure \ref{fig:gammalim}. %, and the method (labeled as ``observational''; purple lines) based on $\{t_\textrm{o},t_\textrm{j}\}$ has a smaller uncertainty. 
We find that 90\% of the sample has $\Gamma_0\theta_\textrm{c}$\,$>$\,$5$, while 35\% of the sample requires $\Gamma_0\theta_\textrm{c}$\,$>$\,$30$,  translating to $\Gamma_0$\,$>$\,$300$ for $\theta_\textrm{c}$\,$=$\,$0.1$ rad.

\begin{figure}
    \centering
    \includegraphics[width=\columnwidth]{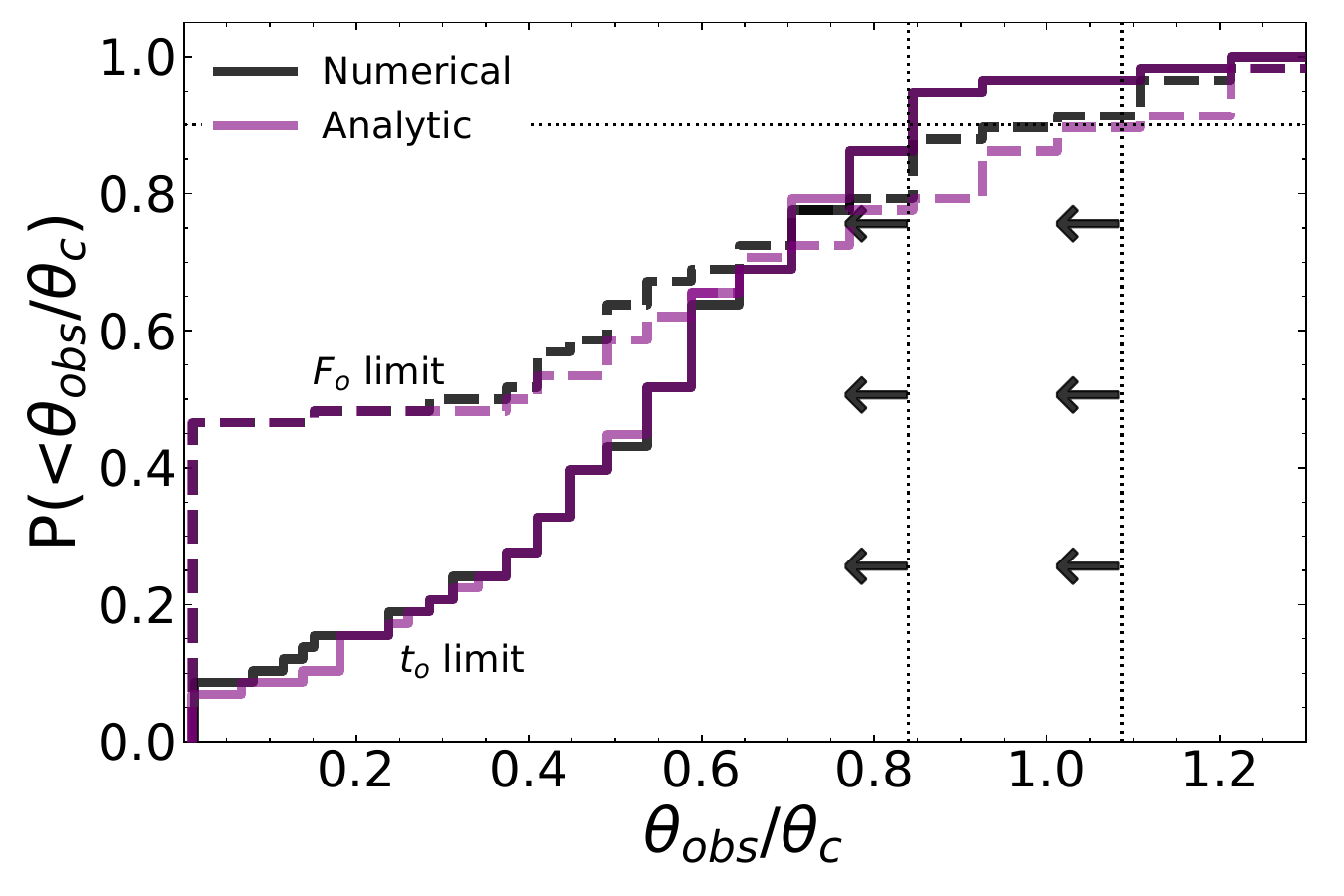}
    \includegraphics[width=\columnwidth]{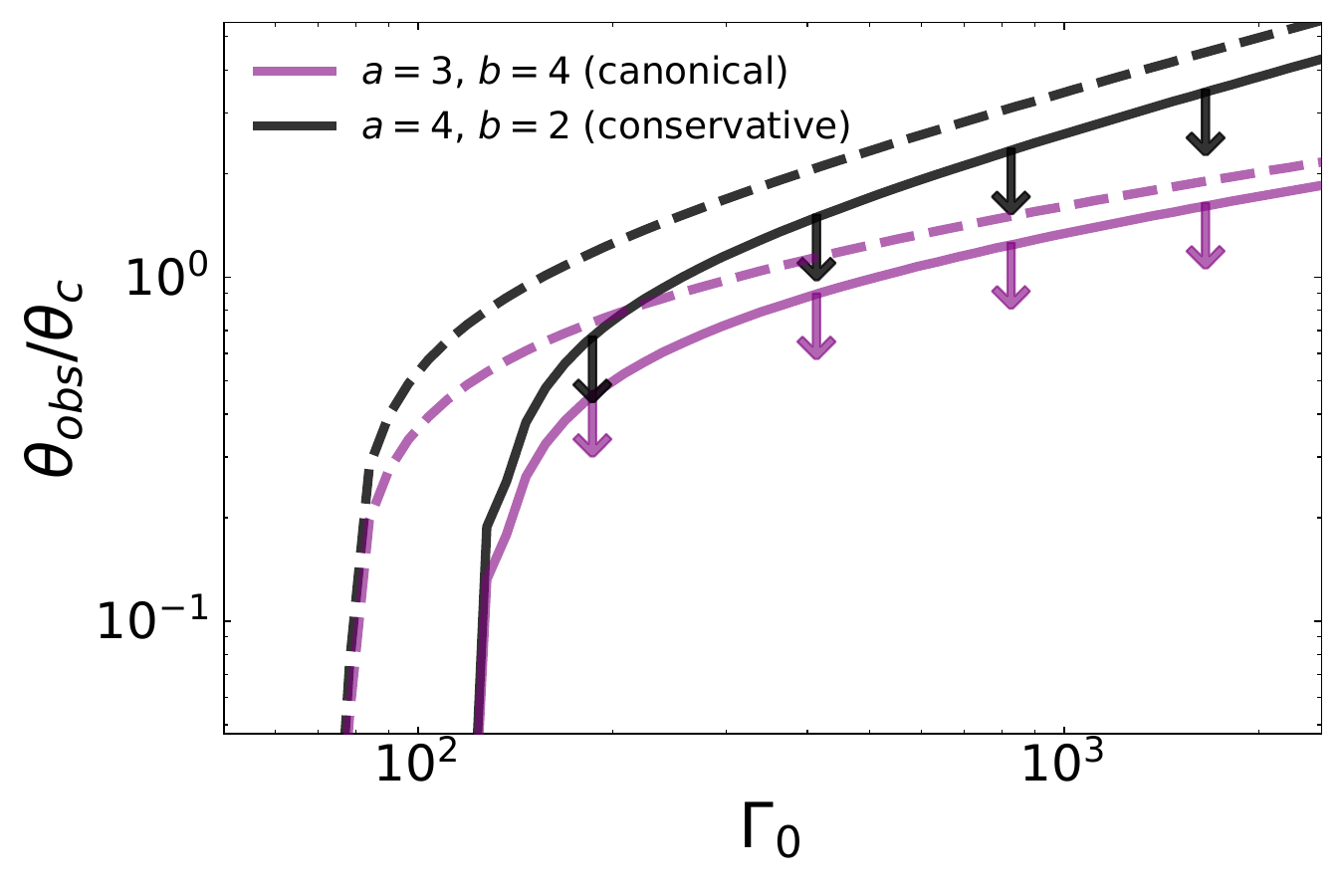}
    \vspace{-0.6cm}
    \caption{\textbf{Top:} 
    Comparison between the cumulative distribution of viewing angles derived using numerical simulations (black) versus analytic prescriptions (purple; \S \ref{methods: jet structure}) for structured jets. This was done assuming the canonical afterglow parameters $n$\,$=$\,$10^{-2}$ cm$^{-3}$, $\varepsilon_e$\,$=$\,$0.1$, $\varepsilon_B$\,$=$\,$10^{-3}$ for a jet structure of $a_\textrm{kin}$\,$=$\,$3$, $a_\gamma$\,$=$\,$3$, and $b$\,$=$\,$4$ with an initial Lorentz factor at the jet's core of $\Gamma_0$\,$=$\,$300$. Solid lines represent the limit based on $t_\textrm{o}$, and dashed lines are based on $F_\textrm{o}$.
    \textbf{Bottom:} Constraints on the viewing angle versus initial Lorentz factor at the jet's core for two jet structures: a canonical structure of $a_\textrm{kin}$\,$=$\,$3$, $a_\gamma$\,$=$\,$3$, and $b$\,$=$\,$4$ and a conservative structure of $a_\textrm{kin}$\,$=$\,$4$, $a_\gamma$\,$=$\,$4$, and $b$\,$=$\,$2$. The upper limits on $\theta_\textrm{obs}/\theta_\textrm{c}$ are the 90th percentile of the population of 58 sGRBs for each Lorentz factor. Solid lines are for the canonical afterglow parameters and dashed lines are for the conservative case (Table \ref{tab: Results_Summary}). These results are independent of assumptions for $\theta_\textrm{c}$.
    }
    \label{fig:structuredlimits}
\end{figure}

\subsection{Off-axis structured jets}
\label{sec: structjetresults}

Here, we extend our study to power-law structured jets (Equation \ref{eqn:appjetstuc}) calibrated using full numerical lightcurve integration \citep{GG2018}. 
We perform our calculation for two jet structures: a canonical structure $\{a_\textrm{kin},b\}$\,$=$\,$\{3,4\}$ and a conservative structure $\{a_\textrm{kin},b\}$\,$=$\,$\{4,2\}$, see Table \ref{tab: structure-summary}. These two structures cover the range of behaviors observed in numerical simulations of sGRB jets \citep[e.g.,][]{Gottlieb2021,Nativi2021}. The main impact of the change in jet structure is to allow for early lightcurve peaks at larger viewing angles for the conservative structure due to the higher line-of-sight Lorentz factor (and therefore earlier deceleration) at larger angles (see \S \ref{methods: jet structure}). 

First, in Figure \ref{fig:structuredlimits} (Top), we display the very good agreement between our analytic scalings (Equations \ref{eqn:structdec} and \ref{eqn:structdec2}) and the full numerical integration. There is a slight deviation for the limits based on the flux $F_\textrm{o}$ at  $\theta_\textrm{obs}/\theta_\textrm{c}$\,$>$\,$0.8$ due to de-beamed emission from smaller angles (e.g., the core) contributing marginally to the observed flux, %\footnote{See \citet{BGG2020,BGG2022} for a discussion of double peaked lightcurves and the ratio of the first to second peak as a function of viewing angle.}, 
which leads to a deviation of the analytic and numerical results at the $5\%$ level at $\theta_\textrm{obs}/\theta_\textrm{c}$\,$=$\,$1$ and increasing at larger viewing angles. We note that this has the impact of allowing for slightly larger viewing angles in the analytic approximation compared to the numerical result, which is conservative for our purposes. 
These deviations do not impact our overall result, and the values of $\theta_\textrm{obs}/\theta_\textrm{c}$ where they occur are specific to the assumed structure and Lorentz factor $\Gamma_0$ (in this case $a_\textrm{kin}$\,$=$\,$3$, $a_\gamma$\,$=$\,$3$, $b$\,$=$\,$4$, and $\Gamma_0$\,$=$\,$300$). For example, for our conservative structure with $a_\textrm{kin}$\,$=$\,$4$, $a_\gamma$\,$=$\,$4$, $b$\,$=$\,$2$, and also assuming $\Gamma_0$\,$=$\,$300$, the analytic approximation for the flux holds at the $\lesssim$\,$5\%$ level out to $\theta_\textrm{obs}/\theta_\textrm{c}$\,$\lesssim$\,$3$. 
For each jet and initial Lorentz factor we similarly confirm whether the analytic approximations hold over the relevant range of viewing angles.

For each of the 58 sGRBs in our sample we compute the most constraining limit (based on either $t_\textrm{o}$ or $F_\textrm{o}$) for a range of Lorentz factors based on the analytic method (see \S \ref{methods: jet structure}). In Figure \ref{fig:structuredlimits} (Bottom) we display the 90th percentile of the distribution of upper limits versus the initial core Lorentz factor. We consider a range of Lorentz factors between $\Gamma_0$\,$=$\,$\{50,2500\}$ for this analytic calculation. We perform this calculation for both our canonical and conservative afterglow parameters (Table \ref{tab: Results_Summary}), varying the kinetic energy and redshift per burst. For sGRBs without a measured redshift we have adopted $z$\,$=$\,$0.5$. 

For our canonical jet structure, we find that the population of sGRBs is either viewed at small angles $\theta_\textrm{obs}/\theta_\textrm{c}$\,$<$\,$1$ or that their initial core Lorentz factors are incredibly high $\Gamma_0$\,$>$\,$500$. 
We emphasize that this result is independent of $\theta_\textrm{c}$.
For our canonical structure and Lorentz factor $\Gamma_0$\,$=$\,$300$, we find $\theta_\textrm{obs}/\theta_\textrm{c}$\,$<$\,$0.74$ ($0.97$) for our canonical (conservative) afterglow parameters. The conservative structure ($b$\,$=$\,$2$) with a shallow Lorentz factor profile $\Gamma(\theta_\textrm{obs})$\,$\propto$\,$[1+(\theta_\textrm{obs}/\theta_\textrm{c})^2]^{-1}$ allows for larger viewing angles out to $\theta_\textrm{obs}/\theta_\textrm{c}$\,$\approx$\,$2$ for high Lorentz factors ($\Gamma_0$\,$\gtrsim$\,$700$). 

The above results (Figure \ref{fig:structuredlimits}) have adopted a constant gamma-ray efficiency with angle ($a_\gamma$\,$=$\,$a_\textrm{kin}$), which is expected to be unrealistic. However, utilizing a steeper value of $a_\gamma$\,$=$\,$6$, such that $a_\gamma$\,$>$\,$a_\textrm{kin}$, only marginally changes the results, and in fact is (slightly) more restrictive on the viewing angle\footnote{Note that the power-law slopes of the kinetic energy and gamma-ray energy profiles work opposite each other as seen in Equation \ref{eqn:boomboom}. A shallow gamma-ray profile and a steep kinetic energy profile are more conservative. As such the most conservative case is $a_\gamma$\,$=$\,$a_\textrm{kin}$.}. This is because a steeper gamma-ray efficiency leads to a larger core energy when converted from the line-of-sight value, which causes the jet to take longer to decelerate (for fixed $\Gamma_0$ and $n$; Equation \ref{eqn:tdec}). As such, the events in our sample can be less off-axis under this condition, though the impact is quite weak. For example, for the canonical case with  $a_\textrm{kin}$\,$=$\,$3$,  $b$\,$=$\,$4$, and $\Gamma_0$\,$=$\,$300$ the constraint on the population of allowed viewing angles changes from only $\theta_\textrm{obs}/\theta_\textrm{c}$\,$<$\,$0.74$ for $a_\gamma$\,$=$\,$3$ to $\theta_\textrm{obs}/\theta_\textrm{c}$\,$<$\,$0.71$ for $a_\gamma$\,$=$\,$6$.

\section{Discussion}
\label{sec: discussion}

\subsection{Implications for the rate of events}
\label{sec:rates}

Due to their collimation, the observed rate of sGRBs is significantly less than the intrinsic rate. The conversion between observed and intrinsic rate utilizes a beaming correction $f_\textrm{b}^{-1}$ that dictates how many more events are unobserved due to their larger viewing angles outside the core of the jet. For a tophat jet that can only be detected on-axis the correction is given by $R_\textrm{true}$\,$=$\,$f_\textrm{b}^{-1}\,R_\textrm{obs}$ where $f_\textrm{b}$\,$=$\,$1-\cos\theta_\textrm{c}$\,$\approx$\,$\theta_\textrm{c}^2/2$. 
If instead a population of tophat jets can be viewed outside the core of their jet then the effective beaming angle should extend to the largest angle that is commonly enough observed to be dominant in the observed rate. If we consider this to be the median allowed angle in our tophat jet model ($\theta_\textrm{obs}/\theta_\textrm{c}$\,$\lesssim$\,$1.04$) the correction is minor and an of order $\lesssim$\,$8\%$ correction to the rate for an assumed core half-opening angle of $\theta_\textrm{c}$\,$=$\,$0.1$ rad.

If instead we consider the more realistic situation of structured jets, our results (Figure \ref{fig:structuredlimits}) provide a typical upper limit to the viewing angle of $\theta_\textrm{obs}/\theta_\textrm{c}$\,$\lesssim$\,$0.7$ assuming our canonical parameters. If we apply the beaming correction for a tophat jet, considering $0.7\theta_\textrm{c}$ to be a typical angle dominating the observed events, this leads to a modification (decrease) by a factor of 2 to the beaming correction, and therefore the inferred rate of events. 
While this approximation holds under the assumption that the probability of detecting all smaller viewing angles is constant, and drops off quickly outside of this angle, the true correction is likely to still be of order unity. For example, for shallower jet structures, or more conservative afterglow parameters, the allowed viewing angles would lead to very similar beaming factors to a standard tophat. In the next section we simulate the detectability of a population of sGRBs over a range of redshifts and energies, and in this case derive more accurate beaming factors that naturally account for the detection probability as a function of angle.

The inferred observed rate of sGRBs depends on both the sample selection and instrument (e.g., \textit{BATSE}, \textit{Swift}, \textit{Fermi}), due both to varying sensitivity and fields of view, but is generally found to be on the order of a few events per Gpc$^3$ yr \citep{Guetta2006,Nakar2006,Coward2012,Wanderman2015,Ghirlanda2016}. The inferred rate has also been shown to depend strongly on the assumed minimum luminosity \citep[see, e.g.,][]{Wanderman2015}, where decreasing the minimum luminosity can drastically increase the expected rate of observed events. 

Assuming emission from tophat jets, past studies have found intrinsic rates in the range of  $R_\textrm{true}$\,$\approx$\,$90$\,$-$\,$1850$ Gpc$^{-3}$ yr$^{-1}$ \citep{Fong2015} and  $R_\textrm{true}$\,$\approx$\,$360$\,$-$\,$1790$ Gpc$^{-3}$ yr$^{-1}$ \citep{RoucoEscorial2022} based on the inferred collimation of the cosmological sample, and $R_\textrm{true}$\,$\approx$\,$60$\,$-$\,$360$ Gpc$^{-3}$ yr$^{-1}$ for low-luminosity events in the local Universe \citep{Dichiara2020}. 
Our results strongly support that sGRBs are viewed within or very close to the core of their jets, implying that past estimates of the beaming correction, and therefore the intrinsic rate, are supported. However, we note that the rate calculation is impacted by the assumed half-opening angles and the assumed jet structure, as well as the efficiency of detecting the jet out to larger viewing angles, which can lead to deviations from the tophat expectation, though, based on the results of this work, we expect the corrections due to jet structure and viewing angle to be of order unity.

The rate of local binary neutron star (BNS) mergers as determined by the LIGO-Virgo collaboration (LVC) is $10$\,$-$\,$1700$ Gpc$^{-3}$ yr$^{-1}$ \citep{Abbott2023gwtc3BNSrate}. The significant overlap between these two predictions sets stringent constraints on both the beaming correction (and therefore jet structure and gamma-ray production; \citealt{Dichiara2020}) and the fraction of BNS mergers producing successful relativistic jets \citep{Beniamini2019structuredjet}. Future LVC observing runs, as well as next-generation GW detectors \citep{ET2010,CE2019,ET2020}, will constrain the rate of BNS mergers, and therefore the rate of sGRB production. These future GW observations may also allow for the localization and characterization of afterglows following BNS mergers, improving our understanding of their angular structure and beaming factors.

\subsection{Implications for the prompt emission}
\label{sec:detectability}

\subsubsection{Motivation}

Our results (\S \ref{sec: results}), and in particular the lack of far off-axis cosmological GRBs observed in the literature to date, indicate that the prompt gamma-ray emission must become inefficient or shut-off outside of the jet's core (e.g., \citealt{BeniaminiNakar2019,Beniamini2019structuredjet,Beniamini2019plateau}; similar constraints on the viewing angle for different jet structures were also derived in \citealt{Gill+20} based on compactness arguments). For example, \citet{BeniaminiNakar2019} argued that the gamma-ray production of long duration GRBs has to shut off outside $\sim$\,$2\theta_\textrm{c}$ in order to be consistent with observations (i.e., to not overproduce lower luminosity events or off-axis afterglow lightcurves). If the efficiency of gamma-ray production is constant with angle, even if the energy profile falls off steeply outside the core, it is difficult to explain the lack of GRBs obviously observed off-axis ($\theta_\textrm{obs}$\,$>$\,$\theta_\textrm{c}$) within the cosmological sample.

%\begin{eqnarray}
%\epsilon(\theta) = \frac{dE}{d\Omega} = \epsilon_0 \left\{ \begin{array}{ll} 1 & \theta < \theta_\textrm{c}, \\
%\left( \frac{\theta}{\theta_\textrm{c}} \right)^{-a} &  \theta \geq \theta_\textrm{c}, 
%\end{array} \right. \\
%\Gamma(\theta) = 1 + (\Gamma_0 -1)  \left\{ \begin{array}{ll} 1 & \theta < \theta_\textrm{c}, \\
%\left( \frac{\theta}{\theta_\textrm{c}} \right)^{-b} &  \theta \geq \theta_\textrm{c}. 
%\end{array} \right.
%\end{eqnarray}

\subsubsection{Simulation methodology}

To test this, we performed Monte Carlo simulations of the detectability of off-axis sGRBs in both gamma-rays and X-rays using the sensitivity of the \textit{Swift} mission. We assume that the jet has a power-law structure (Equation \ref{eqn:appjetstuc}) in isotropic equivalent gamma-ray energy $E_\gamma(\theta_\textrm{obs})$\,$\propto$\,$\Theta^{-a_\gamma}$, isotropic equivalent kinetic energy $E_\textrm{kin}(\theta_\textrm{obs})$\,$\propto$\,$\Theta^{-a_\textrm{kin}}$, and Lorentz factor $\Gamma(\theta_\textrm{obs})$\,$\propto$\,$\Theta^{-b}$. Based on energy budget considerations we require that $a_\gamma$\,$\geq$\,$a_\textrm{kin}$, an inequality that was inferred for GW170817 \citep[e.g.,][]{Beniamini2019structuredjet,Hotokezaka+19,Ghirlanda2019,Mooley2022}). 
%We assume that the jet has a power-law structure in isotropic energy:  %and Lorentz factor: 
%\begin{eqnarray}
%E(\theta) = E_0 \left\{ \begin{array}{ll} 1 & \theta < \theta_\textrm{c}, \\
%\left( \frac{\theta}{\theta_\textrm{c}} \right)^{-a} &  \theta \geq \theta_\textrm{c}. 
%\end{array} \right. %\\
%\Gamma(\theta) = 1 + (\Gamma_0 -1)  \left\{ \begin{array}{ll} 1 & \theta < \theta_\textrm{c}, \\
%\left( \frac{\theta}{\theta_\textrm{c}} \right)^{-b} &  \theta \geq \theta_\textrm{c}. 
%\end{array} \right.
%\end{eqnarray}
%We adopt a canonical value of $a_\gamma$\,$=$\,$6$ for the gamma-ray energy and $a_\textrm{kin}$\,$=$\,$4.5$ for the kinetic energy, consistent with values inferred for GW170817 \citep{Beniamini2019structuredjet}. 
As such, the prompt gamma-ray efficiency varies with angle $\varepsilon_\gamma(\theta_\textrm{obs})$\,$\propto$\,$\Theta^{-c}$, but is normalized to $\varepsilon_{\gamma,0}$\,$=$\,$0.15$ at the core \citep{Nava2014,Beniamini2015,Beniamini2016corr}. The slope of the distribution for $\varepsilon_\gamma$ is roughly given by $c$\,$=$\,$a_\gamma$\,$-$\,$a_\textrm{kin}$. 
As an illustration we adopt canonical values of $a_\gamma$\,$=$\,$6$ for the gamma-ray energy, $a_\textrm{kin}$\,$=$\,$3$ for the kinetic energy, and $b$\,$=$\,$4$ for the Lorentz factor (as used in \S \ref{methods: jet structure} and \S \ref{sec: structjetresults}). This leads to $\varepsilon_\gamma(\theta_\textrm{obs})$\,$\propto$\,$\Theta^{-3}$.

We also considered the addition of a second structured jet model consisting of the power-law model (described above) with an added quasi-isotropic component due to cocoon shock breakout. 
We tested the simple ``cocoon-like'' model outlined by \citet{Beniamini2019structuredjet} (see also \citealt{Bhattacharjee2024}), which is based on \citet{Duffell2018}, but found that it did not impact our results at cosmological distances due to the low luminosity of the cocoon. 
%We similarly adopt a cocoon breakout efficiency (the ratio of the true jet energy to the cocoon energy released in gamma-rays) of $\eta_\textrm{br}$\,$=$\,$10^{-3}$. \citet{Dichiara2020} previously showed that $\eta_\textrm{br}$\,$=$\,$10^{-2}$\,$-$\,$10^{-3}$ was required for consistency between the rate of local sGRBs ($\lesssim$\,$200$ Mpc) and the rate of neutron star mergers. 

For each  GRB in our simulation, we sample a random orientation (assuming an isotropic distribution) relative to the observer's line of sight to determine the viewing angle $\theta_\textrm{obs}$. The isotropic gamma-ray energy of the jet's core is derived from a broken power-law gamma-ray luminosity function \citep{Ghirlanda2016} assuming a rest frame duration of $T_{90}$\,$=$\,$0.2$ s. 
The median core energy for this distribution is $E_{\gamma,0}$\,$=$\,$2\times10^{51}$ erg ($1$\,$-$\,$10,000$ keV; rest frame). From this distribution \citep{Ghirlanda2016}, we sample a core gamma-ray energy $E_{\gamma,0}$ to compute $E_{\gamma}(\theta_\textrm{obs})$ and $E_\textrm{kin}(\theta_\textrm{obs})$. We also randomize the prompt emission spectral parameters according to the distributions derived from \textit{Fermi} short GRBs \citep{Nava2011}: $\log E_\textrm{p,c}$\,$=$\,$2.9\pm0.2$ keV\footnote{Here we have converted the observed distribution of peak energies $E_\textrm{p,c}$ to the central engine frame assuming the median redshift of our sample of events.} and  $\alpha$\,$=$\,$-0.5\pm0.4$ for fixed $\beta$\,$=$\,$-2.25$. We fix the peak energy in the comoving frame such that $E_\textrm{p}(\theta_\textrm{obs})$\,$=$\,$E_\textrm{p,c}\Gamma(\theta_\textrm{obs})/\Gamma_0$  \citep{Bhattacharjee2024}\footnote{This is a model dependent assumption that is not driven directly by observations, though we note it does not strongly impact or modify our results.}. 
The redshift of each simulated sGRB is sampled from the intrinsic distribution derived by \citet{Ghirlanda2016} using $P(z)$\,$=$\,$\frac{1}{1+z}\frac{dV}{dz}\psi(z)$, where $\Psi(z)$ is the comoving rate per unit volume and $\frac{dV}{dz}$ is the differential comoving volume. 

We first compute the detectability in gamma-rays by \textit{Swift}/BAT, requiring that the fluence is $\phi_\gamma(\theta_\textrm{obs})$\,$>$\,$2\times10^{-8}$ erg cm$^{-2}$ in the (observer frame) $15$\,$-$\,$150$ keV energy range \citep{Lien2014}. We only consider the line-of-sight gamma-ray emission, and do not include de-beamed emission (Doppler suppression) from the jet's core \citep[see, e.g.,][]{Beniamini2019structuredjet,Bhattacharjee2024}. This  is because for the assumed jet structure the line-of-sight gamma-ray energy $E_\gamma (\theta_\textrm{obs})$ and spectral peak energy $E_\textrm{p}(\theta_\textrm{obs})$ dominate over the de-beamed component for this jet structure. 

\begin{figure}
    \centering
\includegraphics[width=\columnwidth]{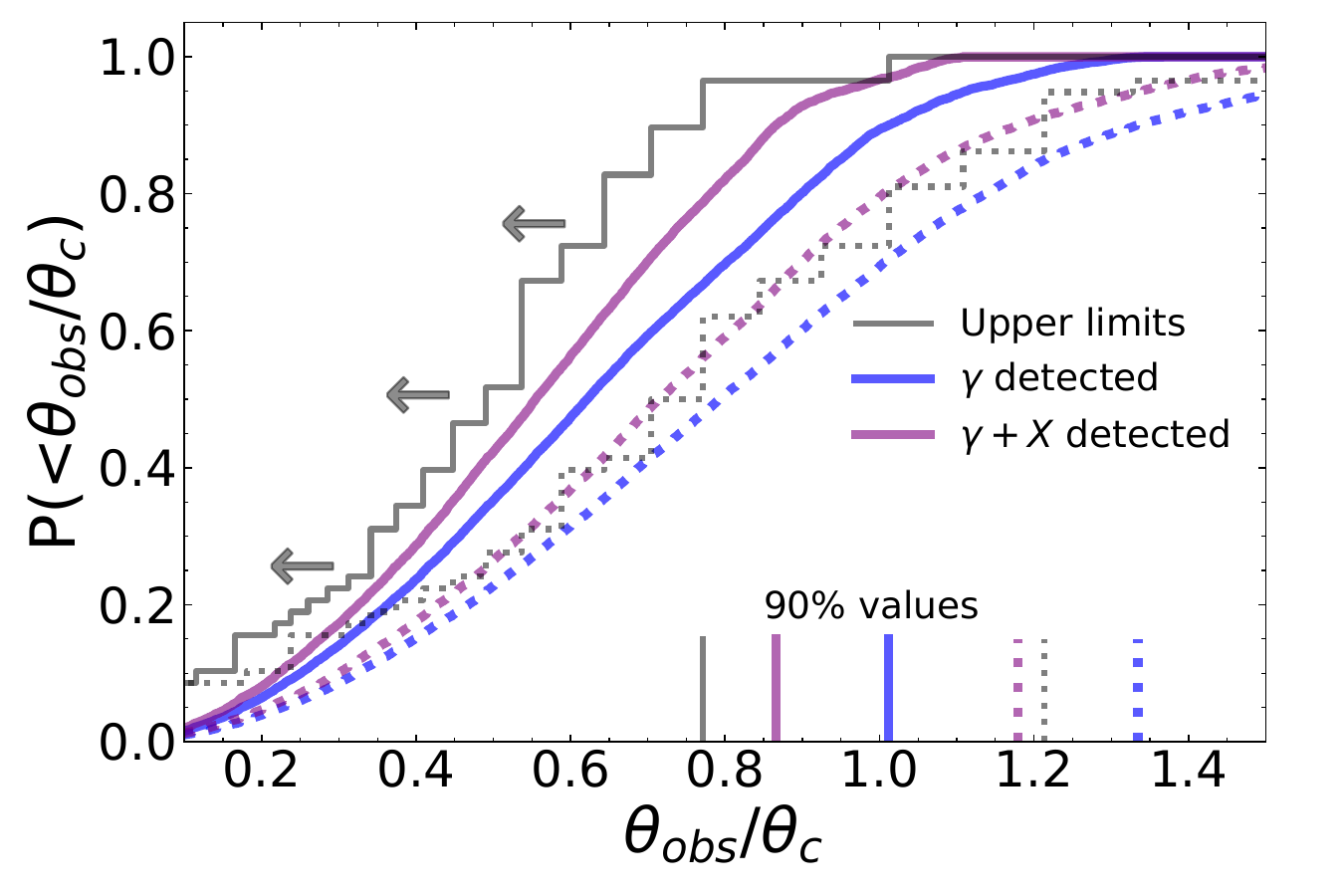}
    \vspace{-0.6cm}
    \caption{Cumulative distribution of viewing angles for simulated sGRBs detectable in gamma-rays (blue) and both gamma-rays and X-rays (purple). 
    Solid lines represent a power-law jet with $a_\gamma$\,$=$\,$10$, $a_\textrm{kin}$\,$=$\,$3$, and $b$\,$=$\,$4$ and dashed lines represent $a_\gamma$\,$=$\,$7$, $a_\textrm{kin}$\,$=$\,$4$, and $b$\,$=$\,$2$. The value of $a_\gamma$ is chosen to re-produce the early \textit{Swift}/XRT completeness of sGRBs. 
    Both models were computed for $n$\,$=$\,$10^{-2}$ cm$^{-3}$, $\varepsilon_e$\,$=$\,$0.1$, $\varepsilon_B$\,$=$\,$10^{-3}$, and $\Gamma_0$\,$=$\,$300$. We also show the upper limit to viewing angle (gray lines) for both jet structures based on the time of the first X-ray detection. 
    Each of these results are independent of assumptions for $\theta_\textrm{c}$. The vertical lines mark the 90th percentile of each distribution.  
    }
    \label{fig:detectability}
\end{figure}

\begin{table}
\centering
\caption{Results of the simulation of sGRB X-ray and gamma-ray detectability requiring $t_\textrm{p}$\,$<$\,$1$ hr and $F_\textrm{p}$\,$>$\,$10^{-12}$ erg cm$^{-2}$ s$^{-1}$. For varying values of $n$ and $\varepsilon_B$ we have determined the slope of the gamma-ray energy distribution that roughly re-produces the early-time sGRB X-ray completeness of $\sim82$\%. We performed this simulation for both our canonical and conservative jet structures (\S \ref{methods: jet structure}). 
}
\begin{tabular}{ccccccc}
\hline
\hline
$n$  & $\varepsilon_B$ & $a_\textrm{kin}$ & $a_\gamma$ & $b$ & X-ray Det. Frac. & $f_\textrm{b}^{-1}$ \\
(cm$^{-3}$) & & & &  & & \\
\hline
$10^{-1}$ & $10^{-2}$ & 3 & 5.5 & 4  & 78\% & 170 \\
$10^{-1}$ & $10^{-3}$ & 3 & 6.5 & 4  &  81\%& 170 \\
$10^{-2}$ & $10^{-2}$ & 3 & 6.5 & 4  & 77\% & 240 \\
$10^{-2}$ & $10^{-3}$ & 3 & 10 & 4  & 79\% & 490 \\
$10^{-3}$ & $10^{-2}$ & 3 & 10.5 & 4  & 79\%& 500 \\
\hline 
$10^{-1}$ & $10^{-2}$ & 4 & 3.5 & 2  & 78\% & 60  \\
$10^{-1}$ & $10^{-3}$ & 4 & 4.5 & 2  & 79\% & 115 \\
$10^{-2}$ & $10^{-2}$ & 4 & 4.5 & 2  & 80\%& 115 \\
$10^{-2}$ & $10^{-3}$ & 4 & 7 & 2  & 82\% & 250 \\
$10^{-3}$ & $10^{-2}$ & 4 & 7 & 2  & 83\% & 250 \\
\hline
\end{tabular}
\label{tab: det_summary}
\end{table}

%\begin{table}
%\centering
%\caption{Results of the simulation of sGRB X-ray and gamma-ray detectability requiring $t_\textrm{p}$\,$<$\,$1$ d and $F_\textrm{p}$\,$>$\,$10^{-13}$ erg cm$^{-2}$ s$^{-1}$. For varying values of $n$ and $\varepsilon_B$ we have determined the slope of the gamma-ray energy distribution that roughly re-produces the observed sGRB X-ray completeness of $\sim69$\,$-$\,$74$\%. We performed this simulation for both our canonical and conservative jet structures (\S \ref{methods: jet structure}). 
%}
%\begin{tabular}{cccccc}
%\hline
%\hline
%$n$  & $\varepsilon_B$ & $a_\textrm{kin}$ & $a_\gamma$ & $b$ & Det. Frac.  \\
%(cm$^{-3}$) & & & &  &  \\
%\hline
%$10^{-1}$ & $10^{-2}$ & 3 & 3.5 & 4  & 68\% \\
%$10^{-1}$ & $10^{-3}$ & 3 & 4.5 & 4  &  75\%\\
%$10^{-2}$ & $10^{-2}$ & 3 & 4.5 & 4  & 73\% \\
%$10^{-2}$ & $10^{-3}$ & 3 & 6 & 4  & 72\% \\
%$10^{-3}$ & $10^{-2}$ & 3 & 5.5 & 4  & 68\% \\
%$10^{-3}$ & $10^{-3}$ & 3 & 8 & 4  & 69\% \\
%\hline 
%$10^{-1}$ & $10^{-2}$ & 4 & 3 & 2  & 75\%  \\
%$10^{-1}$ & $10^{-3}$ & 4 & 3.5 & 2  & 73\%  \\
%$10^{-2}$ & $10^{-2}$ & 4 & 3.5 & 2  & 73\% \\
%$10^{-2}$ & $10^{-3}$ & 4 & 4 & 2  & 78\% \\
%$10^{-3}$ & $10^{-2}$ & 4 & 4.5 & 2  & 80\%  \\
%$10^{-3}$ & $10^{-3}$ & 4 & 6 & 2  &  77\% \\
%\hline
%\end{tabular}
%\label{tab: det_summary}
%\end{table}

We then compute the detectability of the X-ray afterglow peak luminosity for our canonical parameters ($\theta_\textrm{c}$\,$=$\,$0.1$ rad, $n$\,$=$\,$10^{-2}$ cm$^{-3}$, $\varepsilon_e$\,$=$\,$0.1$, $\varepsilon_B$\,$=$\,$10^{-3}$, and $\Gamma_0$\,$=$\,$300$; Table \ref{tab: Results_Summary}). 
We consider X-ray detection criteria that is representative of the lightcurves of sGRBs in our sample (Figure \ref{fig:lightcurves}), corresponding to $t_\textrm{p}$\,$<$\,$1$ hr and $F_\textrm{p}$\,$>$\,$10^{-12}$ erg cm$^{-2}$ s$^{-1}$
%Assuming follow-up  by \textit{Swift}/XRT, we require $t_\textrm{p}$\,$<$\,$1$ d and $F_\textrm{p}$\,$>$\,$10^{-13}$ erg cm$^{-2}$ s$^{-1}$ for the event to be detected. This criteria is such that typical follow-up of a new GRB by \textit{Swift} would likely detect its afterglow (though we note our criteria is rather conservative from the point of view of detecting off-axis events). 
However, if the GRB is not detected by \textit{Swift}/BAT, we do not consider it detectable by \textit{Swift}/XRT, as \textit{Swift} would have been unlikely to slew to the GRB position within $<$1 hr (or even $<$1 d). 
We do not account for other emission mechanisms that may increase the X-ray detectability at early times (less than a few hours) such as early steep decline, internal plateaus, or reverse shock. 

We aim to reproduce the \textit{Swift} sGRB X-ray completeness, which ranges between $69$\,$-$\,$74\%$ depending on the selection criteria \citep{Fong2015,OConnor2020,OConnor2022}. The completeness increases to $82\%$ if considering only events where \textit{Swift} was able to rapidly slew \citep[see the discussion in][]{OConnor2020}. 
We repeat this simulation for $5\times10^6$ events for each jet structure and set of parameters explored below.

\subsubsection{Simulated sGRB detectability}

Our results are displayed in Figure \ref{fig:detectability}. 
The fraction of gamma-ray detected events also detectable in X-rays is $79\%$ (referred to as X-ray Det. Frac. in Table \ref{tab: det_summary}).  This suggests that for this assumed jet structure and afterglow parameters that $a_\gamma$\,$\approx$\,$10$ (corresponding to $\varepsilon_\gamma(\theta_\textrm{obs})$\,$\propto$\,$\Theta^{-c}$ with $c$\,$\approx$\,$7$) can roughly reproduce the main observations.
The distribution of viewing angles (Figure \ref{fig:detectability}) is also broadly consistent with our upper limits for structured jets (\S \ref{sec: structjetresults}), where slight deviations exist because  the assumed detectable value of $t_\textrm{p}$\,$<$\,$1$ hr is at the upper end of our distribution of observed limits to the peak time (Figure \ref{fig:lightcurves}). %and that simulation uses sGRB core energies as core and the upper limits convert LOS energy to core; marginal factor...

As the afterglow detectability depends strongly on the assumed parameters, we tested our assumed values by varying $n$ and $\varepsilon_B$. The results are tabulated in Table \ref{tab: det_summary}. As expected, increasing $n$ and $\varepsilon_B$ requires significantly shallower gamma-ray energy profiles. For the conservative jet structure we find that in order to produce the X-ray completeness fraction we require gamma-ray profiles shallower than even the kinetic energy ($a_\gamma$\,$<$\,$a_\textrm{kin}$), which predicts an increasing gamma-ray efficiency at large angles, incompatible with theoretical expectations. Both $n$ and $\varepsilon_B$ are essentially unknown and can span many order of magnitude as commonly observed in afterglow modeling or population level analyses. In any case, our results strongly imply that either the majority of the sGRB population is occurring in extremely low density environments ($\lesssim$\,$10^{-3}$ cm$^{-3}$), or the Lorentz factor profiles of their jets are substantially steep $b$\,$>$\,$2$.

If instead we adopt different detection thresholds based instead on what \textit{Swift} follow-up is likely to be able to detect ($t_\textrm{p}$\,$<$\,$1$ d and $F_\textrm{p}$\,$>$\,$10^{-13}$ erg cm$^{-2}$ s$^{-1}$) the allowed values of gamma-ray efficiency become substantially shallower. These detectability criteria allow for sGRBs to be detected out to $\theta_\textrm{obs}/\theta_\textrm{c}$\,$\approx$\,$2$ (for either jet structure) and highlight the need for continued X-ray follow-up (e.g., \textit{Swift}, \textit{Chandra}, \textit{XMM-Newton}, \textit{Einstein Probe}) at later times for sGRBs without prompt X-ray detection. 
In particular for the conservative jet structure ($a_\textrm{kin}$\,$=$\,$4$ and $b$\,$=$\,$2$) we similarly find that the gamma-ray efficiency cannot fall off quickly with angle. 
For example, for the same canonical set of afterglow parameters and $a_\gamma$\,$=$\,$4$ (a constant gamma-ray efficiency with angle) we derive an X-ray completeness of $78\%$. If instead we allow for steeper values such as $a_\gamma$\,$\gtrsim$\,$5$, we find a $100\%$ X-ray detection rate, which is incompatible with observations. The inclusion of other X-ray emission mechanisms, such as early steep decline or reverse shock, would only contribute to increase the detectability. 
This simulation suggests that if indeed sGRB jets have shallow Lorentz factor profiles (enabling detections at larger viewing angles) that the gamma-ray efficiency cannot decline very steeply, if at all, with angle. 
This is not completely unreasonable given that the gamma-ray emission is potentially linked with the value of the Lorentz factor \citep[e.g.,][]{lamb2018,BeniaminiNakar2019}, which would also decrease more slowly with angle. Shallow Lorentz factor profiles can also naturally account for compactness constraints as well as for the small scatter in $E_{\gamma}/L_{X}$ observed for long GRBs \citep{BeniaminiNakar2019}.

Based on all of these results, we find that even if all sGRB jets have a similar energy structure to GW170817 they cannot be viewed very far off-axis, which is consistent with our upper limits inferred from their X-ray afterglows.  
The angles at which the gamma-rays are detectable for the cosmological sample are in stark contrast compared to GW170817 which has $\theta_\textrm{obs}/\theta_\textrm{c}$\,$\approx$\,$5$\,$-$\,$6$ when including measurements of the jet's proper motion \citep[e.g.,][]{Mooley2018,Mooley2022,Ghirlanda2022,Hotokezaka+19,Fernandez2022,Makhathini2021ApJ,Govreen-Segal2023,Ryan2023}. This is not surprising considering that the GW trigger enabled a detection of a burst with much lower luminosity in GW170817, as compared to the cosmological sGRB population. 

\subsubsection{Beaming factor for simulated events}

Similar to \citet{Dichiara2020}, we track in our simulations the effective beaming factor $f_\textrm{b}^{-1}$\,$=$\,$N_\textrm{sim}/N_\textrm{det}$ where $N_\textrm{sim}$ is the number of simulated events and $N_\textrm{det}$ is the number of events detected in prompt gamma-rays. We note here that this beaming factor is not strictly a geometrical quantity, as generally used in rate calculations (\S \ref{sec:rates}), but instead accounts for varying detection efficiency with energy, redshift, and viewing angle over all simulated sGRBs. The derived beaming factors are tabulated in Table \ref{tab: det_summary}. These values are independent of assumptions for the X-ray afterglow detectability or afterglow parameters in general.

Even for a fixed core half-opening angles $\theta_\textrm{c}$ the beaming factor depends strongly on the jet structure, but is generally within a factor of 2 of the tophat expectation ($f_\textrm{b}^{-1}$\,$=$\,$200$). If the current tension between the inferred intrinsic rate of sGRBs and the BNS rate derived from gravitational waves grows, then the subtleties regarding the conversion between observed and intrinsic sGRB (local) rates will become more important. We emphasize that the inferred beaming factor from these simulations of cosmological sGRBs is substantially different for simulated events in the local universe for the same jet structures as the nearby distance allows for events to be detected at further angles from the core (e.g., GW170817).

\subsection{Quasi-universal jets}
\label{sec: quasi-universal jets}

The jet structure of gamma-ray bursts is poorly understood with a broad range of distributions observed in hydrodynamic \citep[e.g.,][]{Granot2018,Gottlieb2021} and magnetohydrodynamic \citep[e.g.,][]{Gottlieb2021magneto,Gottlieb2022,Nathanail+20,Nathanail+21} numerical simulations. %Despite this diversity, 
Multiple authors have proposed a quasi-universal jet model \citep[e.g.,][]{Lipunov2001,Rossi02,Salafia2015,Salafia2019,Salafia2023} in which all sGRBs have similar jets. Recently, \citet{Salafia2023} presented an analysis of \textit{Fermi}/GBM sGRBs in the context of a quasi-universal jet model where the prompt emission luminosity at the core can vary slightly, but each sGRB jet has the same angular structure. Their results favor a power-law model for the prompt gamma-ray luminosity structure with slope $L_\gamma(\theta_\textrm{obs})$\,$\propto$\,$\theta_\textrm{obs}^{-4.7}$ outside a narrow core of half-opening angle $\theta_\textrm{c}$\,$\approx$\,$0.036$ rad (2.1 deg; \citealt{Salafia2023}). In this model the observed variation in sGRB gamma-ray luminosity is due to \textit{i}) the scatter in the intrinsic core luminosity, \textit{ii}) viewing angle effects due to the angular structure, and \textit{iii}) variations in the prompt emission spectra (mainly the peak energy $E_p$) with angle. In this context, we can compare our constraints on the viewing angle of sGRBs derived using information from both their prompt gamma-rays and X-ray afterglows from \textit{Swift} to these constraints derived using prompt gamma-rays from \textit{Fermi}. 

It is worth noting that their inferred angular structure $L_\gamma(\theta_\textrm{obs})$\,$\propto$\,$\theta_\textrm{obs}^{-4.7}$ ($a_\gamma$\,$=$\,$\alpha_{L_\gamma}$\,$\approx$\,$3$\,$-$\,$6$ at 90\% confidence; \citealt{Salafia2023}) is slightly steeper than both our canonical and conservative jet structures, which asymptotically approach $E_\textrm{kin}(\theta_\textrm{obs})$\,$\propto$\,$\theta_\textrm{obs}^{-3}$ to $\theta_\textrm{obs}^{-4}$ at large angles. 
However, the relation between the angular structure of $L_\gamma(\theta_\textrm{obs})$ is not necessarily directly related to the kinetic energy profile or Lorentz factor profile. For example, \citet{Beniamini2019structuredjet} argued that efficiency of gamma-ray production decreases with angle, such that a steep structure for gamma-ray luminosity can be produced by shallower kinetic energy profiles (i.e., $a_\gamma$\,$>$\,$a_\textrm{kin}$). In this vein, \citet{Salafia2023} derive the constraints $a_\textrm{kin}$\,$\lesssim$\,$6$ and $b$\,$\lesssim$\,$3$ at the 90\% confidence levels, which are both roughly consistent with our assumed structured jet models. 

 \begin{figure}
    \centering
    \includegraphics[width=\columnwidth]{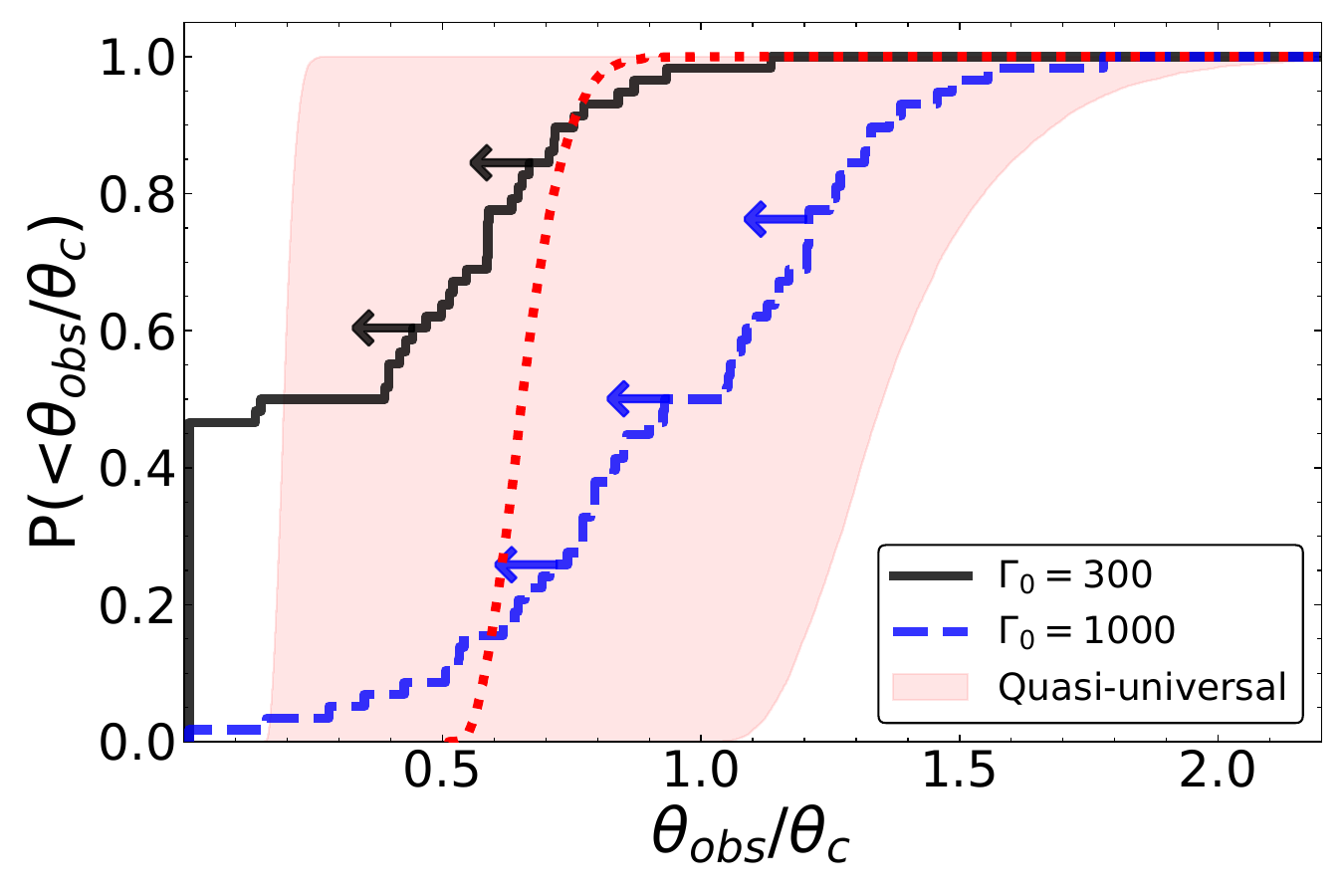}
    \vspace{-0.6cm}
    \caption{Comparison of viewing angle distributions between our work (\S \ref{sec: structjetresults}) and the quasi-universal jet model from \citet{Salafia2023}.
    The mean (dotted red line) and 90\% confidence interval (shaded red region) for their full sample are shown \citep{Salafia2023}. 
    Our results are displayed for the canonical afterglow parameters ($n$\,$=$\,$10^{-2}$ cm$^{-3}$, $\varepsilon_e$\,$=$\,$0.1$, $\varepsilon_B$\,$=$\,$10^{-3}$) for both $\Gamma_0$\,$=$\,$300$ (solid black line) and $\Gamma_0$\,$=$\,$1000$ (dashed blue line). We have utilized a structured jet with $a_\textrm{kin}$\,$=$\,$3$, $a_\gamma$\,$=$\,$3$, and $b$\,$=$\,$4$ as outlined in \S \ref{methods: jet structure}. 
    }
    \label{fig:salafia}
\end{figure}

%\textcolor{purple}{$\alpha_{E_p}\approx 1-3$ at 90\%. $a_\gamma=\alpha_{L_\gamma}\approx 3-6$ at 90\% (really $4.7^{+1.2}_{-1.4}$ and can assume this is $a_\gamma$). $\theta_\textrm{c}=2.1^{+2.4}_{-1.4}=0.7-4.5$ deg 90\% conf = 0.012-0.078 rad...}

A comparison between the distribution of viewing angles derived by \citet{Salafia2023}\footnote{\url{https://github.com/omsharansalafia/grbpop}} and the upper limits derived for our structured jet model (\S \ref{methods: jet structure}) are shown in Figure \ref{fig:salafia}. The distribution of viewing angles in the quasi-universal jet model displayed here are computed assuming the sensitivity of \textit{Swift}/BAT,  requiring a peak photon flux $>$\,$3.5$ cm$^{-2}$ s$^{-1}$ in the $15$\,$-$\,$150$ keV band \citep{Salafia2023}.
We show only the results for our canonical jet structure and canonical afterglow parameters. While we cannot exclude their model's description of the sGRB population, this comparison does imply that including available afterglow data (while difficult for a large sample of \textit{Fermi} GRBs) can improve constraints on the quasi-universal jet model. We do note that, as shown in our simulations in the previous section, the detection of gamma-rays at larger angles (even with a steep gamma-ray efficiency profile) is more efficient than for the X-ray afterglow at those same angles, suggesting that our limits are likely biased in some way to smaller angles than inferred from the gamma-ray data \citep{Salafia2023}. 

That said, our comparison (Figure \ref{fig:salafia})  still implies that either  
\textit{i}) the quasi-universal jet model requires large initial core Lorentz factors in excess of a few hundreds, 
\textit{ii}) the population of sGRBs has larger typical environmental densities than $10^{-2}$ cm$^{-3}$, 
\textit{iii}) or sGRB jets have shallow Lorentz factor profiles (e.g., $b$\,$\approx$\,$2$). We note that a high environmental density of $n$\,$\gtrsim$\,$10^{-2}$ cm$^{-3}$ for the majority of the population is somewhat at odds with observations \citep{Fong2015}, and afterglow modeling of GW170817. 
Moreover, in \S \ref{sec:detectability} we showed that shallow Lorentz factor profiles (consistent with the inferences of \citealt{Salafia2023}) lead to the requirement of a near constant gamma-ray efficiency, which is at odds with observations given the lack of large viewing angles derived for cosmological sGRBs and indeed inferred by \citet{Salafia2023}, see Figure \ref{fig:salafia}. In addition, we showed that shallow Lorentz factor profiles also overpredict the observed X-ray completeness of \textit{Swift} sGRBs. This leaves high Lorentz factors as a potential explanation for the contrast between our upper limits and their model. 

Furthermore, the small distribution of allowed viewing angles (Figure \ref{fig:salafia}) makes constraining the quasi-universal jet model difficult as the main variation in gamma-ray luminosity is supposed to be due to viewing angle effects. We note that the model of \citet{Salafia2023} is steep enough to probe variations due to angle as by $\theta_\textrm{obs}$\,$\approx$\,$\theta_\textrm{c}$ the luminosity $L_\gamma(\theta_\textrm{obs})$ has decreased by a factor of 2 and by $\theta_\textrm{obs}$\,$\approx$\,$2\theta_\textrm{c}$ by a factor of 30. However, this variation is significantly smaller than the observed range of luminosities for sGRBs which spans a factor of $1000$ (excluding GRB 170817A). Therefore, the viewing angle is unlikely to be the dominant effect on the observed scatter in gamma-ray luminosity, which is consistent with our conclusion that cosmological sGRBs cannot be viewed far off-axis. 
The jet structure and opening angles of sGRBs can be further constrained and improved by future GW detections \citep{Farah2020,Biscoveanu2020,Hayes2020,Sarin2022b,Hayes2023}, increasing the sample of far off-axis events that will strongly dictate the allowed jet structure at large angles.

%%%%%%%%%%%%%%%%%%%%%%%%%%%%%%%%%%%%%%%%%%%%%%%%%%

\section{Conclusions}
\label{sec: conclusions}

In this work we investigated the maximum allowed off-axis viewing angle for a sample of 58 cosmological sGRBs with early X-ray lightcurves. The time and flux of the first X-ray detection set robust constraints on the ratio $\theta_\textrm{obs}/\theta_\textrm{c}$. We consider both tophat jets and jets with a power-law angular structure. 
Regardless of jet structure, we find that the population of cosmological sGRBs must be viewed close to the core of their jets. 

In particular, for structured jets (\S \ref{methods: jet structure} and \S \ref{sec: structjetresults}) we obtain the 90th percentile limits $\theta_\textrm{obs}/\theta_\textrm{c}$\,$<$\,$0.74$ ($0.97$) 
for our canonical (conservative) afterglow parameters assuming a structured jet with $a_\gamma$\,$=$\,$3$, $a_\textrm{kin}$\,$=$\,$3$, and $b$\,$=$\,$4$ (Table \ref{tab: structure-summary}) with initial Lorentz factor $\Gamma_0$\,$=$\,$300$. We find that either the majority (or perhaps all) sGRBs are viewed within their jet's core, or else their jets must have extremely fast cores with $\Gamma_0$\,$>$\,$500$ (Figure \ref{fig:structuredlimits}). For a large fraction of sGRBs to be viewed outside of $\theta_\textrm{obs}/\theta_\textrm{c}$\,$>$\,$2$ requires that most sGRBs have extremely large Lorentz factors ($\Gamma_0$\,$>$\,$3000$; for our canonical jet structure).  
We provide a formula fit to the canonical case (solid purple line) of Figure \ref{fig:structuredlimits} showing how the 90th percentile upper limit of $\theta_\textrm{obs}/\theta_\textrm{c}$ depends on the initial Lorentz factor at the jet's core: 
\begin{equation}
\theta_\textrm{obs}/\theta_\textrm{c} = \sqrt{[a(\Gamma_0+b)]^c-1}
\end{equation}
where $a=7.75\times10^{-3}$, $b=99.3$, $c=0.498$. %$a=(7.75\pm0.25)\times10^{-3}$, $b=99.3\pm3.9$, $c=0.498\pm0.006$. 
This fit is valid for $\Gamma_0$\,$>$\,$120$, and below this Lorentz factor our  sample of sGRBs is constrained to have an observer aligned with the symmetry axis of the jet. 

If we consider jets with shallower Lorentz factor profiles $\Gamma(\theta)$ then the constraints on viewing angle become less restrictive, but still in general require $\theta_\textrm{obs}/\theta_\textrm{c}$\,$<$\,$2$ for reasonable Lorentz factors ($\Gamma_0$\,$<$\,$700$). For such shallow jets ($b$\,$=$\,$2$) and high Lorentz factors even a slightly off-axis afterglow would be largely indistinguishable from the observed population as the line-of-sight material dominates over the core (\S \ref{methods: jet structure}). 
However, as shown in \S \ref{sec:detectability} (Figure \ref{fig:detectability}), such jets overpredict the observed X-ray completeness unless the gamma-ray efficiency is constant, or nearly constant, with angle. In any case, events viewed very far off-axis ($\theta_\textrm{obs}$\,$\gg$\,$\theta_\textrm{c}$) would still be distinguishable, and remain undetected in the current population with the significant exception of GW170817. 

These results point towards a potential population of slightly misaligned structured jets, which can potentially explain the observed (external) plateaus \citep{Beniamini2019plateau} or flares \citep{Duque2022} in the early X-ray lightcurves of some \textit{Swift} GRBs. These works focused primarily on long duration GRBs of which approximately half demonstrate X-ray plateaus. 

For the idealized case of tophat jets (\S \ref{methods: tophat jets} and \S \ref{sec: tophatresults}), we constrain the sample to be viewed within $\theta_\textrm{obs}/\theta_\textrm{c}$\,$<$\,$1.06$ and $1.15$ (90\% confidence) for our canonical and conservative afterglow scenarios (Table \ref{tab: Results_Summary}). This result holds over a broad range of assumptions for the afterglow parameters and initial Lorentz factor. A population of tophat jets viewed further off-axis would be noticeably different compared to the observed population with late-peaking afterglows (as opposed to the rapid fading observed in the current population). Moreover, they would create noticeable outliers in correlations between the peak energy and isotropic equivalent gamma-ray energy \citep{Fan2023}. 
In context of a tophat jet, an additional constraint to the viewing angle can be derived based on compactness, see, e.g., \citealt{Matsumoto2019b}. For long GRBs similar constraints %(median viewing angle $\theta_\textrm{obs}/\theta_\textrm{c}$\,$\approx$\,$0.6$) 
have been derived as well through direct X-ray afterglow lightcurve fitting \citep{Ryan15}, and recently as well through multi-wavelength modeling of short GRBs \citep{Wichern2024}. 

Alternatively, if we assume that these events are viewed within the core of a tophat jet  ($\theta_\textrm{obs}$\,$<$\,$\theta_\textrm{c}$) then we can constrain the quantity $\Gamma_0\theta_\textrm{c}$ (Equation \ref{eqn:gamthet}) for a subset with late-time X-ray observations of the jet break \citep{RoucoEscorial2022}. We find $\Gamma_0\theta_\textrm{c}$\,$>$\,$5$  ($18$) at the 90\% (50\%) confidence level. We note that $\Gamma_0\theta_\textrm{c}$\,$>$\,$1$ is required such that the jet is not initially causally connected, and leading to a lack of jet lateral spreading at early times. These constraints quantify the degree of core compactness for sGRBs. 

We note that Equation \ref{eqn:gamthet} can also roughly apply for structured jets if they are not extremely shallow. As the (on-axis) jet break time is not modified and $t_\textrm{p}$ (from the core; Equation \ref{eqn:peaktimeeqn}) is close to $t_\textrm{dec}(\theta_\textrm{obs})$, for nearly on-axis events the modification to $\Gamma_0\theta_\textrm{c}$ is only an order unity factor for structured jets even if not viewed exactly in the core (Figure \ref{fig:structuredlimits}). \citet{BGG2020,BGG2022} have previously shown that the core compactness (defined in their work as $\xi_\textrm{c}$\,$\equiv$\,$(\Gamma_0\theta_\textrm{c})^2$), along with the viewing angle relative to the core, are two of the few critical parameters determining the lightcurve shape of a misaligned structured jet (see Figure 8 of \citealt{BGG2022}).

Moreover, the degree of core compactness is crucial to the success of producing a GRB. Our limit of $\Gamma_0\theta_\textrm{c}$\,$>$\,$5$ constrains the core compactness at jet breakout, and not at the launching of the jet from the central engine. Numerical simulations (or analytic prescriptions) of the jet launching and collimation process during compact binary mergers must be capable of reproducing these observed values \citep[e.g.][]{Bromberg2011,Duffell2018}.

Our results strongly support the notion that cosmological sGRBs cannot be detected far off-axis, indicating that their gamma-ray production may be limited to a narrow range of angles around the core. A larger population of joint GRB-GW detections in future GW observing runs is necessary to settle the debate regarding the gamma-ray production of GRB 170817A compared to the cosmological sample. 

It is clear from this work that gamma-ray selected events are not an ideal probe of jet structure. 
However, in recent years, candidate off-axis or orphan GRBs \citep[][Chand et al. in prep.]{Nakar2002,Huang2002,Dalal2002,Totani2002,Levinson2002,Rhoads2003,Piran2013} have been discovered through a variety of means  \citep{Cenko2013,Law2018,Mooley2022offaxis,Sarin2022,Gupta2022,Perley2024}. Many such events had gamma-ray signatures discovered retrospectively after the initial identification through, e.g., optical surveys \citep{Ho2020,Andreoni2021} and were found to be standard on-axis long GRBs \citep{Cenko2015,Stadler2017,Bhalerao2017ApJ}. These events with retrospective gamma-ray detections are not true orphan afterglows, however there are still some cases of candidate orphan afterglows where a gamma-ray trigger or detection was never identified. 
These cases of potential bona fide orphan GRBs \citep{Cenko2013,Mooley2022offaxis,Sarin2022,Perley2024} represent the best possibility to constrain jet structure and gamma-ray efficiency beyond GW triggered searches. 

The advent of sensitive multi-wavelength wide-field surveys (e.g., \textit{Einstein Probe}, \textit{ULTRASAT}, Rubin Observatory) can potentially increase the detection rate of off-axis events that are discovered through untriggered (non-gamma-ray) searches. For example, the Wide-field X-ray Telescope (WXT) on the \textit{Einstein Probe} \citep{EP2015,EP2022} monitors the soft X-rays ($0.5$\,$-$\,$4$ keV) with an extremely large field of view ($3600$ deg$^2$), and has already detected a number of potential examples since its launch in early 2024. Even if not confirmed to be off-axis, these events represent a previously unexplored parameter space in the GRB phenomena and an exciting opportunity to increase the rate of events using a non-gamma-ray approach. Future soft X-ray GRB detectors, such as the \textit{Gamow Explorer} \citep{White2021,Kann2024}% or \textit{THESEUS} \citep{Amati2021}
, can build on the current and future \textit{Einstein Probe} discoveries, and are critical for the continued study of GRBs.  
The non-detection in $\gamma$-rays of obviously confirmed off-axis (or orphan) events \citep[e.g.,][]{Ho2020,Huang2020,Andreoni2021,Freeburn2024} or fast X-ray transients \citep{Wichern2024} in these sensitive wide-field surveys will have significant implications for the rate of BNS mergers, rate of successful jets launched from BNS mergers, the angular structure of the kinetic energy and Lorentz factor, and the angular profile of the gamma-ray efficiency.

%%%%%%%%%%%%%%%%%%%%%%%%%%%%%%%%%%%%%%%%%%%%%%%%%%
\section*{Acknowledgements}

The authors gratefully acknowledge the referee for their careful review of the manuscript. 
The authors thank Om Salafia and Giancarlo Ghirlanda for sharing their distribution of viewing angles in the quasi-universal jet model. 

B.O. is supported by the McWilliams Postdoctoral Fellowship at Carnegie Mellon University. B.O. gratefully acknowledges the staff of the Pittsburgh Supercomputing Center (PSC) for operating the Vera cluster for the McWilliams Center for Cosmology and Astrophysics at Carnegie Mellon University. 
P.B. is supported by a grant (no. 2020747) from the United States-Israel Binational Science Foundation (BSF), Jerusalem, Israel, by a grant (no. 1649/23) from the Israel Science Foundation and by a grant (no. 80NSSC 24K0770) from the NASA astrophysics theory program.
R.G. is supported by PAPIIT-2023 (IA105823) grant. 
This work made use of data supplied by the UK \textit{Swift} Science Data Centre at the University of Leicester.

%%%%%%%%%%%%%%%%%%%%%%%%%%%%%%%%%%%%%%%%%%%%%%%%%%
\section*{Data Availability}
The data underlying this study is publicly available from the UK \textit{Swift} Science Data Centre at the University of Leicester. %\footnote{\url{https://www.swift.ac.uk/archive/index.php}}. %produced in this study will be shared on reasonable request to the authors. 
%%%%%%%%%%%%%%%%%%%%%%%%%%%%%%%%%%%%%%%%%%%%%%%%%%

\bibliographystyle{mnras}
\bibliography{refs}

\begin{thebibliography}{}
\makeatletter
\relax
\def\mn@urlcharsother{\let\do\@makeother \do\$\do\&\do\#\do\^\do\_\do\%\do\~}
\def\mn@doi{\begingroup\mn@urlcharsother \@ifnextchar [ {\mn@doi@}
  {\mn@doi@[]}}
\def\mn@doi@[#1]#2{\def\@tempa{#1}\ifx\@tempa\@empty \href
  {http://dx.doi.org/#2} {doi:#2}\else \href {http://dx.doi.org/#2} {#1}\fi
  \endgroup}
\def\mn@eprint#1#2{\mn@eprint@#1:#2::\@nil}
\def\mn@eprint@arXiv#1{\href {http://arxiv.org/abs/#1} {{\tt arXiv:#1}}}
\def\mn@eprint@dblp#1{\href {http://dblp.uni-trier.de/rec/bibtex/#1.xml}
  {dblp:#1}}
\def\mn@eprint@#1:#2:#3:#4\@nil{\def\@tempa {#1}\def\@tempb {#2}\def\@tempc
  {#3}\ifx \@tempc \@empty \let \@tempc \@tempb \let \@tempb \@tempa \fi \ifx
  \@tempb \@empty \def\@tempb {arXiv}\fi \@ifundefined
  {mn@eprint@\@tempb}{\@tempb:\@tempc}{\expandafter \expandafter \csname
  mn@eprint@\@tempb\endcsname \expandafter{\@tempc}}}

\bibitem[\protect\citeauthoryear{{Abbott} et~al.,}{{Abbott}
  et~al.}{2017}]{Abbott+17-GW170817A-MMO}
{Abbott} B.~P.,  et~al., 2017, \mn@doi [\apjl] {10.3847/2041-8213/aa91c9},
  \href {http://adsabs.harvard.edu/abs/2017ApJ...848L..12A} {848, L12}

\bibitem[\protect\citeauthoryear{{Abbott} et~al.,}{{Abbott}
  et~al.}{2023}]{Abbott2023gwtc3BNSrate}
{Abbott} R.,  et~al., 2023, \mn@doi [Physical Review X]
  {10.1103/PhysRevX.13.011048}, \href
  {https://ui.adsabs.harvard.edu/abs/2023PhRvX..13a1048A} {13, 011048}

\bibitem[\protect\citeauthoryear{{Alexander} et~al.,}{{Alexander}
  et~al.}{2018}]{Alexander2018}
{Alexander} K.~D.,  et~al., 2018, \mn@doi [\apj] {10.3847/2041-8213/aad637},
  \href {https://ui.adsabs.harvard.edu/\#abs/2018ApJ...863L..18A} {863, L18}

\bibitem[\protect\citeauthoryear{{Andreoni} et~al.,}{{Andreoni}
  et~al.}{2021}]{Andreoni2021}
{Andreoni} I.,  et~al., 2021, \mn@doi [\apj] {10.3847/1538-4357/ac0bc7}, \href
  {https://ui.adsabs.harvard.edu/abs/2021ApJ...918...63A} {918, 63}

\bibitem[\protect\citeauthoryear{{Balasubramanian} et~al.,}{{Balasubramanian}
  et~al.}{2022}]{Balasubramanian2022}
{Balasubramanian} A.,  et~al., 2022, \mn@doi [\apj] {10.3847/1538-4357/ac9133},
  \href {https://ui.adsabs.harvard.edu/abs/2022ApJ...938...12B} {938, 12}

\bibitem[\protect\citeauthoryear{{Band} et~al.,}{{Band}
  et~al.}{1993}]{Band1993}
{Band} D.,  et~al., 1993, \mn@doi [\apj] {10.1086/172995}, \href
  {https://ui.adsabs.harvard.edu/abs/1993ApJ...413..281B} {413, 281}

\bibitem[\protect\citeauthoryear{{Barthelmy} et~al.,}{{Barthelmy}
  et~al.}{2005}]{Barthelmy2005}
{Barthelmy} S.~D.,  et~al., 2005, \mn@doi [\ssr] {10.1007/s11214-005-5096-3},
  \href {https://ui.adsabs.harvard.edu/abs/2005SSRv..120..143B} {120, 143}

\bibitem[\protect\citeauthoryear{{Beniamini} \& {Nakar}}{{Beniamini} \&
  {Nakar}}{2019}]{BeniaminiNakar2019}
{Beniamini} P.,  {Nakar} E.,  2019, \mn@doi [\mnras] {10.1093/mnras/sty3110},
  \href {https://ui.adsabs.harvard.edu/abs/2019MNRAS.482.5430B} {482, 5430}

\bibitem[\protect\citeauthoryear{{Beniamini}, {Nava}, {Duran}  \&
  {Piran}}{{Beniamini} et~al.}{2015}]{Beniamini2015}
{Beniamini} P.,  {Nava} L.,  {Duran} R.~B.,   {Piran} T.,  2015, \mn@doi
  [\mnras] {10.1093/mnras/stv2033}, \href
  {https://ui.adsabs.harvard.edu/abs/2015MNRAS.454.1073B} {454, 1073}

\bibitem[\protect\citeauthoryear{{Beniamini}, {Nava}  \& {Piran}}{{Beniamini}
  et~al.}{2016}]{Beniamini2016corr}
{Beniamini} P.,  {Nava} L.,   {Piran} T.,  2016, \mn@doi [\mnras]
  {10.1093/mnras/stw1331}, \href
  {https://ui.adsabs.harvard.edu/abs/2016MNRAS.461...51B} {461, 51}

\bibitem[\protect\citeauthoryear{{Beniamini}, {Giannios}  \&
  {Metzger}}{{Beniamini} et~al.}{2017}]{BGM2017}
{Beniamini} P.,  {Giannios} D.,   {Metzger} B.~D.,  2017, \mn@doi [\mnras]
  {10.1093/mnras/stx2095}, \href
  {https://ui.adsabs.harvard.edu/abs/2017MNRAS.472.3058B} {472, 3058}

\bibitem[\protect\citeauthoryear{{Beniamini}, {Petropoulou}, {Barniol Duran}
  \& {Giannios}}{{Beniamini} et~al.}{2019}]{Beniamini2019structuredjet}
{Beniamini} P.,  {Petropoulou} M.,  {Barniol Duran} R.,   {Giannios} D.,  2019,
  \mn@doi [\mnras] {10.1093/mnras/sty3093}, \href
  {https://ui.adsabs.harvard.edu/abs/2019MNRAS.483..840B} {483, 840}

\bibitem[\protect\citeauthoryear{{Beniamini}, {Duque}, {Daigne}  \&
  {Mochkovitch}}{{Beniamini} et~al.}{2020a}]{Beniamini2019plateau}
{Beniamini} P.,  {Duque} R.,  {Daigne} F.,   {Mochkovitch} R.,  2020a, \mn@doi
  [\mnras] {10.1093/mnras/staa070}, \href
  {https://ui.adsabs.harvard.edu/abs/2020MNRAS.492.2847B} {492, 2847}

\bibitem[\protect\citeauthoryear{{Beniamini}, {Granot}  \& {Gill}}{{Beniamini}
  et~al.}{2020b}]{BGG2020}
{Beniamini} P.,  {Granot} J.,   {Gill} R.,  2020b, \mn@doi [\mnras]
  {10.1093/mnras/staa538}, \href
  {https://ui.adsabs.harvard.edu/abs/2020MNRAS.493.3521B} {493, 3521}

\bibitem[\protect\citeauthoryear{{Beniamini}, {Duran}, {Petropoulou}  \&
  {Giannios}}{{Beniamini} et~al.}{2020c}]{BBPG2020}
{Beniamini} P.,  {Duran} R.~B.,  {Petropoulou} M.,   {Giannios} D.,  2020c,
  \mn@doi [\apjl] {10.3847/2041-8213/ab9223}, \href
  {https://ui.adsabs.harvard.edu/abs/2020ApJ...895L..33B} {895, L33}

\bibitem[\protect\citeauthoryear{{Beniamini}, {Gill}  \& {Granot}}{{Beniamini}
  et~al.}{2022}]{BGG2022}
{Beniamini} P.,  {Gill} R.,   {Granot} J.,  2022, \mn@doi [\mnras]
  {10.1093/mnras/stac1821}, \href
  {https://ui.adsabs.harvard.edu/abs/2022MNRAS.515..555B} {515, 555}

\bibitem[\protect\citeauthoryear{{Beniamini}, {Piran}  \&
  {Matsumoto}}{{Beniamini} et~al.}{2023}]{Beniamini2023TDE}
{Beniamini} P.,  {Piran} T.,   {Matsumoto} T.,  2023, \mn@doi [\mnras]
  {10.1093/mnras/stad1950}, \href
  {https://ui.adsabs.harvard.edu/abs/2023MNRAS.524.1386B} {524, 1386}

\bibitem[\protect\citeauthoryear{{Bhalerao} et~al.,}{{Bhalerao}
  et~al.}{2017}]{Bhalerao2017ApJ}
{Bhalerao} V.,  et~al., 2017, \mn@doi [\apj] {10.3847/1538-4357/aa81d2}, \href
  {https://ui.adsabs.harvard.edu/abs/2017ApJ...845..152B} {845, 152}

\bibitem[\protect\citeauthoryear{{Bhattacharjee} et~al.,}{{Bhattacharjee}
  et~al.}{2024}]{Bhattacharjee2024}
{Bhattacharjee} S.,  et~al., 2024, \mn@doi [\mnras] {10.1093/mnras/stae284},
  \href {https://ui.adsabs.harvard.edu/abs/2024MNRAS.528.4255B} {528, 4255}

\bibitem[\protect\citeauthoryear{{Biscoveanu}, {Thrane}  \&
  {Vitale}}{{Biscoveanu} et~al.}{2020}]{Biscoveanu2020}
{Biscoveanu} S.,  {Thrane} E.,   {Vitale} S.,  2020, \mn@doi [\apj]
  {10.3847/1538-4357/ab7eaf}, \href
  {https://ui.adsabs.harvard.edu/abs/2020ApJ...893...38B} {893, 38}

\bibitem[\protect\citeauthoryear{{Blandford} \& {McKee}}{{Blandford} \&
  {McKee}}{1976}]{Blandford1976}
{Blandford} R.~D.,  {McKee} C.~F.,  1976, \mn@doi [Physics of Fluids]
  {10.1063/1.861619}, \href
  {https://ui.adsabs.harvard.edu/abs/1976PhFl...19.1130B} {19, 1130}

\bibitem[\protect\citeauthoryear{{Bloom}, {Frail}  \& {Kulkarni}}{{Bloom}
  et~al.}{2003}]{Bloom2003}
{Bloom} J.~S.,  {Frail} D.~A.,   {Kulkarni} S.~R.,  2003, \mn@doi [\apj]
  {10.1086/377125}, \href
  {https://ui.adsabs.harvard.edu/abs/2003ApJ...594..674B} {594, 674}

\bibitem[\protect\citeauthoryear{{Bostanc{\i}}, {Kaneko}  \&
  {G{\"o}{\u{g}}{\"u}{\c{s}}}}{{Bostanc{\i}} et~al.}{2013}]{Bostanci2013}
{Bostanc{\i}} Z.~F.,  {Kaneko} Y.,   {G{\"o}{\u{g}}{\"u}{\c{s}}} E.,  2013,
  \mn@doi [\mnras] {10.1093/mnras/sts157}, \href
  {https://ui.adsabs.harvard.edu/abs/2013MNRAS.428.1623B} {428, 1623}

\bibitem[\protect\citeauthoryear{{Bromberg}, {Nakar}, {Piran}  \&
  {Sari}}{{Bromberg} et~al.}{2011}]{Bromberg2011}
{Bromberg} O.,  {Nakar} E.,  {Piran} T.,   {Sari} R.,  2011, \mn@doi [\apj]
  {10.1088/0004-637X/740/2/100}, \href
  {https://ui.adsabs.harvard.edu/abs/2011ApJ...740..100B} {740, 100}

\bibitem[\protect\citeauthoryear{{Bromberg}, {Tchekhovskoy}, {Gottlieb},
  {Nakar}  \& {Piran}}{{Bromberg} et~al.}{2018}]{Bromberg2018}
{Bromberg} O.,  {Tchekhovskoy} A.,  {Gottlieb} O.,  {Nakar} E.,   {Piran} T.,
  2018, \mn@doi [\mnras] {10.1093/mnras/stx3316}, \href
  {https://ui.adsabs.harvard.edu/abs/2018MNRAS.475.2971B} {475, 2971}

\bibitem[\protect\citeauthoryear{{Burrows} et~al.,}{{Burrows}
  et~al.}{2005}]{Burrows2005}
{Burrows} D.~N.,  et~al., 2005, \mn@doi [\ssr] {10.1007/s11214-005-5097-2},
  \href {https://ui.adsabs.harvard.edu/abs/2005SSRv..120..165B} {120, 165}

\bibitem[\protect\citeauthoryear{{Burrows} et~al.,}{{Burrows}
  et~al.}{2006}]{Burrows2006}
{Burrows} D.~N.,  et~al., 2006, \mn@doi [\apj] {10.1086/508740}, \href
  {https://ui.adsabs.harvard.edu/abs/2006ApJ...653..468B} {653, 468}

\bibitem[\protect\citeauthoryear{{Cenko} et~al.,}{{Cenko}
  et~al.}{2013}]{Cenko2013}
{Cenko} S.~B.,  et~al., 2013, \mn@doi [\apj] {10.1088/0004-637X/769/2/130},
  \href {https://ui.adsabs.harvard.edu/abs/2013ApJ...769..130C} {769, 130}

\bibitem[\protect\citeauthoryear{{Cenko} et~al.,}{{Cenko}
  et~al.}{2015}]{Cenko2015}
{Cenko} S.~B.,  et~al., 2015, \mn@doi [\apjl] {10.1088/2041-8205/803/2/L24},
  \href {https://ui.adsabs.harvard.edu/abs/2015ApJ...803L..24C} {803, L24}

\bibitem[\protect\citeauthoryear{{Chevalier} \& {Li}}{{Chevalier} \&
  {Li}}{2000}]{ChevalierLi2000}
{Chevalier} R.~A.,  {Li} Z.-Y.,  2000, \mn@doi [\apj] {10.1086/308914}, \href
  {https://ui.adsabs.harvard.edu/abs/2000ApJ...536..195C} {536, 195}

\bibitem[\protect\citeauthoryear{{Chrimes} et~al.,}{{Chrimes}
  et~al.}{2022}]{Chrimes2022}
{Chrimes} A.~A.,  et~al., 2022, \mn@doi [\mnras] {10.1093/mnras/stac1796},
  \href {https://ui.adsabs.harvard.edu/abs/2022MNRAS.515.2591C} {515, 2591}

\bibitem[\protect\citeauthoryear{{Colombo}, {Salafia}, {Gabrielli},
  {Ghirlanda}, {Giacomazzo}, {Perego}  \& {Colpi}}{{Colombo}
  et~al.}{2022}]{Colombo2022}
{Colombo} A.,  {Salafia} O.~S.,  {Gabrielli} F.,  {Ghirlanda} G.,  {Giacomazzo}
  B.,  {Perego} A.,   {Colpi} M.,  2022, \mn@doi [\apj]
  {10.3847/1538-4357/ac8d00}, \href
  {https://ui.adsabs.harvard.edu/abs/2022ApJ...937...79C} {937, 79}

\bibitem[\protect\citeauthoryear{{Coward} et~al.,}{{Coward}
  et~al.}{2012}]{Coward2012}
{Coward} D.~M.,  et~al., 2012, \mn@doi [\mnras]
  {10.1111/j.1365-2966.2012.21604.x}, \href
  {https://ui.adsabs.harvard.edu/abs/2012MNRAS.425.2668C} {425, 2668}

\bibitem[\protect\citeauthoryear{{D'Avanzo} et~al.,}{{D'Avanzo}
  et~al.}{2018}]{D'Avanzo2018}
{D'Avanzo} P.,  et~al., 2018, \mn@doi [\aap] {10.1051/0004-6361/201832664},
  \href {https://ui.adsabs.harvard.edu/abs/2018A&A...613L...1D} {613, L1}

\bibitem[\protect\citeauthoryear{{Dalal}, {Griest}  \& {Pruet}}{{Dalal}
  et~al.}{2002}]{Dalal2002}
{Dalal} N.,  {Griest} K.,   {Pruet} J.,  2002, \mn@doi [\apj] {10.1086/324142},
  \href {https://ui.adsabs.harvard.edu/abs/2002ApJ...564..209D} {564, 209}

\bibitem[\protect\citeauthoryear{{Della Valle} et~al.,}{{Della Valle}
  et~al.}{2006}]{DellaValle2006}
{Della Valle} M.,  et~al., 2006, \mn@doi [\nat] {10.1038/nature05374}, \href
  {https://ui.adsabs.harvard.edu/abs/2006Natur.444.1050D} {444, 1050}

\bibitem[\protect\citeauthoryear{{Dichiara}, {Troja}, {O'Connor}, {Marshall},
  {Beniamini}, {Cannizzo}, {Lien}  \& {Sakamoto}}{{Dichiara}
  et~al.}{2020}]{Dichiara2020}
{Dichiara} S.,  {Troja} E.,  {O'Connor} B.,  {Marshall} F.~E.,  {Beniamini} P.,
   {Cannizzo} J.~K.,  {Lien} A.~Y.,   {Sakamoto} T.,  2020, \mn@doi [\mnras]
  {10.1093/mnras/staa124}, \href
  {https://ui.adsabs.harvard.edu/abs/2020MNRAS.492.5011D} {492, 5011}

\bibitem[\protect\citeauthoryear{{Dichiara}, {Tsang}, {Troja}, {Neill},
  {Norris}  \& {Yang}}{{Dichiara} et~al.}{2023}]{Dichiara2023}
{Dichiara} S.,  {Tsang} D.,  {Troja} E.,  {Neill} D.,  {Norris} J.~P.,   {Yang}
  Y.~H.,  2023, \mn@doi [\apjl] {10.3847/2041-8213/acf21d}, \href
  {https://ui.adsabs.harvard.edu/abs/2023ApJ...954L..29D} {954, L29}

\bibitem[\protect\citeauthoryear{{Duffell}, {Quataert}, {Kasen}  \&
  {Klion}}{{Duffell} et~al.}{2018}]{Duffell2018}
{Duffell} P.~C.,  {Quataert} E.,  {Kasen} D.,   {Klion} H.,  2018, \mn@doi
  [\apj] {10.3847/1538-4357/aae084}, \href
  {https://ui.adsabs.harvard.edu/abs/2018ApJ...866....3D} {866, 3}

\bibitem[\protect\citeauthoryear{{Duque}, {Beniamini}, {Daigne}  \&
  {Mochkovitch}}{{Duque} et~al.}{2020}]{Duque2020}
{Duque} R.,  {Beniamini} P.,  {Daigne} F.,   {Mochkovitch} R.,  2020, \mn@doi
  [\aap] {10.1051/0004-6361/201937115}, \href
  {https://ui.adsabs.harvard.edu/abs/2020A&A...639A..15D} {639, A15}

\bibitem[\protect\citeauthoryear{{Duque}, {Beniamini}, {Daigne}  \&
  {Mochkovitch}}{{Duque} et~al.}{2022}]{Duque2022}
{Duque} R.,  {Beniamini} P.,  {Daigne} F.,   {Mochkovitch} R.,  2022, \mn@doi
  [\mnras] {10.1093/mnras/stac938}, \href
  {https://ui.adsabs.harvard.edu/abs/2022MNRAS.513..951D} {513, 951}

\bibitem[\protect\citeauthoryear{{Eichler} \& {Granot}}{{Eichler} \&
  {Granot}}{2006}]{EichlerGranot2006}
{Eichler} D.,  {Granot} J.,  2006, \mn@doi [\apjl] {10.1086/503667}, \href
  {https://ui.adsabs.harvard.edu/abs/2006ApJ...641L...5E} {641, L5}

\bibitem[\protect\citeauthoryear{{Farah}, {Essick}, {Doctor}, {Fishbach}  \&
  {Holz}}{{Farah} et~al.}{2020}]{Farah2020}
{Farah} A.,  {Essick} R.,  {Doctor} Z.,  {Fishbach} M.,   {Holz} D.~E.,  2020,
  \mn@doi [\apj] {10.3847/1538-4357/ab8d26}, \href
  {https://ui.adsabs.harvard.edu/abs/2020ApJ...895..108F} {895, 108}

\bibitem[\protect\citeauthoryear{{Fern{\'a}ndez}, {Kobayashi}  \&
  {Lamb}}{{Fern{\'a}ndez} et~al.}{2022}]{Fernandez2022}
{Fern{\'a}ndez} J.~J.,  {Kobayashi} S.,   {Lamb} G.~P.,  2022, \mn@doi [\mnras]
  {10.1093/mnras/stab2879}, \href
  {https://ui.adsabs.harvard.edu/abs/2022MNRAS.509..395F} {509, 395}

\bibitem[\protect\citeauthoryear{{Fong} et~al.,}{{Fong}
  et~al.}{2012}]{Fong2012}
{Fong} W.,  et~al., 2012, \mn@doi [\apj] {10.1088/0004-637X/756/2/189}, \href
  {https://ui.adsabs.harvard.edu/abs/2012ApJ...756..189F} {756, 189}

\bibitem[\protect\citeauthoryear{{Fong} et~al.,}{{Fong}
  et~al.}{2013}]{Fong2013}
{Fong} W.,  et~al., 2013, \mn@doi [\apj] {10.1088/0004-637X/769/1/56}, \href
  {https://ui.adsabs.harvard.edu/abs/2013ApJ...769...56F} {769, 56}

\bibitem[\protect\citeauthoryear{{Fong}, {Berger}, {Margutti}  \&
  {Zauderer}}{{Fong} et~al.}{2015}]{Fong2015}
{Fong} W.,  {Berger} E.,  {Margutti} R.,   {Zauderer} B.~A.,  2015, \mn@doi
  [\apj] {10.1088/0004-637X/815/2/102}, \href
  {https://ui.adsabs.harvard.edu/abs/2015ApJ...815..102F} {815, 102}

\bibitem[\protect\citeauthoryear{{Fong} et~al.,}{{Fong}
  et~al.}{2021}]{Fong2021kn}
{Fong} W.,  et~al., 2021, \mn@doi [\apj] {10.3847/1538-4357/abc74a}, \href
  {https://ui.adsabs.harvard.edu/abs/2021ApJ...906..127F} {906, 127}

\bibitem[\protect\citeauthoryear{{Fong} et~al.,}{{Fong}
  et~al.}{2022}]{Fong2022}
{Fong} W.-f.,  et~al., 2022, \mn@doi [\apj] {10.3847/1538-4357/ac91d0}, \href
  {https://ui.adsabs.harvard.edu/abs/2022ApJ...940...56F} {940, 56}

\bibitem[\protect\citeauthoryear{{Frail} et~al.,}{{Frail}
  et~al.}{2001}]{Frail2001}
{Frail} D.~A.,  et~al., 2001, \mn@doi [\apjl] {10.1086/338119}, \href
  {https://ui.adsabs.harvard.edu/abs/2001ApJ...562L..55F} {562, L55}

\bibitem[\protect\citeauthoryear{{Freeburn} et~al.,}{{Freeburn}
  et~al.}{2024}]{Freeburn2024}
{Freeburn} J.,  et~al., 2024, arXiv e-prints, \href
  {https://ui.adsabs.harvard.edu/abs/2024arXiv240511949F} {p. arXiv:2405.11949}

\bibitem[\protect\citeauthoryear{{Gal-Yam} et~al.,}{{Gal-Yam}
  et~al.}{2006}]{Galyam2006}
{Gal-Yam} A.,  et~al., 2006, \mn@doi [\nat] {10.1038/nature05373}, \href
  {https://ui.adsabs.harvard.edu/abs/2006Natur.444.1053G} {444, 1053}

\bibitem[\protect\citeauthoryear{{Gehrels} et~al.,}{{Gehrels}
  et~al.}{2004}]{Gehrels2004}
{Gehrels} N.,  et~al., 2004, \mn@doi [\apj] {10.1086/422091}, \href
  {https://ui.adsabs.harvard.edu/abs/2004ApJ...611.1005G} {611, 1005}

\bibitem[\protect\citeauthoryear{{Gehrels} et~al.,}{{Gehrels}
  et~al.}{2006}]{Gehrels2006}
{Gehrels} N.,  et~al., 2006, \mn@doi [\nat] {10.1038/nature05376}, \href
  {https://ui.adsabs.harvard.edu/abs/2006Natur.444.1044G} {444, 1044}

\bibitem[\protect\citeauthoryear{{Ghirlanda} \& {Salvaterra}}{{Ghirlanda} \&
  {Salvaterra}}{2022}]{Ghirlanda2022}
{Ghirlanda} G.,  {Salvaterra} R.,  2022, \mn@doi [\apj]
  {10.3847/1538-4357/ac6e43}, \href
  {https://ui.adsabs.harvard.edu/abs/2022ApJ...932...10G} {932, 10}

\bibitem[\protect\citeauthoryear{{Ghirlanda} et~al.,}{{Ghirlanda}
  et~al.}{2012}]{Ghirlanda2012}
{Ghirlanda} G.,  et~al., 2012, \mn@doi [\mnras]
  {10.1111/j.1365-2966.2012.20815.x}, \href
  {https://ui.adsabs.harvard.edu/abs/2012MNRAS.422.2553G} {422, 2553}

\bibitem[\protect\citeauthoryear{{Ghirlanda} et~al.,}{{Ghirlanda}
  et~al.}{2016}]{Ghirlanda2016}
{Ghirlanda} G.,  et~al., 2016, \mn@doi [\aap] {10.1051/0004-6361/201628993},
  \href {https://ui.adsabs.harvard.edu/abs/2016A&A...594A..84G} {594, A84}

\bibitem[\protect\citeauthoryear{{Ghirlanda} et~al.,}{{Ghirlanda}
  et~al.}{2018}]{Ghirlanda2018}
{Ghirlanda} G.,  et~al., 2018, \mn@doi [\aap] {10.1051/0004-6361/201731598},
  \href {https://ui.adsabs.harvard.edu/abs/2018A&A...609A.112G} {609, A112}

\bibitem[\protect\citeauthoryear{{Ghirlanda} et~al.,}{{Ghirlanda}
  et~al.}{2019}]{Ghirlanda2019}
{Ghirlanda} G.,  et~al., 2019, \mn@doi [Science] {10.1126/science.aau8815},
  \href {https://ui.adsabs.harvard.edu/abs/2019Sci...363..968G} {363, 968}

\bibitem[\protect\citeauthoryear{{Ghisellini}, {Ghirlanda}, {Nava}  \&
  {Celotti}}{{Ghisellini} et~al.}{2010}]{Ghisellini2010}
{Ghisellini} G.,  {Ghirlanda} G.,  {Nava} L.,   {Celotti} A.,  2010, \mn@doi
  [\mnras] {10.1111/j.1365-2966.2009.16171.x}, \href
  {https://ui.adsabs.harvard.edu/abs/2010MNRAS.403..926G} {403, 926}

\bibitem[\protect\citeauthoryear{{Gill} \& {Granot}}{{Gill} \&
  {Granot}}{2018}]{GG2018}
{Gill} R.,  {Granot} J.,  2018, \mn@doi [\mnras] {10.1093/mnras/sty1214}, \href
  {https://ui.adsabs.harvard.edu/abs/2018MNRAS.478.4128G} {478, 4128}

\bibitem[\protect\citeauthoryear{{Gill} \& {Granot}}{{Gill} \&
  {Granot}}{2023}]{GG2023}
{Gill} R.,  {Granot} J.,  2023, \mn@doi [\mnras] {10.1093/mnrasl/slad075},
  \href {https://ui.adsabs.harvard.edu/abs/2023MNRAS.524L..78G} {524, L78}

\bibitem[\protect\citeauthoryear{{Gill}, {Granot}, {De Colle}  \&
  {Urrutia}}{{Gill} et~al.}{2019}]{Gill2019}
{Gill} R.,  {Granot} J.,  {De Colle} F.,   {Urrutia} G.,  2019, \mn@doi [\apj]
  {10.3847/1538-4357/ab3577}, \href
  {https://ui.adsabs.harvard.edu/abs/2019ApJ...883...15G} {883, 15}

\bibitem[\protect\citeauthoryear{{Gill}, {Granot}  \& {Kumar}}{{Gill}
  et~al.}{2020a}]{Gill2020}
{Gill} R.,  {Granot} J.,   {Kumar} P.,  2020a, \mn@doi [\mnras]
  {10.1093/mnras/stz2976}, \href
  {https://ui.adsabs.harvard.edu/abs/2020MNRAS.491.3343G} {491, 3343}

\bibitem[\protect\citeauthoryear{{Gill}, {Granot}  \& {Kumar}}{{Gill}
  et~al.}{2020b}]{Gill+20}
{Gill} R.,  {Granot} J.,   {Kumar} P.,  2020b, \mn@doi [\mnras]
  {10.1093/mnras/stz2976}, \href
  {https://ui.adsabs.harvard.edu/abs/2020MNRAS.491.3343G} {491, 3343}

\bibitem[\protect\citeauthoryear{{Gillanders} et~al.,}{{Gillanders}
  et~al.}{2023}]{Gillanders2023}
{Gillanders} J.~H.,  et~al., 2023, \mn@doi [arXiv e-prints]
  {10.48550/arXiv.2308.00633}, \href
  {https://ui.adsabs.harvard.edu/abs/2023arXiv230800633G} {p. arXiv:2308.00633}

\bibitem[\protect\citeauthoryear{{Goldstein} et~al.,}{{Goldstein}
  et~al.}{2017}]{Goldstein2017}
{Goldstein} A.,  et~al., 2017, \mn@doi [\apjl] {10.3847/2041-8213/aa8f41},
  \href {https://ui.adsabs.harvard.edu/abs/2017ApJ...848L..14G} {848, L14}

\bibitem[\protect\citeauthoryear{{Gompertz}, {Fruchter}  \& {Pe'er}}{{Gompertz}
  et~al.}{2018}]{Gompertz2018grb}
{Gompertz} B.~P.,  {Fruchter} A.~S.,   {Pe'er} A.,  2018, \mn@doi [\apj]
  {10.3847/1538-4357/aadba8}, \href
  {https://ui.adsabs.harvard.edu/abs/2018ApJ...866..162G} {866, 162}

\bibitem[\protect\citeauthoryear{{Gompertz} et~al.,}{{Gompertz}
  et~al.}{2023}]{Gompertz2023}
{Gompertz} B.~P.,  et~al., 2023, \mn@doi [Nature Astronomy]
  {10.1038/s41550-022-01819-4}, \href
  {https://ui.adsabs.harvard.edu/abs/2023NatAs...7...67G} {7, 67}

\bibitem[\protect\citeauthoryear{{Gottlieb} \& {Globus}}{{Gottlieb} \&
  {Globus}}{2021}]{Gottlieb2021magneto}
{Gottlieb} O.,  {Globus} N.,  2021, \mn@doi [\apjl] {10.3847/2041-8213/ac05c5},
  \href {https://ui.adsabs.harvard.edu/abs/2021ApJ...915L...4G} {915, L4}

\bibitem[\protect\citeauthoryear{{Gottlieb}, {Nakar}, {Piran}  \&
  {Hotokezaka}}{{Gottlieb} et~al.}{2018}]{Gottlieb2018}
{Gottlieb} O.,  {Nakar} E.,  {Piran} T.,   {Hotokezaka} K.,  2018, \mn@doi
  [\mnras] {10.1093/mnras/sty1462}, \href
  {https://ui.adsabs.harvard.edu/abs/2018MNRAS.479..588G} {479, 588}

\bibitem[\protect\citeauthoryear{{Gottlieb}, {Nakar}  \& {Bromberg}}{{Gottlieb}
  et~al.}{2021}]{Gottlieb2021}
{Gottlieb} O.,  {Nakar} E.,   {Bromberg} O.,  2021, \mn@doi [\mnras]
  {10.1093/mnras/staa350110.48550/arXiv.2006.02466}, \href
  {https://ui.adsabs.harvard.edu/abs/2021MNRAS.500.3511G} {500, 3511}

\bibitem[\protect\citeauthoryear{{Gottlieb}, {Liska}, {Tchekhovskoy},
  {Bromberg}, {Lalakos}, {Giannios}  \& {M{\"o}sta}}{{Gottlieb}
  et~al.}{2022}]{Gottlieb2022}
{Gottlieb} O.,  {Liska} M.,  {Tchekhovskoy} A.,  {Bromberg} O.,  {Lalakos} A.,
  {Giannios} D.,   {M{\"o}sta} P.,  2022, \mn@doi [\apjl]
  {10.3847/2041-8213/ac753010.48550/arXiv.2204.12501}, \href
  {https://ui.adsabs.harvard.edu/abs/2022ApJ...933L...9G} {933, L9}

\bibitem[\protect\citeauthoryear{{Govreen-Segal} \& {Nakar}}{{Govreen-Segal} \&
  {Nakar}}{2023}]{Govreen-Segal2023}
{Govreen-Segal} T.,  {Nakar} E.,  2023, arXiv e-prints, \href
  {https://ui.adsabs.harvard.edu/abs/2023arXiv231109297G} {p. arXiv:2311.09297}

\bibitem[\protect\citeauthoryear{{Granot}}{{Granot}}{2005}]{Granot05}
{Granot} J.,  2005, \mn@doi [\apj] {10.1086/432676}, \href
  {https://ui.adsabs.harvard.edu/abs/2005ApJ...631.1022G} {631, 1022}

\bibitem[\protect\citeauthoryear{{Granot} \& {Piran}}{{Granot} \&
  {Piran}}{2012}]{Granot2012}
{Granot} J.,  {Piran} T.,  2012, \mn@doi [\mnras]
  {10.1111/j.1365-2966.2011.20335.x}, \href
  {https://ui.adsabs.harvard.edu/abs/2012MNRAS.421..570G} {421, 570}

\bibitem[\protect\citeauthoryear{{Granot} \& {Sari}}{{Granot} \&
  {Sari}}{2002}]{Granot2002}
{Granot} J.,  {Sari} R.,  2002, \mn@doi [\apj] {10.1086/338966}, \href
  {https://ui.adsabs.harvard.edu/abs/2002ApJ...568..820G} {568, 820}

\bibitem[\protect\citeauthoryear{{Granot}, {Miller}, {Piran}, {Suen}  \&
  {Hughes}}{{Granot} et~al.}{2001}]{Granot2001}
{Granot} J.,  {Miller} M.,  {Piran} T.,  {Suen} W.~M.,   {Hughes} P.~A.,  2001,
  in {Costa} E.,  {Frontera} F.,   {Hjorth} J.,  eds, Gamma-ray Bursts in the
  Afterglow Era. p.~312 (\mn@eprint {arXiv} {astro-ph/0103038}),
  \mn@doi{10.1007/10853853_82}

\bibitem[\protect\citeauthoryear{{Granot}, {Panaitescu}, {Kumar}  \&
  {Woosley}}{{Granot} et~al.}{2002}]{Granot2002jet}
{Granot} J.,  {Panaitescu} A.,  {Kumar} P.,   {Woosley} S.~E.,  2002, \mn@doi
  [\apjl] {10.1086/340991}, \href
  {https://ui.adsabs.harvard.edu/abs/2002ApJ...570L..61G} {570, L61}

\bibitem[\protect\citeauthoryear{{Granot}, {Guetta}  \& {Gill}}{{Granot}
  et~al.}{2017}]{Granot2017}
{Granot} J.,  {Guetta} D.,   {Gill} R.,  2017, \mn@doi [\apjl]
  {10.3847/2041-8213/aa991d}, \href
  {https://ui.adsabs.harvard.edu/abs/2017ApJ...850L..24G} {850, L24}

\bibitem[\protect\citeauthoryear{{Granot}, {De Colle}  \&
  {Ramirez-Ruiz}}{{Granot} et~al.}{2018}]{Granot2018}
{Granot} J.,  {De Colle} F.,   {Ramirez-Ruiz} E.,  2018, \mn@doi [\mnras]
  {10.1093/mnras/sty2454}, \href
  {https://ui.adsabs.harvard.edu/abs/2018MNRAS.481.2711G} {481, 2711}

\bibitem[\protect\citeauthoryear{{Guetta} \& {Piran}}{{Guetta} \&
  {Piran}}{2006}]{Guetta2006}
{Guetta} D.,  {Piran} T.,  2006, \mn@doi [\aap] {10.1051/0004-6361:20054498},
  \href {https://ui.adsabs.harvard.edu/abs/2006A&A...453..823G} {453, 823}

\bibitem[\protect\citeauthoryear{{Gupta} et~al.,}{{Gupta}
  et~al.}{2022}]{Gupta2022}
{Gupta} R.,  et~al., 2022, \mn@doi [Journal of Astrophysics and Astronomy]
  {10.1007/s12036-021-09794-4}, \href
  {https://ui.adsabs.harvard.edu/abs/2022JApA...43...11G} {43, 11}

\bibitem[\protect\citeauthoryear{{Hajela} et~al.,}{{Hajela}
  et~al.}{2022}]{Hajela2022}
{Hajela} A.,  et~al., 2022, \mn@doi [\apjl] {10.3847/2041-8213/ac504a}, \href
  {https://ui.adsabs.harvard.edu/abs/2022ApJ...927L..17H} {927, L17}

\bibitem[\protect\citeauthoryear{{Hayes}, {Heng}, {Veitch}  \&
  {Williams}}{{Hayes} et~al.}{2020}]{Hayes2020}
{Hayes} F.,  {Heng} I.~S.,  {Veitch} J.,   {Williams} D.,  2020, \mn@doi [\apj]
  {10.3847/1538-4357/ab72fc}, \href
  {https://ui.adsabs.harvard.edu/abs/2020ApJ...891..124H} {891, 124}

\bibitem[\protect\citeauthoryear{{Hayes}, {Heng}, {Lamb}, {Lin}, {Veitch}  \&
  {Williams}}{{Hayes} et~al.}{2023}]{Hayes2023}
{Hayes} F.,  {Heng} I.~S.,  {Lamb} G.,  {Lin} E.-T.,  {Veitch} J.,   {Williams}
  M.~J.,  2023, \mn@doi [\apj] {10.3847/1538-4357/ace899}, \href
  {https://ui.adsabs.harvard.edu/abs/2023ApJ...954...92H} {954, 92}

\bibitem[\protect\citeauthoryear{{Ho} et~al.,}{{Ho} et~al.}{2020}]{Ho2020}
{Ho} A. Y.~Q.,  et~al., 2020, \mn@doi [\apj] {10.3847/1538-4357/abc34d}, \href
  {https://ui.adsabs.harvard.edu/abs/2020ApJ...905...98H} {905, 98}

\bibitem[\protect\citeauthoryear{{Hotokezaka}, {Nakar}, {Gottlieb}, {Nissanke},
  {Masuda}, {Hallinan}, {Mooley}  \& {Deller}}{{Hotokezaka}
  et~al.}{2019}]{Hotokezaka+19}
{Hotokezaka} K.,  {Nakar} E.,  {Gottlieb} O.,  {Nissanke} S.,  {Masuda} K.,
  {Hallinan} G.,  {Mooley} K.~P.,   {Deller} A.~T.,  2019, \mn@doi [Nature
  Astronomy] {10.1038/s41550-019-0820-1}, \href
  {https://ui.adsabs.harvard.edu/abs/2019NatAs...3..940H} {3, 940}

\bibitem[\protect\citeauthoryear{{Howell}, {Ackley}, {Rowlinson}  \&
  {Coward}}{{Howell} et~al.}{2019}]{Howell2019}
{Howell} E.~J.,  {Ackley} K.,  {Rowlinson} A.,   {Coward} D.,  2019, \mn@doi
  [\mnras] {10.1093/mnras/stz455}, \href
  {https://ui.adsabs.harvard.edu/abs/2019MNRAS.485.1435H} {485, 1435}

\bibitem[\protect\citeauthoryear{{Huang}, {Dai}  \& {Lu}}{{Huang}
  et~al.}{2002}]{Huang2002}
{Huang} Y.~F.,  {Dai} Z.~G.,   {Lu} T.,  2002, \mn@doi [\mnras]
  {10.1046/j.1365-8711.2002.05334.x}, \href
  {https://ui.adsabs.harvard.edu/abs/2002MNRAS.332..735H} {332, 735}

\bibitem[\protect\citeauthoryear{{Huang} et~al.,}{{Huang}
  et~al.}{2020}]{Huang2020}
{Huang} Y.-J.,  et~al., 2020, \mn@doi [\apj] {10.3847/1538-4357/ab8f9a}, \href
  {https://ui.adsabs.harvard.edu/abs/2020ApJ...897...69H} {897, 69}

\bibitem[\protect\citeauthoryear{{Ioka} \& {Nakamura}}{{Ioka} \&
  {Nakamura}}{2018}]{Ioka2018}
{Ioka} K.,  {Nakamura} T.,  2018, \mn@doi [Progress of Theoretical and
  Experimental Physics] {10.1093/ptep/pty036}, \href
  {https://ui.adsabs.harvard.edu/abs/2018PTEP.2018d3E02I} {2018, 043E02}

\bibitem[\protect\citeauthoryear{{Jacovich}, {Beniamini}  \& {van der
  Horst}}{{Jacovich} et~al.}{2021}]{Jacovich2020}
{Jacovich} T.~E.,  {Beniamini} P.,   {van der Horst} A.~J.,  2021, \mn@doi
  [\mnras] {10.1093/mnras/stab911}, \href
  {https://ui.adsabs.harvard.edu/abs/2021MNRAS.504..528J} {504, 528}

\bibitem[\protect\citeauthoryear{{Jin} et~al.,}{{Jin} et~al.}{2018}]{Jin2018}
{Jin} Z.-P.,  et~al., 2018, \mn@doi [\apj] {10.3847/1538-4357/aab76d}, \href
  {https://ui.adsabs.harvard.edu/abs/2018ApJ...857..128J} {857, 128}

\bibitem[\protect\citeauthoryear{{Kaneko}, {Bostanc{\i}},
  {G{\"o}{\u{g}}{\"u}{\c{s}}}  \& {Lin}}{{Kaneko} et~al.}{2015}]{Kaneko2015}
{Kaneko} Y.,  {Bostanc{\i}} Z.~F.,  {G{\"o}{\u{g}}{\"u}{\c{s}}} E.,   {Lin} L.,
   2015, \mn@doi [\mnras] {10.1093/mnras/stv1286}, \href
  {https://ui.adsabs.harvard.edu/abs/2015MNRAS.452..824K} {452, 824}

\bibitem[\protect\citeauthoryear{{Kann} et~al.,}{{Kann}
  et~al.}{2024}]{Kann2024}
{Kann} D.~A.,  et~al., 2024, \mn@doi [arXiv e-prints]
  {10.48550/arXiv.2403.00101}, \href
  {https://ui.adsabs.harvard.edu/abs/2024arXiv240300101K} {p. arXiv:2403.00101}

\bibitem[\protect\citeauthoryear{{Kasliwal} et~al.,}{{Kasliwal}
  et~al.}{2017}]{Kasliwal2017}
{Kasliwal} M.~M.,  et~al., 2017, \mn@doi [Science] {10.1126/science.aap9455},
  \href {https://ui.adsabs.harvard.edu/abs/2017Sci...358.1559K} {358, 1559}

\bibitem[\protect\citeauthoryear{{Kathirgamaraju}, {Barniol Duran}  \&
  {Giannios}}{{Kathirgamaraju} et~al.}{2018}]{Kathirgamaraju2018}
{Kathirgamaraju} A.,  {Barniol Duran} R.,   {Giannios} D.,  2018, \mn@doi
  [\mnras] {10.1093/mnrasl/slx175}, \href
  {https://ui.adsabs.harvard.edu/abs/2018MNRAS.473L.121K} {473, L121}

\bibitem[\protect\citeauthoryear{{Kouveliotou}, {Meegan}, {Fishman}, {Bhat},
  {Briggs}, {Koshut}, {Paciesas}  \& {Pendleton}}{{Kouveliotou}
  et~al.}{1993}]{Kouveliotou1993}
{Kouveliotou} C.,  {Meegan} C.~A.,  {Fishman} G.~J.,  {Bhat} N.~P.,  {Briggs}
  M.~S.,  {Koshut} T.~M.,  {Paciesas} W.~S.,   {Pendleton} G.~N.,  1993,
  \mn@doi [\apjl] {10.1086/186969}, \href
  {https://ui.adsabs.harvard.edu/abs/1993ApJ...413L.101K} {413, L101}

\bibitem[\protect\citeauthoryear{{Kumar} \& {Granot}}{{Kumar} \&
  {Granot}}{2003}]{Kumar2003}
{Kumar} P.,  {Granot} J.,  2003, \mn@doi [\apj] {10.1086/375186}, \href
  {https://ui.adsabs.harvard.edu/abs/2003ApJ...591.1075K} {591, 1075}

\bibitem[\protect\citeauthoryear{{Kumar} \& {Zhang}}{{Kumar} \&
  {Zhang}}{2015}]{KumarZhang2015}
{Kumar} P.,  {Zhang} B.,  2015, \mn@doi [\physrep]
  {10.1016/j.physrep.2014.09.008}, \href
  {https://ui.adsabs.harvard.edu/abs/2015PhR...561....1K} {561, 1}

\bibitem[\protect\citeauthoryear{{LHAASO Collaboration} et~al.,}{{LHAASO
  Collaboration} et~al.}{2023}]{LHAASO221009A}
{LHAASO Collaboration} et~al., 2023, \mn@doi [Science]
  {10.1126/science.adg9328}, \href
  {https://ui.adsabs.harvard.edu/abs/2023Sci...380.1390L} {380, 1390}

\bibitem[\protect\citeauthoryear{{LIGO Scientific Collaboration} \& {Virgo
  Collaboration}}{{LIGO Scientific Collaboration} \& {Virgo
  Collaboration}}{2017}]{Abbott2017}
{LIGO Scientific Collaboration} {Virgo Collaboration} 2017, \mn@doi [\prl]
  {10.1103/PhysRevLett.119.161101}, \href
  {https://ui.adsabs.harvard.edu/abs/2017PhRvL.119p1101A} {119, 161101}

\bibitem[\protect\citeauthoryear{{Lamb} \& {Kobayashi}}{{Lamb} \&
  {Kobayashi}}{2016}]{LambKobayashi2016}
{Lamb} G.~P.,  {Kobayashi} S.,  2016, \mn@doi [\apj]
  {10.3847/0004-637X/829/2/112}, \href
  {https://ui.adsabs.harvard.edu/abs/2016ApJ...829..112L} {829, 112}

\bibitem[\protect\citeauthoryear{{Lamb} \& {Kobayashi}}{{Lamb} \&
  {Kobayashi}}{2017}]{Lamb2017jet}
{Lamb} G.~P.,  {Kobayashi} S.,  2017, \mn@doi [\mnras] {10.1093/mnras/stx2345},
  \href {https://ui.adsabs.harvard.edu/abs/2017MNRAS.472.4953L} {472, 4953}

\bibitem[\protect\citeauthoryear{{Lamb} \& {Kobayashi}}{{Lamb} \&
  {Kobayashi}}{2018}]{lamb2018}
{Lamb} G.~P.,  {Kobayashi} S.,  2018, \mn@doi [\mnras] {10.1093/mnras/sty1108},
  \href {https://ui.adsabs.harvard.edu/\#abs/2018MNRAS.478..733L} {478, 733}

\bibitem[\protect\citeauthoryear{{Lamb}, {Mandel}  \& {Resmi}}{{Lamb}
  et~al.}{2018}]{Lamb2018jet}
{Lamb} G.~P.,  {Mandel} I.,   {Resmi} L.,  2018, \mn@doi [\mnras]
  {10.1093/mnras/sty2196}, \href
  {https://ui.adsabs.harvard.edu/abs/2018MNRAS.481.2581L} {481, 2581}

\bibitem[\protect\citeauthoryear{{Lamb} et~al.,}{{Lamb}
  et~al.}{2019}]{Lamb2019grb160821B}
{Lamb} G.~P.,  et~al., 2019, \mn@doi [\apj] {10.3847/1538-4357/ab38bb}, \href
  {https://ui.adsabs.harvard.edu/abs/2019ApJ...883...48L} {883, 48}

\bibitem[\protect\citeauthoryear{{Lamb} et~al.,}{{Lamb}
  et~al.}{2021}]{Lamb2021rev}
{Lamb} G.~P.,  et~al., 2021, \mn@doi [Universe] {10.3390/universe7090329},
  \href {https://ui.adsabs.harvard.edu/abs/2021Univ....7..329L} {7, 329}

\bibitem[\protect\citeauthoryear{{Lamb}, {Nativi}, {Rosswog}, {Kann}, {Levan},
  {Lundman}  \& {Tanvir}}{{Lamb} et~al.}{2022}]{Lamb2022}
{Lamb} G.~P.,  {Nativi} L.,  {Rosswog} S.,  {Kann} D.~A.,  {Levan} A.,
  {Lundman} C.,   {Tanvir} N.,  2022, arXiv e-prints, \href
  {https://ui.adsabs.harvard.edu/abs/2022arXiv220109796L} {p. arXiv:2201.09796}

\bibitem[\protect\citeauthoryear{{Laskar} et~al.,}{{Laskar}
  et~al.}{2022}]{Laskar2022}
{Laskar} T.,  et~al., 2022, \mn@doi [\apjl] {10.3847/2041-8213/ac8421}, \href
  {https://ui.adsabs.harvard.edu/abs/2022ApJ...935L..11L} {935, L11}

\bibitem[\protect\citeauthoryear{{Laskar} et~al.,}{{Laskar}
  et~al.}{2023}]{Laskar2023}
{Laskar} T.,  et~al., 2023, \mn@doi [\apjl] {10.3847/2041-8213/acbfad}, \href
  {https://ui.adsabs.harvard.edu/abs/2023ApJ...946L..23L} {946, L23}

\bibitem[\protect\citeauthoryear{{Law}, {Gaensler}, {Metzger}, {Ofek}  \&
  {Sironi}}{{Law} et~al.}{2018}]{Law2018}
{Law} C.~J.,  {Gaensler} B.~M.,  {Metzger} B.~D.,  {Ofek} E.~O.,   {Sironi} L.,
   2018, \mn@doi [\apjl] {10.3847/2041-8213/aae5f3}, \href
  {https://ui.adsabs.harvard.edu/abs/2018ApJ...866L..22L} {866, L22}

\bibitem[\protect\citeauthoryear{{Lazzati}, {Deich}, {Morsony}  \&
  {Workman}}{{Lazzati} et~al.}{2017}]{Lazzati2017}
{Lazzati} D.,  {Deich} A.,  {Morsony} B.~J.,   {Workman} J.~C.,  2017, \mn@doi
  [\mnras] {10.1093/mnras/stx1683}, \href
  {https://ui.adsabs.harvard.edu/abs/2017MNRAS.471.1652L} {471, 1652}

\bibitem[\protect\citeauthoryear{{Lazzati}, {Perna}, {Morsony}, {Lopez-Camara},
  {Cantiello}, {Ciolfi}, {Giacomazzo}  \& {Workman}}{{Lazzati}
  et~al.}{2018}]{Lazzati2018}
{Lazzati} D.,  {Perna} R.,  {Morsony} B.~J.,  {Lopez-Camara} D.,  {Cantiello}
  M.,  {Ciolfi} R.,  {Giacomazzo} B.,   {Workman} J.~C.,  2018, \mn@doi [\prl]
  {10.1103/PhysRevLett.120.241103}, \href
  {https://ui.adsabs.harvard.edu/abs/2018PhRvL.120x1103L} {120, 241103}

\bibitem[\protect\citeauthoryear{{Levan} et~al.,}{{Levan}
  et~al.}{2023}]{Levan2023}
{Levan} A.,  et~al., 2023, \mn@doi [arXiv e-prints]
  {10.48550/arXiv.2307.02098}, \href
  {https://ui.adsabs.harvard.edu/abs/2023arXiv230702098L} {p. arXiv:2307.02098}

\bibitem[\protect\citeauthoryear{{Levinson}, {Ofek}, {Waxman}  \&
  {Gal-Yam}}{{Levinson} et~al.}{2002}]{Levinson2002}
{Levinson} A.,  {Ofek} E.~O.,  {Waxman} E.,   {Gal-Yam} A.,  2002, \mn@doi
  [\apj] {10.1086/341866}, \href
  {https://ui.adsabs.harvard.edu/abs/2002ApJ...576..923L} {576, 923}

\bibitem[\protect\citeauthoryear{{Lien}, {Sakamoto}, {Gehrels}, {Palmer},
  {Barthelmy}, {Graziani}  \& {Cannizzo}}{{Lien} et~al.}{2014}]{Lien2014}
{Lien} A.,  {Sakamoto} T.,  {Gehrels} N.,  {Palmer} D.~M.,  {Barthelmy} S.~D.,
  {Graziani} C.,   {Cannizzo} J.~K.,  2014, \mn@doi [\apj]
  {10.1088/0004-637X/783/1/24}, \href
  {https://ui.adsabs.harvard.edu/abs/2014ApJ...783...24L} {783, 24}

\bibitem[\protect\citeauthoryear{{Lien} et~al.,}{{Lien}
  et~al.}{2016}]{Lien2016}
{Lien} A.,  et~al., 2016, \mn@doi [\apj] {10.3847/0004-637X/829/1/7}, \href
  {https://ui.adsabs.harvard.edu/abs/2016ApJ...829....7L} {829, 7}

\bibitem[\protect\citeauthoryear{{Lipunov}, {Postnov}  \&
  {Prokhorov}}{{Lipunov} et~al.}{2001}]{Lipunov2001}
{Lipunov} V.~M.,  {Postnov} K.~A.,   {Prokhorov} M.~E.,  2001, \mn@doi
  [Astronomy Reports] {10.1134/1.1353364}, \href
  {https://ui.adsabs.harvard.edu/abs/2001ARep...45..236L} {45, 236}

\bibitem[\protect\citeauthoryear{{Maggiore} et~al.,}{{Maggiore}
  et~al.}{2020}]{ET2020}
{Maggiore} M.,  et~al., 2020, \mn@doi [\jcap] {10.1088/1475-7516/2020/03/050},
  \href {https://ui.adsabs.harvard.edu/abs/2020JCAP...03..050M} {2020, 050}

\bibitem[\protect\citeauthoryear{{Makhathini} et~al.,}{{Makhathini}
  et~al.}{2021}]{Makhathini2021ApJ}
{Makhathini} S.,  et~al., 2021, \mn@doi [\apj] {10.3847/1538-4357/ac1ffc},
  \href {https://ui.adsabs.harvard.edu/abs/2021ApJ...922..154M} {922, 154}

\bibitem[\protect\citeauthoryear{{Margutti} et~al.,}{{Margutti}
  et~al.}{2018}]{Margutti2018}
{Margutti} R.,  et~al., 2018, \mn@doi [\apjl] {10.3847/2041-8213/aab2ad}, \href
  {https://ui.adsabs.harvard.edu/abs/2018ApJ...856L..18M} {856, L18}

\bibitem[\protect\citeauthoryear{{Matsumoto}, {Nakar}  \& {Piran}}{{Matsumoto}
  et~al.}{2019a}]{Matsumoto2019a}
{Matsumoto} T.,  {Nakar} E.,   {Piran} T.,  2019a, \mn@doi [\mnras]
  {10.1093/mnras/sty3200}, \href
  {https://ui.adsabs.harvard.edu/abs/2019MNRAS.483.1247M} {483, 1247}

\bibitem[\protect\citeauthoryear{{Matsumoto}, {Nakar}  \& {Piran}}{{Matsumoto}
  et~al.}{2019b}]{Matsumoto2019b}
{Matsumoto} T.,  {Nakar} E.,   {Piran} T.,  2019b, \mn@doi [\mnras]
  {10.1093/mnras/stz923}, \href
  {https://ui.adsabs.harvard.edu/abs/2019MNRAS.486.1563M} {486, 1563}

\bibitem[\protect\citeauthoryear{{McCarthy} \& {Laskar}}{{McCarthy} \&
  {Laskar}}{2024}]{McCarthy2024}
{McCarthy} G.~A.,  {Laskar} T.,  2024, \mn@doi [arXiv e-prints]
  {10.48550/arXiv.2405.11309}, \href
  {https://ui.adsabs.harvard.edu/abs/2024arXiv240511309M} {p. arXiv:2405.11309}

\bibitem[\protect\citeauthoryear{{M{\'e}sz{\'a}ros} \&
  {Rees}}{{M{\'e}sz{\'a}ros} \& {Rees}}{1997}]{Meszaros1997}
{M{\'e}sz{\'a}ros} P.,  {Rees} M.~J.,  1997, \mn@doi [\apj] {10.1086/303625},
  \href {https://ui.adsabs.harvard.edu/abs/1997ApJ...476..232M} {476, 232}

\bibitem[\protect\citeauthoryear{{M{\'e}sz{\'a}ros}, {Rees}  \&
  {Wijers}}{{M{\'e}sz{\'a}ros} et~al.}{1998}]{Meszaros1998}
{M{\'e}sz{\'a}ros} P.,  {Rees} M.~J.,   {Wijers} R.~A.~M.~J.,  1998, \mn@doi
  [\apj] {10.1086/305635}, \href
  {https://ui.adsabs.harvard.edu/abs/1998ApJ...499..301M} {499, 301}

\bibitem[\protect\citeauthoryear{{Metzger}, {Giannios}, {Thompson},
  {Bucciantini}  \& {Quataert}}{{Metzger} et~al.}{2011}]{Metzger2011}
{Metzger} B.~D.,  {Giannios} D.,  {Thompson} T.~A.,  {Bucciantini} N.,
  {Quataert} E.,  2011, \mn@doi [\mnras] {10.1111/j.1365-2966.2011.18280.x},
  \href {https://ui.adsabs.harvard.edu/abs/2011MNRAS.413.2031M} {413, 2031}

\bibitem[\protect\citeauthoryear{{Metzger}, {Beniamini}  \&
  {Giannios}}{{Metzger} et~al.}{2018}]{Metzger2018}
{Metzger} B.~D.,  {Beniamini} P.,   {Giannios} D.,  2018, \mn@doi [\apj]
  {10.3847/1538-4357/aab70c}, \href
  {https://ui.adsabs.harvard.edu/abs/2018ApJ...857...95M} {857, 95}

\bibitem[\protect\citeauthoryear{{Mogushi}, {Cavagli{\`a}}  \&
  {Siellez}}{{Mogushi} et~al.}{2019}]{Mogushi2019}
{Mogushi} K.,  {Cavagli{\`a}} M.,   {Siellez} K.,  2019, \mn@doi [\apj]
  {10.3847/1538-4357/ab1f76}, \href
  {https://ui.adsabs.harvard.edu/abs/2019ApJ...880...55M} {880, 55}

\bibitem[\protect\citeauthoryear{{Molinari} et~al.,}{{Molinari}
  et~al.}{2007}]{Molinari2007}
{Molinari} E.,  et~al., 2007, \mn@doi [\aap] {10.1051/0004-6361:20077388},
  \href {https://ui.adsabs.harvard.edu/abs/2007A&A...469L..13M} {469, L13}

\bibitem[\protect\citeauthoryear{{Mooley} et~al.,}{{Mooley}
  et~al.}{2018}]{Mooley2018}
{Mooley} K.~P.,  et~al., 2018, \mn@doi [\nat] {10.1038/s41586-018-0486-3},
  \href {https://ui.adsabs.harvard.edu/abs/2018Natur.561..355M} {561, 355}

\bibitem[\protect\citeauthoryear{{Mooley}, {Anderson}  \& {Lu}}{{Mooley}
  et~al.}{2022a}]{Mooley2022}
{Mooley} K.~P.,  {Anderson} J.,   {Lu} W.,  2022a, \mn@doi [\nat]
  {10.1038/s41586-022-05145-7}, \href
  {https://ui.adsabs.harvard.edu/abs/2022Natur.610..273M} {610, 273}

\bibitem[\protect\citeauthoryear{{Mooley} et~al.,}{{Mooley}
  et~al.}{2022b}]{Mooley2022offaxis}
{Mooley} K.~P.,  et~al., 2022b, \mn@doi [\apj] {10.3847/1538-4357/ac3330},
  \href {https://ui.adsabs.harvard.edu/abs/2022ApJ...924...16M} {924, 16}

\bibitem[\protect\citeauthoryear{{Nakar} \& {Piran}}{{Nakar} \&
  {Piran}}{2021}]{Nakar-Piran-21}
{Nakar} E.,  {Piran} T.,  2021, \mn@doi [\apj] {10.3847/1538-4357/abd6cd},
  \href {https://ui.adsabs.harvard.edu/abs/2021ApJ...909..114N} {909, 114}

\bibitem[\protect\citeauthoryear{{Nakar}, {Piran}  \& {Granot}}{{Nakar}
  et~al.}{2002}]{Nakar2002}
{Nakar} E.,  {Piran} T.,   {Granot} J.,  2002, \mn@doi [\apj] {10.1086/342791},
  \href {https://ui.adsabs.harvard.edu/abs/2002ApJ...579..699N} {579, 699}

\bibitem[\protect\citeauthoryear{{Nakar}, {Gal-Yam}  \& {Fox}}{{Nakar}
  et~al.}{2006}]{Nakar2006}
{Nakar} E.,  {Gal-Yam} A.,   {Fox} D.~B.,  2006, \mn@doi [\apj]
  {10.1086/505855}, \href
  {https://ui.adsabs.harvard.edu/abs/2006ApJ...650..281N} {650, 281}

\bibitem[\protect\citeauthoryear{{Nakar}, {Gottlieb}, {Piran}, {Kasliwal}  \&
  {Hallinan}}{{Nakar} et~al.}{2018}]{Nakar2018}
{Nakar} E.,  {Gottlieb} O.,  {Piran} T.,  {Kasliwal} M.~M.,   {Hallinan} G.,
  2018, \mn@doi [\apj] {10.3847/1538-4357/aae205}, \href
  {https://ui.adsabs.harvard.edu/abs/2018ApJ...867...18N} {867, 18}

\bibitem[\protect\citeauthoryear{{Nappo}, {Ghisellini}, {Ghirlanda},
  {Melandri}, {Nava}  \& {Burlon}}{{Nappo} et~al.}{2014}]{Nappo2014}
{Nappo} F.,  {Ghisellini} G.,  {Ghirlanda} G.,  {Melandri} A.,  {Nava} L.,
  {Burlon} D.,  2014, \mn@doi [\mnras] {10.1093/mnras/stu1832}, \href
  {https://ui.adsabs.harvard.edu/abs/2014MNRAS.445.1625N} {445, 1625}

\bibitem[\protect\citeauthoryear{{Nathanail}, {Gill}, {Porth}, {Fromm}  \&
  {Rezzolla}}{{Nathanail} et~al.}{2020}]{Nathanail+20}
{Nathanail} A.,  {Gill} R.,  {Porth} O.,  {Fromm} C.~M.,   {Rezzolla} L.,
  2020, \mn@doi [\mnras] {10.1093/mnras/staa1454}, \href
  {https://ui.adsabs.harvard.edu/abs/2020MNRAS.495.3780N} {495, 3780}

\bibitem[\protect\citeauthoryear{{Nathanail}, {Gill}, {Porth}, {Fromm}  \&
  {Rezzolla}}{{Nathanail} et~al.}{2021}]{Nathanail+21}
{Nathanail} A.,  {Gill} R.,  {Porth} O.,  {Fromm} C.~M.,   {Rezzolla} L.,
  2021, \mn@doi [\mnras] {10.1093/mnras/stab115}, \href
  {https://ui.adsabs.harvard.edu/abs/2021MNRAS.502.1843N} {502, 1843}

\bibitem[\protect\citeauthoryear{{Nativi}, {Bulla}, {Rosswog}, {Lundman},
  {Kowal}, {Gizzi}, {Lamb}  \& {Perego}}{{Nativi} et~al.}{2021}]{Nativi2021}
{Nativi} L.,  {Bulla} M.,  {Rosswog} S.,  {Lundman} C.,  {Kowal} G.,  {Gizzi}
  D.,  {Lamb} G.~P.,   {Perego} A.,  2021, \mn@doi [\mnras]
  {10.1093/mnras/staa3337}, \href
  {https://ui.adsabs.harvard.edu/abs/2021MNRAS.500.1772N} {500, 1772}

\bibitem[\protect\citeauthoryear{{Nava}, {Ghirlanda}, {Ghisellini}  \&
  {Celotti}}{{Nava} et~al.}{2011}]{Nava2011}
{Nava} L.,  {Ghirlanda} G.,  {Ghisellini} G.,   {Celotti} A.,  2011, \mn@doi
  [\aap] {10.1051/0004-6361/201016270}, \href
  {https://ui.adsabs.harvard.edu/abs/2011A&A...530A..21N} {530, A21}

\bibitem[\protect\citeauthoryear{{Nava}, {Sironi}, {Ghisellini}, {Celotti}  \&
  {Ghirlanda}}{{Nava} et~al.}{2013}]{Nava2013}
{Nava} L.,  {Sironi} L.,  {Ghisellini} G.,  {Celotti} A.,   {Ghirlanda} G.,
  2013, \mn@doi [\mnras] {10.1093/mnras/stt872}, \href
  {https://ui.adsabs.harvard.edu/abs/2013MNRAS.433.2107N} {433, 2107}

\bibitem[\protect\citeauthoryear{{Nava} et~al.,}{{Nava}
  et~al.}{2014}]{Nava2014}
{Nava} L.,  et~al., 2014, \mn@doi [\mnras] {10.1093/mnras/stu1451}, \href
  {https://ui.adsabs.harvard.edu/abs/2014MNRAS.443.3578N} {443, 3578}

\bibitem[\protect\citeauthoryear{{Norris} \& {Bonnell}}{{Norris} \&
  {Bonnell}}{2006}]{Norris2006}
{Norris} J.~P.,  {Bonnell} J.~T.,  2006, \mn@doi [\apj] {10.1086/502796}, \href
  {https://ui.adsabs.harvard.edu/abs/2006ApJ...643..266N} {643, 266}

\bibitem[\protect\citeauthoryear{{Norris}, {Gehrels}  \& {Scargle}}{{Norris}
  et~al.}{2010}]{Norris2010}
{Norris} J.~P.,  {Gehrels} N.,   {Scargle} J.~D.,  2010, \mn@doi [\apj]
  {10.1088/0004-637X/717/1/411}, \href
  {https://ui.adsabs.harvard.edu/abs/2010ApJ...717..411N} {717, 411}

\bibitem[\protect\citeauthoryear{{Nugent} et~al.,}{{Nugent}
  et~al.}{2022}]{Nugent2022}
{Nugent} A.~E.,  et~al., 2022, \mn@doi [\apj] {10.3847/1538-4357/ac91d1}, \href
  {https://ui.adsabs.harvard.edu/abs/2022ApJ...940...57N} {940, 57}

\bibitem[\protect\citeauthoryear{{O'Connor}, {Beniamini}  \&
  {Kouveliotou}}{{O'Connor} et~al.}{2020}]{OConnor2020}
{O'Connor} B.,  {Beniamini} P.,   {Kouveliotou} C.,  2020, \mn@doi [\mnras]
  {10.1093/mnras/staa1433}, \href
  {https://ui.adsabs.harvard.edu/abs/2020MNRAS.495.4782O} {495, 4782}

\bibitem[\protect\citeauthoryear{{O'Connor} et~al.,}{{O'Connor}
  et~al.}{2021}]{OConnor2021kn}
{O'Connor} B.,  et~al., 2021, \mn@doi [\mnras] {10.1093/mnras/stab132}, \href
  {https://ui.adsabs.harvard.edu/abs/2021MNRAS.502.1279O} {502, 1279}

\bibitem[\protect\citeauthoryear{{O'Connor} et~al.,}{{O'Connor}
  et~al.}{2022}]{OConnor2022}
{O'Connor} B.,  et~al., 2022, \mn@doi [\mnras] {10.1093/mnras/stac1982}, \href
  {https://ui.adsabs.harvard.edu/abs/2022MNRAS.515.4890O} {515, 4890}

\bibitem[\protect\citeauthoryear{{O'Connor} et~al.,}{{O'Connor}
  et~al.}{2023}]{OConnor2023}
{O'Connor} B.,  et~al., 2023, \mn@doi [Science Advances]
  {10.1126/sciadv.adi1405}, \href
  {https://ui.adsabs.harvard.edu/abs/2023SciA....9I1405O} {9, eadi1405}

\bibitem[\protect\citeauthoryear{{Panaitescu} \& {Kumar}}{{Panaitescu} \&
  {Kumar}}{2002}]{Panaitescu2002}
{Panaitescu} A.,  {Kumar} P.,  2002, \mn@doi [\apj] {10.1086/340094}, \href
  {https://ui.adsabs.harvard.edu/abs/2002ApJ...571..779P} {571, 779}

\bibitem[\protect\citeauthoryear{{Panaitescu} \& {Kumar}}{{Panaitescu} \&
  {Kumar}}{2003}]{Panaitescu2003}
{Panaitescu} A.,  {Kumar} P.,  2003, \mn@doi [\apj] {10.1086/375563}, \href
  {https://ui.adsabs.harvard.edu/abs/2003ApJ...592..390P} {592, 390}

\bibitem[\protect\citeauthoryear{{Patricelli}, {Bernardini}, {Mapelli},
  {D'Avanzo}, {Santoliquido}, {Cella}, {Razzano}  \& {Cuoco}}{{Patricelli}
  et~al.}{2022}]{Patricelli2022}
{Patricelli} B.,  {Bernardini} M.~G.,  {Mapelli} M.,  {D'Avanzo} P.,
  {Santoliquido} F.,  {Cella} G.,  {Razzano} M.,   {Cuoco} E.,  2022, \mn@doi
  [\mnras] {10.1093/mnras/stac1167}, \href
  {https://ui.adsabs.harvard.edu/abs/2022MNRAS.513.4159P} {513, 4159}

\bibitem[\protect\citeauthoryear{{Pe'er}}{{Pe'er}}{2015}]{Pe'er2015}
{Pe'er} A.,  2015, \mn@doi [Advances in Astronomy] {10.1155/2015/907321}, \href
  {https://ui.adsabs.harvard.edu/abs/2015AdAst2015E..22P} {2015, 907321}

\bibitem[\protect\citeauthoryear{{Perley} et~al.,}{{Perley}
  et~al.}{2024}]{Perley2024}
{Perley} D.~A.,  et~al., 2024, \mn@doi [arXiv e-prints]
  {10.48550/arXiv.2401.16470}, \href
  {https://ui.adsabs.harvard.edu/abs/2024arXiv240116470P} {p. arXiv:2401.16470}

\bibitem[\protect\citeauthoryear{{Piran}}{{Piran}}{2004}]{Piran2004}
{Piran} T.,  2004, \mn@doi [Reviews of Modern Physics]
  {10.1103/RevModPhys.76.1143}, \href
  {https://ui.adsabs.harvard.edu/abs/2004RvMP...76.1143P} {76, 1143}

\bibitem[\protect\citeauthoryear{{Piran}, {Nakar}  \& {Rosswog}}{{Piran}
  et~al.}{2013}]{Piran2013}
{Piran} T.,  {Nakar} E.,   {Rosswog} S.,  2013, \mn@doi [\mnras]
  {10.1093/mnras/stt037}, \href
  {https://ui.adsabs.harvard.edu/abs/2013MNRAS.430.2121P} {430, 2121}

\bibitem[\protect\citeauthoryear{{Planck Collaboration} et~al.,}{{Planck
  Collaboration} et~al.}{2020}]{Planck2018}
{Planck Collaboration} et~al., 2020, \mn@doi [\aap]
  {10.1051/0004-6361/201833910}, \href
  {https://ui.adsabs.harvard.edu/abs/2020A&A...641A...6P} {641, A6}

\bibitem[\protect\citeauthoryear{{Punturo} et~al.,}{{Punturo}
  et~al.}{2010}]{ET2010}
{Punturo} M.,  et~al., 2010, \mn@doi [Classical and Quantum Gravity]
  {10.1088/0264-9381/27/19/194002}, \href
  {https://ui.adsabs.harvard.edu/abs/2010CQGra..27s4002P} {27, 194002}

\bibitem[\protect\citeauthoryear{{Rastinejad} et~al.,}{{Rastinejad}
  et~al.}{2022}]{Rastinejad2022}
{Rastinejad} J.~C.,  et~al., 2022, \mn@doi [\nat] {10.1038/s41586-022-05390-w},
  \href {https://ui.adsabs.harvard.edu/abs/2022Natur.612..223R} {612, 223}

\bibitem[\protect\citeauthoryear{{Reitze} et~al.,}{{Reitze}
  et~al.}{2019}]{CE2019}
{Reitze} D.,  et~al., 2019, in Bulletin of the American Astronomical Society.
  p.~35 (\mn@eprint {arXiv} {1907.04833}), \mn@doi{10.48550/arXiv.1907.04833}

\bibitem[\protect\citeauthoryear{{Resmi} et~al.,}{{Resmi}
  et~al.}{2018}]{Resmi2018}
{Resmi} L.,  et~al., 2018, \mn@doi [\apj] {10.3847/1538-4357/aae1a6}, \href
  {https://ui.adsabs.harvard.edu/abs/2018ApJ...867...57R} {867, 57}

\bibitem[\protect\citeauthoryear{{Rhoads}}{{Rhoads}}{1999}]{Rhoads1999}
{Rhoads} J.~E.,  1999, \mn@doi [\apj] {10.1086/307907}, \href
  {https://ui.adsabs.harvard.edu/abs/1999ApJ...525..737R} {525, 737}

\bibitem[\protect\citeauthoryear{{Rhoads}}{{Rhoads}}{2003}]{Rhoads2003}
{Rhoads} J.~E.,  2003, \mn@doi [\apj] {10.1086/368125}, \href
  {https://ui.adsabs.harvard.edu/abs/2003ApJ...591.1097R} {591, 1097}

\bibitem[\protect\citeauthoryear{{Ronchini} et~al.,}{{Ronchini}
  et~al.}{2022}]{Ronchini2022}
{Ronchini} S.,  et~al., 2022, \mn@doi [\aap] {10.1051/0004-6361/202243705},
  \href {https://ui.adsabs.harvard.edu/abs/2022A&A...665A..97R} {665, A97}

\bibitem[\protect\citeauthoryear{{Rossi}, {Lazzati}  \& {Rees}}{{Rossi}
  et~al.}{2002}]{Rossi02}
{Rossi} E.,  {Lazzati} D.,   {Rees} M.~J.,  2002, \mn@doi [\mnras]
  {10.1046/j.1365-8711.2002.05363.x}, \href
  {https://ui.adsabs.harvard.edu/abs/2002MNRAS.332..945R} {332, 945}

\bibitem[\protect\citeauthoryear{{Rossi}, {Lazzati}, {Salmonson}  \&
  {Ghisellini}}{{Rossi} et~al.}{2004}]{Rossi2004}
{Rossi} E.~M.,  {Lazzati} D.,  {Salmonson} J.~D.,   {Ghisellini} G.,  2004,
  \mn@doi [\mnras] {10.1111/j.1365-2966.2004.08165.x}, \href
  {https://ui.adsabs.harvard.edu/abs/2004MNRAS.354...86R} {354, 86}

\bibitem[\protect\citeauthoryear{{Rouco Escorial} et~al.,}{{Rouco Escorial}
  et~al.}{2022}]{RoucoEscorial2022}
{Rouco Escorial} A.,  et~al., 2022, \mn@doi [arXiv e-prints]
  {10.48550/arXiv.2210.05695}, \href
  {https://ui.adsabs.harvard.edu/abs/2022arXiv221005695R} {p. arXiv:2210.05695}

\bibitem[\protect\citeauthoryear{{Ryan}, {van Eerten}, {MacFadyen}  \&
  {Zhang}}{{Ryan} et~al.}{2015}]{Ryan15}
{Ryan} G.,  {van Eerten} H.,  {MacFadyen} A.,   {Zhang} B.-B.,  2015, \mn@doi
  [\apj] {10.1088/0004-637X/799/1/3}, \href
  {https://ui.adsabs.harvard.edu/abs/2015ApJ...799....3R} {799, 3}

\bibitem[\protect\citeauthoryear{{Ryan}, {van Eerten}, {Piro}  \&
  {Troja}}{{Ryan} et~al.}{2020}]{Ryan2020}
{Ryan} G.,  {van Eerten} H.,  {Piro} L.,   {Troja} E.,  2020, \mn@doi [\apj]
  {10.3847/1538-4357/ab93cf}, \href
  {https://ui.adsabs.harvard.edu/abs/2020ApJ...896..166R} {896, 166}

\bibitem[\protect\citeauthoryear{{Ryan}, {van Eerten}, {Troja}, {Piro},
  {O'Connor}  \& {Ricci}}{{Ryan} et~al.}{2023}]{Ryan2023}
{Ryan} G.,  {van Eerten} H.,  {Troja} E.,  {Piro} L.,  {O'Connor} B.,   {Ricci}
  R.,  2023, \mn@doi [arXiv e-prints] {10.48550/arXiv.2310.02328}, \href
  {https://ui.adsabs.harvard.edu/abs/2023arXiv231002328R} {p. arXiv:2310.02328}

\bibitem[\protect\citeauthoryear{{Salafia}, {Ghisellini}, {Pescalli},
  {Ghirlanda}  \& {Nappo}}{{Salafia} et~al.}{2015}]{Salafia2015}
{Salafia} O.~S.,  {Ghisellini} G.,  {Pescalli} A.,  {Ghirlanda} G.,   {Nappo}
  F.,  2015, \mn@doi [\mnras] {10.1093/mnras/stv766}, \href
  {https://ui.adsabs.harvard.edu/abs/2015MNRAS.450.3549S} {450, 3549}

\bibitem[\protect\citeauthoryear{{Salafia}, {Ghisellini}  \&
  {Ghirlanda}}{{Salafia} et~al.}{2018}]{Salafia2018}
{Salafia} O.~S.,  {Ghisellini} G.,   {Ghirlanda} G.,  2018, \mn@doi [\mnras]
  {10.1093/mnrasl/slx189}, \href
  {https://ui.adsabs.harvard.edu/abs/2018MNRAS.474L...7S} {474, L7}

\bibitem[\protect\citeauthoryear{{Salafia}, {Ghirlanda}, {Ascenzi}  \&
  {Ghisellini}}{{Salafia} et~al.}{2019}]{Salafia2019}
{Salafia} O.~S.,  {Ghirlanda} G.,  {Ascenzi} S.,   {Ghisellini} G.,  2019,
  \mn@doi [\aap] {10.1051/0004-6361/201935831}, \href
  {https://ui.adsabs.harvard.edu/abs/2019A&A...628A..18S} {628, A18}

\bibitem[\protect\citeauthoryear{{Salafia}, {Ravasio}, {Ghirlanda}  \&
  {Mandel}}{{Salafia} et~al.}{2023}]{Salafia2023}
{Salafia} O.~S.,  {Ravasio} M.~E.,  {Ghirlanda} G.,   {Mandel} I.,  2023,
  \mn@doi [\aap] {10.1051/0004-6361/202347298}, \href
  {https://ui.adsabs.harvard.edu/abs/2023A&A...680A..45S} {680, A45}

\bibitem[\protect\citeauthoryear{{Saleem}}{{Saleem}}{2020}]{Saleem2020}
{Saleem} M.,  2020, \mn@doi [\mnras] {10.1093/mnras/staa303}, \href
  {https://ui.adsabs.harvard.edu/abs/2020MNRAS.493.1633S} {493, 1633}

\bibitem[\protect\citeauthoryear{{Sari} \& {Esin}}{{Sari} \&
  {Esin}}{2001}]{Sari2001}
{Sari} R.,  {Esin} A.~A.,  2001, \mn@doi [\apj] {10.1086/319003}, \href
  {https://ui.adsabs.harvard.edu/abs/2001ApJ...548..787S} {548, 787}

\bibitem[\protect\citeauthoryear{{Sari} \& {Piran}}{{Sari} \&
  {Piran}}{1999}]{Sari1999}
{Sari} R.,  {Piran} T.,  1999, \mn@doi [\apj] {10.1086/307508}, \href
  {https://ui.adsabs.harvard.edu/abs/1999ApJ...520..641S} {520, 641}

\bibitem[\protect\citeauthoryear{{Sari}, {Piran}  \& {Narayan}}{{Sari}
  et~al.}{1998}]{Sari1998}
{Sari} R.,  {Piran} T.,   {Narayan} R.,  1998, \mn@doi [\apjl]
  {10.1086/311269}, \href
  {https://ui.adsabs.harvard.edu/abs/1998ApJ...497L..17S} {497, L17}

\bibitem[\protect\citeauthoryear{{Sari}, {Piran}  \& {Halpern}}{{Sari}
  et~al.}{1999}]{SariPiranHalpern1999}
{Sari} R.,  {Piran} T.,   {Halpern} J.~P.,  1999, \mn@doi [\apjl]
  {10.1086/312109}, \href
  {https://ui.adsabs.harvard.edu/abs/1999ApJ...519L..17S} {519, L17}

\bibitem[\protect\citeauthoryear{{Sarin}, {Lasky}, {Vivanco}, {Stevenson},
  {Chattopadhyay}, {Smith}  \& {Thrane}}{{Sarin} et~al.}{2022a}]{Sarin2022b}
{Sarin} N.,  {Lasky} P.~D.,  {Vivanco} F.~H.,  {Stevenson} S.~P.,
  {Chattopadhyay} D.,  {Smith} R.,   {Thrane} E.,  2022a, \mn@doi [\prd]
  {10.1103/PhysRevD.105.083004}, \href
  {https://ui.adsabs.harvard.edu/abs/2022PhRvD.105h3004S} {105, 083004}

\bibitem[\protect\citeauthoryear{{Sarin}, {Hamburg}, {Burns}, {Ashton}, {Lasky}
   \& {Lamb}}{{Sarin} et~al.}{2022b}]{Sarin2022}
{Sarin} N.,  {Hamburg} R.,  {Burns} E.,  {Ashton} G.,  {Lasky} P.~D.,   {Lamb}
  G.~P.,  2022b, \mn@doi [\mnras] {10.1093/mnras/stac601}, \href
  {https://ui.adsabs.harvard.edu/abs/2022MNRAS.512.1391S} {512, 1391}

\bibitem[\protect\citeauthoryear{{Savchenko} et~al.,}{{Savchenko}
  et~al.}{2017}]{Savchenko2017}
{Savchenko} V.,  et~al., 2017, \mn@doi [\apjl] {10.3847/2041-8213/aa8f94},
  \href {https://ui.adsabs.harvard.edu/abs/2017ApJ...848L..15S} {848, L15}

\bibitem[\protect\citeauthoryear{{Schulze} et~al.,}{{Schulze}
  et~al.}{2011}]{Schulze2011}
{Schulze} S.,  et~al., 2011, \mn@doi [\aap] {10.1051/0004-6361/201015581},
  \href {https://ui.adsabs.harvard.edu/abs/2011A&A...526A..23S} {526, A23}

\bibitem[\protect\citeauthoryear{{Soderberg} et~al.,}{{Soderberg}
  et~al.}{2006}]{Soderberg2006}
{Soderberg} A.~M.,  et~al., 2006, \mn@doi [\apj] {10.1086/506429}, \href
  {https://ui.adsabs.harvard.edu/abs/2006ApJ...650..261S} {650, 261}

\bibitem[\protect\citeauthoryear{{Srinivasaragavan}, {Dainotti}, {Fraija},
  {Hernandez}, {Nagataki}, {Lenart}, {Bowden}  \& {Wagner}}{{Srinivasaragavan}
  et~al.}{2020}]{Srinivasaragavan2020}
{Srinivasaragavan} G.~P.,  {Dainotti} M.~G.,  {Fraija} N.,  {Hernandez} X.,
  {Nagataki} S.,  {Lenart} A.,  {Bowden} L.,   {Wagner} R.,  2020, \mn@doi
  [\apj] {10.3847/1538-4357/abb702}, \href
  {https://ui.adsabs.harvard.edu/abs/2020ApJ...903...18S} {903, 18}

\bibitem[\protect\citeauthoryear{{Stalder} et~al.,}{{Stalder}
  et~al.}{2017}]{Stadler2017}
{Stalder} B.,  et~al., 2017, \mn@doi [\apj] {10.3847/1538-4357/aa95c1}, \href
  {https://ui.adsabs.harvard.edu/abs/2017ApJ...850..149S} {850, 149}

\bibitem[\protect\citeauthoryear{{Takahashi} \& {Ioka}}{{Takahashi} \&
  {Ioka}}{2021}]{Takahashi2021}
{Takahashi} K.,  {Ioka} K.,  2021, \mn@doi [\mnras] {10.1093/mnras/stab032},
  \href {https://ui.adsabs.harvard.edu/abs/2021MNRAS.501.5746T} {501, 5746}

\bibitem[\protect\citeauthoryear{{Thompson}, {Chang}  \& {Quataert}}{{Thompson}
  et~al.}{2004}]{Thompson04}
{Thompson} T.~A.,  {Chang} P.,   {Quataert} E.,  2004, \mn@doi [\apj]
  {10.1086/421969}, \href
  {https://ui.adsabs.harvard.edu/abs/2004ApJ...611..380T} {611, 380}

\bibitem[\protect\citeauthoryear{{Totani} \& {Panaitescu}}{{Totani} \&
  {Panaitescu}}{2002}]{Totani2002}
{Totani} T.,  {Panaitescu} A.,  2002, \mn@doi [\apj] {10.1086/341738}, \href
  {https://ui.adsabs.harvard.edu/abs/2002ApJ...576..120T} {576, 120}

\bibitem[\protect\citeauthoryear{{Troja} et~al.,}{{Troja}
  et~al.}{2016}]{Troja2016jetbreak}
{Troja} E.,  et~al., 2016, \mn@doi [\apj] {10.3847/0004-637X/827/2/102}, \href
  {https://ui.adsabs.harvard.edu/abs/2016ApJ...827..102T} {827, 102}

\bibitem[\protect\citeauthoryear{{Troja} et~al.,}{{Troja}
  et~al.}{2017}]{Troja2017}
{Troja} E.,  et~al., 2017, \mn@doi [\nat] {10.1038/nature24290}, \href
  {https://ui.adsabs.harvard.edu/abs/2017Natur.551...71T} {551, 71}

\bibitem[\protect\citeauthoryear{{Troja} et~al.,}{{Troja}
  et~al.}{2019}]{Troja2019b}
{Troja} E.,  et~al., 2019, \mn@doi [\mnras] {10.1093/mnras/stz2255}, \href
  {https://ui.adsabs.harvard.edu/abs/2019MNRAS.489.2104T} {489, 2104}

\bibitem[\protect\citeauthoryear{{Troja} et~al.,}{{Troja}
  et~al.}{2020}]{Troja2020}
{Troja} E.,  et~al., 2020, \mn@doi [\mnras] {10.1093/mnras/staa2626}, \href
  {https://ui.adsabs.harvard.edu/abs/2020MNRAS.498.5643T} {498, 5643}

\bibitem[\protect\citeauthoryear{{Troja} et~al.,}{{Troja}
  et~al.}{2022}]{Troja2022}
{Troja} E.,  et~al., 2022, \mn@doi [\nat] {10.1038/s41586-022-05327-3}, \href
  {https://ui.adsabs.harvard.edu/abs/2022Natur.612..228T} {612, 228}

\bibitem[\protect\citeauthoryear{{Usov}}{{Usov}}{1992}]{Usov1992}
{Usov} V.~V.,  1992, \mn@doi [\nat] {10.1038/357472a0}, \href
  {https://ui.adsabs.harvard.edu/abs/1992Natur.357..472U} {357, 472}

\bibitem[\protect\citeauthoryear{{Wanderman} \& {Piran}}{{Wanderman} \&
  {Piran}}{2015}]{Wanderman2015}
{Wanderman} D.,  {Piran} T.,  2015, \mn@doi [\mnras] {10.1093/mnras/stv123},
  \href {https://ui.adsabs.harvard.edu/abs/2015MNRAS.448.3026W} {448, 3026}

\bibitem[\protect\citeauthoryear{{Wang}, {Zhang}, {Liang}, {Lu}, {Lin}, {Li}
  \& {Li}}{{Wang} et~al.}{2018}]{Wang2018}
{Wang} X.-G.,  {Zhang} B.,  {Liang} E.-W.,  {Lu} R.-J.,  {Lin} D.-B.,  {Li} J.,
    {Li} L.,  2018, \mn@doi [\apj] {10.3847/1538-4357/aabc13}, \href
  {https://ui.adsabs.harvard.edu/abs/2018ApJ...859..160W} {859, 160}

\bibitem[\protect\citeauthoryear{{White} et~al.,}{{White}
  et~al.}{2021}]{White2021}
{White} N.~E.,  et~al., 2021, in {Siegmund} O.~H.,  ed.,  Society of
  Photo-Optical Instrumentation Engineers (SPIE) Conference Series Vol. 11821,
  UV, X-Ray, and Gamma-Ray Space Instrumentation for Astronomy XXII. p. 1182109
  (\mn@eprint {arXiv} {2111.06497}), \mn@doi{10.1117/12.2599293}

\bibitem[\protect\citeauthoryear{{Wichern}, {Ravasio}, {Jonker},
  {Quirola-V{\'a}squez}, {Levan}, {Bauer}  \& {Kann}}{{Wichern}
  et~al.}{2024}]{Wichern2024}
{Wichern} H.~C.~I.,  {Ravasio} M.~E.,  {Jonker} P.~G.,  {Quirola-V{\'a}squez}
  J.~A.,  {Levan} A.~J.,  {Bauer} F.~E.,   {Kann} D.~A.,  2024, arXiv e-prints,
  \href {https://ui.adsabs.harvard.edu/abs/2024arXiv240706371W} {p.
  arXiv:2407.06371}

\bibitem[\protect\citeauthoryear{{Wijers} \& {Galama}}{{Wijers} \&
  {Galama}}{1999}]{Wijers1999}
{Wijers} R.~A.~M.~J.,  {Galama} T.~J.,  1999, \mn@doi [\apj] {10.1086/307705},
  \href {https://ui.adsabs.harvard.edu/abs/1999ApJ...523..177W} {523, 177}

\bibitem[\protect\citeauthoryear{{Williams} et~al.,}{{Williams}
  et~al.}{2023}]{Williams2023}
{Williams} M.~A.,  et~al., 2023, \mn@doi [\apjl] {10.3847/2041-8213/acbcd1},
  \href {https://ui.adsabs.harvard.edu/abs/2023ApJ...946L..24W} {946, L24}

\bibitem[\protect\citeauthoryear{{Xie}, {Zrake}  \& {MacFadyen}}{{Xie}
  et~al.}{2018}]{Xie2018}
{Xie} X.,  {Zrake} J.,   {MacFadyen} A.,  2018, \mn@doi [\apj]
  {10.3847/1538-4357/aacf9c}, \href
  {https://ui.adsabs.harvard.edu/abs/2018ApJ...863...58X} {863, 58}

\bibitem[\protect\citeauthoryear{{Xu}, {Huang}, {Geng}, {Wu}, {Li}  \&
  {Zhang}}{{Xu} et~al.}{2023}]{Fan2023}
{Xu} F.,  {Huang} Y.-F.,  {Geng} J.-J.,  {Wu} X.-F.,  {Li} X.-J.,   {Zhang}
  Z.-B.,  2023, \mn@doi [\aap] {10.1051/0004-6361/202245414}, \href
  {https://ui.adsabs.harvard.edu/abs/2023A&A...673A..20X} {673, A20}

\bibitem[\protect\citeauthoryear{{Yang} et~al.,}{{Yang}
  et~al.}{2015}]{Yang2015}
{Yang} B.,  et~al., 2015, \mn@doi [Nature Communications] {10.1038/ncomms8323},
  \href {https://ui.adsabs.harvard.edu/abs/2015NatCo...6.7323Y} {6, 7323}

\bibitem[\protect\citeauthoryear{{Yang} et~al.,}{{Yang}
  et~al.}{2022}]{Yang2022kn211211A}
{Yang} J.,  et~al., 2022, \mn@doi [\nat] {10.1038/s41586-022-05403-8}, \href
  {https://ui.adsabs.harvard.edu/abs/2022Natur.612..232Y} {612, 232}

\bibitem[\protect\citeauthoryear{{Yang} et~al.,}{{Yang}
  et~al.}{2023}]{Yang2023}
{Yang} Y.-H.,  et~al., 2023, \mn@doi [arXiv e-prints]
  {10.48550/arXiv.2308.00638}, \href
  {https://ui.adsabs.harvard.edu/abs/2023arXiv230800638Y} {p. arXiv:2308.00638}

\bibitem[\protect\citeauthoryear{{Yost}, {Harrison}, {Sari}  \& {Frail}}{{Yost}
  et~al.}{2003}]{Yost2003}
{Yost} S.~A.,  {Harrison} F.~A.,  {Sari} R.,   {Frail} D.~A.,  2003, \mn@doi
  [\apj] {10.1086/378288}, \href
  {https://ui.adsabs.harvard.edu/abs/2003ApJ...597..459Y} {597, 459}

\bibitem[\protect\citeauthoryear{{Yu} et~al.,}{{Yu} et~al.}{2021}]{Yu2021}
{Yu} J.,  et~al., 2021, \mn@doi [\apj] {10.3847/1538-4357/ac0628}, \href
  {https://ui.adsabs.harvard.edu/abs/2021ApJ...916...54Y} {916, 54}

\bibitem[\protect\citeauthoryear{{Yuan} et~al.,}{{Yuan} et~al.}{2015}]{EP2015}
{Yuan} W.,  et~al., 2015, \mn@doi [arXiv e-prints] {10.48550/arXiv.1506.07735},
  \href {https://ui.adsabs.harvard.edu/abs/2015arXiv150607735Y} {p.
  arXiv:1506.07735}

\bibitem[\protect\citeauthoryear{{Yuan}, {Zhang}, {Chen}  \& {Ling}}{{Yuan}
  et~al.}{2022}]{EP2022}
{Yuan} W.,  {Zhang} C.,  {Chen} Y.,   {Ling} Z.,  2022, in , Handbook of X-ray
  and Gamma-ray Astrophysics.
p.~86, \mn@doi{10.1007/978-981-16-4544-0_151-1}

\bibitem[\protect\citeauthoryear{{Zhang} \& {MacFadyen}}{{Zhang} \&
  {MacFadyen}}{2009}]{Zhang2009}
{Zhang} W.,  {MacFadyen} A.,  2009, \mn@doi [\apj]
  {10.1088/0004-637X/698/2/1261}, \href
  {https://ui.adsabs.harvard.edu/abs/2009ApJ...698.1261Z} {698, 1261}

\bibitem[\protect\citeauthoryear{{Zhang} \& {M{\'e}sz{\'a}ros}}{{Zhang} \&
  {M{\'e}sz{\'a}ros}}{2002}]{Zhang2002}
{Zhang} B.,  {M{\'e}sz{\'a}ros} P.,  2002, \mn@doi [\apj] {10.1086/339981},
  \href {https://ui.adsabs.harvard.edu/abs/2002ApJ...571..876Z} {571, 876}

\bibitem[\protect\citeauthoryear{{Zou}, {Fan}  \& {Piran}}{{Zou}
  et~al.}{2009}]{Zou2009}
{Zou} Y.-C.,  {Fan} Y.-Z.,   {Piran} T.,  2009, \mn@doi [\mnras]
  {10.1111/j.1365-2966.2009.14779.x}, \href
  {https://ui.adsabs.harvard.edu/abs/2009MNRAS.396.1163Z} {396, 1163}

\bibitem[\protect\citeauthoryear{{van Eerten} \& {MacFadyen}}{{van Eerten} \&
  {MacFadyen}}{2011}]{vanEerten2011}
{van Eerten} H.~J.,  {MacFadyen} A.~I.,  2011, \mn@doi [\apjl]
  {10.1088/2041-8205/733/2/L37}, \href
  {https://ui.adsabs.harvard.edu/abs/2011ApJ...733L..37V} {733, L37}

\bibitem[\protect\citeauthoryear{{van Eerten}, {Zhang}  \& {MacFadyen}}{{van
  Eerten} et~al.}{2010}]{vanEerten2010}
{van Eerten} H.,  {Zhang} W.,   {MacFadyen} A.,  2010, \mn@doi [\apj]
  {10.1088/0004-637X/722/1/235}, \href
  {https://ui.adsabs.harvard.edu/abs/2010ApJ...722..235V} {722, 235}

\makeatother
\end{thebibliography}

\appendix

\section{Short gamma-ray burst sample}

In Table \ref{Table_SGRB_DATA}, we present the sample of short GRBs used in this work.

\begin{table*}
\centering
\caption{Tabulated observational data for the 58 sGRBs in our sample. Error on $\phi_{\gamma}$ is 90\% confidence level.  
%We present the redshift $z$, \textit{Swift}/BAT gamma-ray fluence $\phi_{\gamma,-7}=\phi_\gamma/10^{-7}$ erg cm$^{-2}$ (and 90\% confidence interval), upper limit on peak time $t_\textrm{o}$, and the lower limit on peak unabsorbed flux $F_{\rm{o},-10}=F_{\rm{o}}/10^{-10}$ erg cm$^{-2}$ s$^{-1}$ (and $1\sigma$ error) for sGRBs in our sample \citep[see also][]{OConnor2020}. 
$^a$ host galaxy photometric redshift.  
}
\vspace{-0.2cm}
\begin{tabular}{|c|c|c|c|c|c|}
\hline
\hline
\vspace{-0.05cm}
GRB & $z$  & $\phi_{\gamma,-7}$ (erg cm$^{-2}$)  & $t_\textrm{o}$ (s) & $F_{\rm{o},-10}$ (erg cm$^{-2}$ s$^{-1}$)   \\[-0.5mm]
\hline
\hline
051221A & 0.546 & $12.0\pm0.4$  & 95$\pm3$ & $2.9\pm0.6$  \\[0.37mm]
060801 & 1.131 & $0.8\pm0.1$ & 109$^{+5}_{-7}$ & $2.3\pm0.5$   \\[0.37mm]
061006 & 0.438 & $14.2^{+1.4}_{-1.3}$ & 168$^{+20}_{-13}$ & $0.38\pm0.09$    \\[0.37mm]
070429B & 0.902 & $0.63\pm0.09$ & 520$_{-270}^{+350}$ & $0.035\pm 0.001$    \\[0.37mm]
070714B & 0.923 & $7.2\pm0.09$ & 68.6$\pm0.9$ & 14$_{-1.6}^{+2.0}$ &    \\[0.37mm]
070724A & 0.457 & $0.34\pm0.08$ & 386$_{-130}^{+540}$ & $0.11\pm0.02$ &   \\[0.37mm]
070809 & 0.473 & 1.0$\pm0.2$ & 126$_{-50}^{+100}$ & 0.18$\pm0.04$ \\[0.37mm]
071227 & 0.381 & 6.0$\pm2.0$ & 464$_{-140}^{+340}$ & 0.05$\pm0.01$ &   \\[0.37mm] 
080123 &  0.495 & 2.8$\pm1.7$ & 550$^{+230}_{-160}$ & 0.022$\pm0.006$   \\[0.37mm]
080905A & 0.122 & 1.4$\pm0.2$ & 1053$^{+190}_{-130}$ & 0.014$\pm0.003$   \\[0.37mm]
090426A & 2.609 & 1.8$_{-0.2}^{+0.3}$ & 255$\pm20$ & 0.32$\pm0.07$ &   \\[0.37mm] 
090510 & 0.903& $4.08\pm0.07$& 100$\pm2$ & 3.9$\pm0.8$    \\[0.37mm] 
100625A & 0.452 & $8.8\pm3.3$ & 175$_{-20}^{+30}$ & 2.7$\pm0.06$ \\[0.37mm] 
100724A & 1.288 & 1.6$\pm0.2$ & 79$\pm4$ & 1.4$\pm0.4$    \\[0.37mm] 
101219A & 0.718 & 4.3$\pm0.25$ & 83$_{-6}^{+7}$ & 3.5$\pm0.8$  \\[0.37mm]
111117A & 2.211 & 1.4$\pm0.2$ & 108$_{-18}^{+19}$ & 1.04$\pm0.13$  \\[0.37mm]
120305A & 0.225 & 2.0$\pm0.1$ & 370$^{+240}_{-60}$ & 0.14$\pm0.03$    \\[0.37mm]
120630A & 0.6$^a$ & 0.6$\pm0.2$ & 175$^{+120}_{-80}$ & 0.06$\pm0.01$   \\[0.37mm]
120804A & 1.3$^a$ & 8.8$\pm0.5$ & 322$_{-11}^{+9}$ & 3.0$\pm0.7$   \\[0.37mm]
121226A & 1.4$^a$ & 1.4$\pm0.2$ & 207$\pm20$ & 1.6$\pm0.4$    \\[0.37mm]
130515A & 0.8$^a$ & 1.5$\pm0.2$ & 377$^{+420}_{-130}$ & 0.015$\pm0.004$    \\[0.37mm]
130603B & 0.356 & 6.3$\pm0.3$ & 416$^{+70}_{-80}$ & 1.0$\pm0.2$   \\[0.37mm]
130912A & 0.717 & 1.7$\pm0.2$ & 155$^{+15}_{-13}$ & 2.4$\pm0.5$    \\[0.37mm]
140129B & 0.43 & 0.7$\pm0.1$ & 355$\pm10$ & 0.9$\pm0.2$    \\[0.37mm]
140903A & 0.351 & 1.4$\pm0.1$ & 196$\pm61$ & 0.25$\pm0.05$    \\[0.37mm]
140930B & 1.465 & 4.2$\pm0.4$ & 187$\pm4$ & 4.0$\pm0.5$   \\[0.37mm]
141212A & 0.596 & 0.73$\pm0.12$ & 180$_{-100}^{+320}$ & 0.020$\pm0.005$  \\[0.37mm]
150423A & 1.394 & 0.7$\pm0.1$ & 109$\pm30$ & 0.16$\pm0.04$   \\[0.37mm]
160821B & 0.161 & 1.2$\pm0.1$ & 379$_{-40}^{+120}$ & 0.11$\pm 0.02$    \\[0.37mm]
170428A & 0.453 & 2.8$\pm0.2$ & 800$^{+140}_{-100}$ & 0.1$\pm0.03$    \\[0.37mm]
181123B & 1.754 & 1.2$\pm0.2$ & 125$^{+20}_{-30}$ & 0.32$\pm0.07$    \\[0.37mm]
200411A & 0.6$^a$ & $1.0\pm0.1$ & 350$\pm70$ & 0.1$\pm0.002$ \\[0.37mm]
200522A & 0.553  & $1.1\pm0.1$ &  495$^{+150}_{-90}$ & 0.05$\pm0.01$ \\[0.37mm]
 211023B & 0.862  & $1.7\pm0.3$ & 85$^{+4}_{-5}$ & $0.8\pm0.2$\\[0.37mm]
 231117A & 0.257 & $22.7\pm0.6$ & 390$^{+90}_{-80}$ & $0.11\pm0.03$ \\[-0.5mm]
\hline
\hline
060313 & -- & 11.2$\pm0.5$ & 196$_{-6}^{+11}$ & 1.8$\pm0.5$   \\[0.37mm]
061201 & -- & 3.3$\pm0.3$ & 90$\pm4$ & 2.2$\pm0.4$   \\[0.37mm]
080426 & -- & 3.7$\pm0.2$ & 269$\pm40$ & 0.45$\pm0.1$   \\[0.37mm]
080702A & -- & 0.3$\pm0.1$ & 126$\pm50$ & 0.27$\pm0.06$   \\[0.37mm]
080919 & -- & 0.72$\pm0.1$ & 81$^{+4}_{-7}$ & 1.5$\pm0.3$   \\[0.37mm]
081024A & -- & 1.2$\pm0.16$ & 78$\pm2$ & 26$\pm4$   \\[0.37mm]
081226A & -- & 1.0$\pm0.2$ & 179$^{+80}_{-70}$ & 0.6$\pm0.14$   \\[0.37mm]
090621B & -- & 0.7$\pm0.1$ & 336$^{+500}_{-250}$ & 0.04$\pm0.01$   \\[0.37mm]
091109B & -- & 2.0$\pm0.2$ & 155$^{+120}_{-70}$ & 0.1$\pm0.02$    \\[0.37mm]
110112A &--  & 0.3$\pm0.1$ & 114$\pm30$ & 0.2$\pm0.04$  \\[0.37mm]
111020A &--  & 0.7$\pm0.1$ & 124$^{+50}_{-40}$ & 0.4$\pm0.1$   \\[0.37mm]
111121A & -- & 22$\pm1.5$ & 484$^{+70}_{-50}$ & 0.6$\pm0.1$     \\[0.37mm]
%150301A & -- & 0.7$\pm0.1$ & 52$\pm1$ & 59$\pm9$  \\[0.37mm]
151127A & -- & 0.23$\pm0.06$ & 174$^{+90}_{-80}$ & 0.08$\pm0.02$   \\[0.37mm]
160408A & -- & 1.6$\pm0.2$ & 850$^{+570}_{-200}$ & 0.015$\pm0.003$   \\[0.37mm]
160411A &--  & 0.8$\pm0.2$ & 3350$^{+270}_{-210}$& 0.018$\pm0.004$   \\[0.37mm]
160525B & -- & 0.32$\pm0.08$ & 73$\pm3$ & 1.3$\pm0.3$    \\[0.37mm]
160601A & -- & 0.7$\pm0.1$ & 270$^{+420}_{-190}$ & 0.012$\pm0.003$   \\[0.37mm]
160927A & -- & 1.4 $\pm0.2$& 109$\pm20$ & 0.23$\pm0.5$   \\[0.37mm]
170127B & -- & 1.0$\pm0.2$ & 130$^{+50}_{-30}$ & 0.6$\pm0.1$    \\[0.37mm]
180402A & -- & 1.4$\pm0.2$ & 87$^{+2}_{-3}$ & 5$\pm1$   \\[0.37mm]
201006A & -- & $1.3\pm0.2$ & $150\pm60$ & $0.2\pm0.04$ \\[0.37mm]
210726A & -- & $0.5\pm0.1$ & 520$^{+140}_{-110}$ & $0.04\pm0.01$ \\[0.37mm]
230228A & -- & $3.6\pm0.4$ & 270$^{+50}_{-40}$ & $0.3\pm0.06$ \\[-0.37mm]
\hline
\end{tabular}
\label{Table_SGRB_DATA}
\end{table*}

\section{Other synchrotron PLS}
\label{appendix:PLS}

\subsection{Characteristic synchrotron frequencies}

In the globally slow cooling regime ($\nu_\textrm{m}$\,$<$\,$\nu_\textrm{c}$), the (on-axis) cooling frequency $\nu_\textrm{c}$ and the (on-axis) injection frequency $\nu_\textrm{m}$ are \citep{Granot2002} 
\begin{align}
    & \nu_\textrm{c}=8.6\times 10^{22}\, \varepsilon_{B,-4}^{-3/2}n^{-1}_{-4}E_{\textrm{kin,0},52}^{-1/2}t_{\textrm{day}}^{-1/2}(1+z)^{-1/2}(1+Y)^{-2}\; \textrm{Hz}, \label{eqn: slow cool frequ.} \\
    & \nu_\textrm{m}=1.6\times 10^{10}\,E_{\textrm{kin,0},52}^{1/2}\varepsilon_{B,-4}^{1/2}\varepsilon_{e,-1}^{2}t_{\textrm{day}}^{-3/2}(1+z)^{1/2}\; \textrm{Hz},
\end{align}
where we have included Inverse Compton (IC)  corrections \citep{Sari2001,Zou2009,Beniamini2015} to the cooling frequency. Following \citet{Beniamini2015} (see also \citealt{Jacovich2020,McCarthy2024}), we compute the Compton parameter in the Thomson regime for $\nu_\textrm{m}$\,$<$\,$\nu_\textrm{c}$ as 
\begin{eqnarray}
Y = \left\{ \begin{array}{ll} 
\frac{\varepsilon_e}{\varepsilon_B (3-p)}\left( \frac{\nu_\textrm{m}}{\nu_{c,\textrm{syn}}} \right)^{(p-2)/2} & Y \ll 1, \\
\left[ \frac{\varepsilon_e}{\varepsilon_B (3-p)}\left( \frac{\nu_\textrm{m}}{\nu_{c,\textrm{syn}}} \right)^{(p-2)/2}\right]^{1/(4-p)} & Y \gg 1 
\end{array} \right.
\end{eqnarray}
where $\nu_\textrm{c}$\,$=$\,$\nu_{c,\textrm{syn}}(1+Y)^{-2}$. For our canonical parameters the Compton parameter is in the regime $Y$\,$\gg$\,$1$, provided by
\begin{align}
\label{eqn:Ysimplified}
Y=10\,\varepsilon_{e,-1}^{2/3}\varepsilon_{B,-4}^{-4/9}\Big(\frac{E_{\textrm{kin,0},52}\,n_{-4}\,(1+z)}{t_{\textrm{day}}}\Big)^{1/20},
\end{align}
where we have adopted $p$\,$=$\,$2.2$. 

In order to derive the values for $\nu_\textrm{m}$ and $\nu_\textrm{c}$ at the deceleration time (relevant to \S \ref{methods: jet structure} and Equations \ref{eqn:freqstruct} and \ref{eqn:freqstruct2}) we simply insert $t_\textrm{day}$\,$\rightarrow$\,$t_\textrm{dec,0}$. For simplicity Equation \ref{eqn:freqstruct} does not include IC corrections in the presented analytic scaling relation for $\nu_\textrm{c,dec}(\theta_\textrm{obs})$. Numerically, it is possible to include IC corrections by adding a factor of $\frac{(1+Y_\textrm{dec}(\theta_\textrm{obs}))^2}{(1+Y_\textrm{dec,0})^2}$ to Equation \ref{eqn:freqstruct} (where $Y_\textrm{dec,0}$ denotes the on-axis value at deceleration), and we do so within our calculations reported in \S \ref{sec: structjetresults}. In general, computing these values for off-axis angles can be performed by simply inserting the line-of-sight values for the kinetic energy $E_\textrm{kin}(\theta_\textrm{obs})$ and Lorentz factor $\Gamma(\theta_\textrm{obs})$ into the on-axis equations presented here (this is true as well for the luminosity equations below). 

\subsection{Luminosity equations for PLS G}

As in \citet{OConnor2020}, we integrate over the $0.3$\,$-$\,$10$ keV energy range of \textit{Swift}/XRT to calculate the total observed X-ray luminosity (here, we have assumed $\nu_\textrm{m}$\,$<$\,$\nu_\textrm{X}$\,$<$\,$\nu_\textrm{c}$; PLS G): 
\begin{align}
   %L_\textrm{X}= 7.9\times 10^{42}\, \varepsilon_{e,-1}^{6/5}\varepsilon_{B,-4}^{4/5} E_{\textrm{kin,0},52}^{1.3}n_{-3}^{1/2}t_{\textrm{day}}^{-0.9}(1+z)^{1.3}\, \textrm{erg/s}, \label{eqn: Lx not peak yet}
   L_\textrm{X}= 2.5\times 10^{42}\, \varepsilon_{e,-1}^{6/5}\varepsilon_{B,-4}^{4/5} E_{\textrm{kin,0},52}^{1.3}n_{-4}^{1/2}t_{\textrm{day}}^{-0.9}(1+z)^{1.3}\, \textrm{erg/s}, \label{eqn: Lx not peak yet}
\end{align}
where we have applied $p$\,$=$\,$2.2$. 

For an on-axis observer, the observed luminosity peaks at the deceleration time $t_\textrm{dec,0}$ (Equation \ref{eqn:tdec}), which yields 
\begin{align}
%L_\textrm{dec,0}=3.5\times 10^{44}\, \varepsilon_{e,-1}^{6/5}\varepsilon_{B,-4}^{4/5} E_{\textrm{kin,0},52} n_{-3}^{4/5}\Gamma_2^{12/5}(1+z)^{2/5}
L_\textrm{dec,0}=5.6\times 10^{43}\, \varepsilon_{e,-1}^{6/5}\varepsilon_{B,-4}^{4/5} E_{\textrm{kin,0},52} n_{-4}^{4/5}\Gamma_{0,2}^{12/5}(1+z)^{2/5}
\textrm{erg/s}.
\label{eqn:Ldec}
\end{align}
The luminosity due to material along the observer's line-of-sight can be derived using Equation \ref{eqn:structdec2}. In the following sections we present the on-axis peak luminosity in other synchrotron PLS. Similar scalings to the ones provided in Equation \ref{eqn:structdec2} can be derived for these other PLS by inserting the line-of-sight values for kinetic energy $E_\textrm{kin}(\theta_\textrm{obs})$ and Lorentz factor $\Gamma(\theta_\textrm{obs})$. Therefore, we only provide their on-axis values below.  

\subsection{Luminosity equations for PLS H}
\label{sec:PLS_H}

In order to compute the luminosity for PLS H ($\nu_\textrm{m}$\,$<$\,$\nu_\textrm{c}$\,$<$\,$\nu_\textrm{X}$), we similarly compute the X-ray luminosity in the $0.3$\,$-$\,$10$ keV energy range:
\begin{align}
   L_\textrm{X}=& \int^{10\, \textrm{keV}}_{0.3\, \textrm{keV}} \Big(\frac{\nu}{\nu_\textrm{c}} \Big)^{-p/2}\, L_{\nu_\textrm{c}}d\nu, \nonumber \\
 =&\,6.7\times 10^{43}\, \varepsilon_{e,-1}^{8/15}\varepsilon_{B,-4}^{89/180} E_{\textrm{kin,0},52} n_{-4}^{-1/20}t_\textrm{day}^{-1.1}(1+z)\, \textrm{erg/s}, 
\end{align}
where we have applied Equation \ref{eqn:Ysimplified} for the Compton parameter and adopted $p=2.2$.

Inserting the deceleration time, we find an on-axis peak flux of 
\begin{align}
   L_\textrm{dec,0}=&3.0\times 10^{45}\, \frac{\varepsilon_{e,-1}^{8/15}\varepsilon_{B,-4}^{89/180} E_{\textrm{kin,0},52}^{19/30} n_{-4}^{19/60}\Gamma_{0,2}^{44/15}}{(1+z)^{1/10}} \textrm{erg/s}. 
\end{align}

\subsection{Luminosity equations for PLS D}
\label{sec:PLS_D}

For frequencies below the injection frequency (PLS D; $\nu_\textrm{X}$\,$<$\,$\nu_\textrm{m}$\,$<$\,$\nu_\textrm{c}$), which can be applicable at very early times to the X-ray band (relevant to this work for large $\Gamma_0$), we can parameterize our calculation in terms of the spectral luminosity at the injection frequency $L_{\nu_\textrm{m}}$: 
\begin{align}
    L_{\nu_\textrm{m}} = 2.9\times10^{28}\,(1+z)\,E_{\textrm{kin,0},52}\,\varepsilon_{B,-4}^{1/2} \,n_{-4}^{1/2} \,\textrm{erg/s/Hz}.
\end{align}

The X-ray luminosity in this regime is
\begin{align}
   L_\textrm{X}=& \int^{10\, \textrm{keV}}_{0.3\, \textrm{keV}} \Big(\frac{\nu}{\nu_\textrm{m}} \Big)^{1/3}\, L_{\nu_\textrm{m}} d\nu, \nonumber \\
 =&\,1.2\times 10^{47}\; \varepsilon_{e,-1}^{2/3}\varepsilon_{B,-4}^{1/3}\, E_{\textrm{kin,0},52} ^{5/6}\,n_{-4}^{1/2}t_\textrm{day}^{1/2}(1+z)^{5/6}\, \textrm{erg/s}, 
\end{align}
where we also can compute the luminosity at deceleration as 
\begin{align}
   L_\textrm{dec,0}=&\,2.1\times 10^{46}\,\varepsilon_{e,-1}^{2/3}\varepsilon_{B,-4}^{1/3} E_{\textrm{kin,0},52} n_{-4}^{1/3}\Gamma_{2}^{-4/3}(1+z)^{4/3}\, \textrm{erg/s}. 
\end{align}

% Don't change these lines
\bsp	% typesetting comment
\label{lastpage}
\end{document}